\documentclass[12pt,preprint]{aastex}

\begin{document}

\shorttitle{Spitzer Observations of NGC 2362}
\shortauthors{Currie, T., et al.}

\title{The Last Gasp of Gas Giant Planet Formation: \\A Spitzer 
Study of the 5 Myr-old Cluster NGC 2362}
\author{Thayne Currie\altaffilmark{1}, Charles J. Lada\altaffilmark{1}, Peter Plavchan\altaffilmark{2}\\ 
   Thomas P. Robitaille\altaffilmark{1}, Jonathan Irwin\altaffilmark{1,3}, and Scott J. Kenyon\altaffilmark{1}}
\altaffiltext{1}{Harvard-Smithsonian Center for Astrophysics, 60 Garden St. Cambridge, MA 02140}
\altaffiltext{2}{NASA Exoplanet Science Institute, California Institute of Technology}
\altaffiltext{3}{Institute of Astronomy, University of Cambridge, Cambridge, UK}
\email{tcurrie@cfa.harvard.edu, clada@cfa.harvard.edu, plavchan@ipac.caltech.edu, 
trobitaille@cfa.harvard.edu, jirwin@cfa.harvard.edu, skenyon@cfa.harvard.edu}
\begin{abstract}
Expanding upon the IRAC survey from \citet{Dh07},
we describe Spitzer IRAC and MIPS observations of the populous, 5 Myr-old open cluster NGC 2362.
We analyze the mid-IR colors of cluster members and compared their 
spectral energy distributions to star+circumstellar disk models to constrain 
the disk morphologies and evolutionary states.
Early/intermediate-type confirmed/candidate cluster members either have photospheric mid-IR emission 
or weak, optically-thin infrared excess emission at $\lambda$ $\ge$ 24 $\mu m$ 
consistent with debris disks.  Few late-type, solar/subsolar-mass stars have primordial disks.
  The disk population around late-type stars is dominated by 
 disks with inner holes (canonical 'transition disks') and 'homologously depleted' disks.  
Both types of disks represent an intermediate stage between primordial disks and debris disks.  
Thus, in agreement with previous results, we find that 
multiple paths for the primordial-to-debris disk transition exist.  Because 
these 'evolved primordial disks' greatly outnumber primordial disks, our results 
undermine standard arguments in favor of a $\lesssim$ 10$^{5}$ year timescale for 
the transition based on data from Taurus-Auriga.  Because the typical transition timescale is 
far longer than 10$^{5}$ yr, these data also appear to rule out standard 
UV photoevaporation scenarios as the primary mechanism to explain the transition.
Combining our data with other Spitzer surveys, we investigate the evolution of 
debris disks around high/intermediate-mass stars and investigate timescales for 
 giant planet formation.  Consistent with \citet{Cu08a}, the luminosity of 24 $\mu m$ 
emission in debris disks due to planet formation
 peaks at $\approx$ 10--20 Myr.  If the gas and dust in disks evolve 
on similar timescales, the formation timescale for gas giant planets surrounding 
early-type, high/intermediate-mass ($\gtrsim$ 1.4 M$_{\odot}$) 
stars is likely 1--5 Myr.  Most solar/subsolar-mass 
stars detected by Spitzer have SEDs that indicate their disks 
may be actively leaving the primordial disk phase.  Thus, gas giant 
planet formation may also occur by $\sim$ 5 Myr around solar/subsolar-mass 
stars as well.
\end{abstract}
\textit{Facility: }\facility{Spitzer Space Telescope}
\section{Introduction}
Primordial circumstellar disks of gas and small dust grains 
surround the vast majority of $\approx$ 1 Myr-old stars 
and are the building blocks of planets \citep[][]{La92, Kh95}.  
Observable as emission in excess of the stellar photosphere at 
infrared wavelengths, primordial disks accrete gas onto the 
host star and have strong, optically-thick emission comparable 
in luminosity to the stellar photosphere (L$_{d}$/L$_{\star}$ $\sim$ 0.1--1).  
Small dust grains in primordial disks comprise the seed 
material for the cores of planets; the gaseous envelopes of 
gas giant and ice giant planets form from the circumstellar gas in these disks.

Primordial disks disappear on $\sim$ 3--10 Myr timescales.  In clusters with 
ages of $\sim$ 3 Myr, many disks have much less emission than 
typical primordial disks at wavelengths $\lesssim$ 10 $\mu m$ \citep[e.g.][]{La06, He07}.
 Primordial disks disappear around nearly all 
 stars by $\sim$ 10--15 Myr \citep[e.g.][]{Cu07a}.  Disks surrounding 
$\ge$ 10 Myr-old stars are typically optically-thin 
\textit{debris disks} with weaker infrared emission \citep{Ri05, Ch05, Cu07a, Cu08a, Cpk08, For08, Hi08, Pl08} and 
little evidence for substantial reservoirs of circumstellar gas \citep[e.g.][]{Da05, Cu07c}.  
Because dust from debris disks can be removed by stellar radiation 
on very short ($\lesssim$ 0.1 Myr) timescales \citep[e.g.][]{Bu79, Tak01, Pl05}, the dust requires 
an active replenishment source from collisions between larger objects.  Thus, 
debris disks are signposts for active planet formation \citep[e.g.][]{Bp93, Kb04, Kb08}.

While debris disks may trace active terrestrial and icy planet formation, they 
do not trace the final stages of active gas giant planet formation, where massive 
cores accrete large gaseous envelopes.  
Debris disk have very low masses of residual gas, 
$<<$ 1 M$_{\oplus}$, \citep[e.g.][]{ Zu95, Ck04, Ro07, Fr07}.  This 
mass is too small to contribute to a gaseous envelope with a mass
 comparable to those found in Jupiter and Saturn.  

The timescale for the disappearance of primordial disks and the subsequent dominance of 
debris disks has important implications for gas giant planet formation. 
  In the core accretion theory for gas 
giant planet formation \citep[e.g.][]{Po96}, $\gtrsim$ Mars-mass 
protoplanetary cores can accrete a gaseous envelope, which is in hydrostatic 
equilibrium.  However, once the cores reach $\approx$ 5--10 M$_{\oplus}$ 
in mass, they can rapidly accrete a much more massive envelope.  
If the nebular gas disperses before 
the critical core mass is reached, gas giants fail to form.  Typical timescales 
for massive cores to form range from $\approx$ 1 Myr \citep{Raf04, Cu05, Kb08b} to 
10 Myr \citep{Po96, Rice03}.  Because debris disks are gas poor, the time  
 when most primordial disks evolve into debris disks sets an empirical upper limit 
for the formation timescale of gas giant planets.

Identifying \textit{how} primordial disks evolve into 
debris disks may also have important implications for planet formation.  
Early work from IRAS and ground-based data
 (specifically Taurus-Auriga) suggested that $\sim$ 1--10\% of $\approx$ 
1 Myr-old stars have so-called 'transition disks': disks with 
levels of mid-infrared emission in between 
those of stars surrounded by primordial disks and stars lacking a disk as 
determined from color-color diagrams
 \citep[e.g.][]{Har90,Sk90,Sp95, Ww96}.  Later studies of 'transition' 
disks began to associate the term with a specific disk morphology: disks 
with inner regions that are optically thin and thus depleted or devoid 
of gas and dust but colder, outer regions with optically thick emission 
\citep[e.g.][]{Cal05}. If all disks follow make this transition, then disks 
are cleared from the inside out with a transition 
time of $\sim$ 10$^{4}$--10$^{5}$ years.  Plausible physical processes 
for removing primordial disks must then reproduce this rapid transition timescale.  
 UV photoevaporation models are successful in clearing disks within 
$\sim$ 0.01--0.1 Myr \citep{Cl01,Al06}.

However, recent spectral energy distribution (SED) analysis of some 2--3 Myr-old stars 
indicates that they may follow a different 
evolutionary sequence from primordial disks to debris disks \citep{La06, Ck08}.
   Their near-to-mid IR fluxes are consistent with a 'homologous depletion' in disk emission at all wavelengths 
through 24 $\mu m$ instead of an 'inside out' evolution, indicative of a reduced mass of small ($\lesssim$ 100 $\mu m$ -- 
1 mm) dust grains at all stellocentric distances \citep{Wo02} that could result from the coagulation of dust into 
1 mm -- 1km-sized planetesimals and depletion of solid material by accretion onto the star.  
If many disks evolve from primordial disks to debris disks along 
 this alternate path, then the primordial-to-debris disk transition timescale
 estimated from the frequency of canonical 'transition disks' will 
be underestimated.  Furthermore, if most primordial disks last longer than $\sim$ 1 Myr, the frequency of 
'transition disks' around 1 Myr-old stars may be low either because the transition timescale is fast \textit{or} 
 because most disks simply haven't had enough time to start leaving the primordial disk 
phase.  If the latter is the correct interpretation, then the transition timescale 
for many disks could be much longer.  Observations of disks at slightly greater, $\sim$ 4--5 Myr 
ages can further test whether there are multiple pathways from primordial disks to debris disks 
and can better constrain the primordial-to-debris disk transition timescale.

In this paper, we investigate the primordial-to-debris disk transition in 
the open cluster NGC 2362 in Canis Major ($\alpha_{2000}$ = 
7$^{h}$18$^{m}$48$^{s}$, $\delta_{2000}$ = -24$^{o}$57'00.0").  
  With an age of $\sim$ 4--5 Myr \citep{Dahm05, Mn08}, NGC 2362 lies between 
the epoch when most stars have optically-thick primordial disks ($\sim$ 1 Myr) and 
the epoch where primordial disks are exceedingly rare ($\sim$ 10 Myr).  
While the cluster is very distant ($\sim$ 1.48 kpc), 
it is also massive ($>$ 500 M$_{\odot}$). 
Because of its mass, even nondetections of cluster 
stars with the MIPS 24 $\mu m$ channel can help constrain the source 
spectral energy distributions (SEDs) and thus the disk evolutionary state 
\citep{Cu07b}.

This paper builds upon the study of NGC 2362 by \citet{Dh07} who focus 
on IRAC observations of 337 candidate/confirmed cluster members and 
find evidence for disks surrounding many late-type stars.
Here we add MIPS 24 $\mu m$ data to extend the previous Spitzer 
wavelength coverage of cluster members.
We also rereduce and reanalyze IRAC data and expand the membership 
sample so that we can include candidate cluster members not studied by \citet{Dh07}.
 MIPS 24 $\mu m$ data probes disk emission that can come from dust colder 
than the ice sublimation temperature (T $\sim$ 170 K). Therefore, 
 our knowldedge of the disk population now extends
from the warm, inner terrestrial zone regions to cooler, outer regions where 
gas giant planets may form.  The combination of MIPS and IRAC data allow us to better discriminate 
between disk evolutionary states.  

We organize our study as follows.  In \S 2, we describe IRAC and MIPS photometry of the
NGC 2362 field and ancillary data used to identify Spitzer counterparts to known and 
candidate cluster members. \S 3 focuses on the IRAC+MIPS colors of 
(candidate) cluster members, providing a first-order constraint on the 
mid-IR properties of disks in NGC 2362.  To constrain the evolutionary state 
of MIPS-detected cluster members, we compare the observed SEDs 
to predictions for (evolved) primordial disks and debris disks in \S 4.  \S 5 draws 
inferences about the timescale for gas giant planet formation and the primordial-to-debris 
disk transition from these data by comparing them with Spitzer results for other 
$\lesssim$ 10--15 Myr-old clusters.  We conclude with a brief summary of the paper, discuss 
our results in the context of other recent work, and discuss future 
work that can build on our results.

\section{Data}
\subsection{Spitzer Observations, Image Processing, and Photometry}
\subsubsection{IRAC Data}
 NGC 2362 was observed with the Infrared Array Camera 
\citep[IRAC;][]{Faz04} on March 3, 2004 as a part of the 
\textit{Guaranteed Time Observations} program (Program ID 37).
IRAC data cover an area of $\sim$ 0.3 square degrees centered 
near $\alpha_{2000}$ $\sim$ 07$^{h}$19$^{m}$00$^{s}$,
$\delta_{2000}$ $\sim$ -24$^{o}$27'00".
IRAC observations were taken in 
high-dynamic range mode with 0.4s and 10.4.s exposures.  Sources were 
observed at two dither points yielding an average integration time/pixel of 
$\sim$ 20.8s for the long exposures.  

The IRAC Basic Calibration Data (BCD) require additional post-BCD processing before mosaicing.  
Specifically, the BCD frames at [3.6] and [4.5] show strong evidence for 
column pulldown and 'striping' effects, which affect pixel flux densities by 
up to $\approx$ 0.5 MJy/sr.  The [5.8] and [8] channel BCD frames show evidence for 
cosmic-ray hits, which are especially numerous in the [5.8] filter.
Finally, the automatically produced post-BCD mosaics at [3.6] and [4.5] contain
 background level mismatch of up to $\sim$ 0.1 MJy/sr near bright stars, 
which causes a noticeable 'tiling' effect.

We cleaned the IRAC data according to the methods described 
by \citet{Cpk08}, applying the array-dependent correction for point sources \citep{Qu04} 
and mitigating image artifacts using modified versions of the \textit{muxstriping} and \textit{column pulldown} 
algorithms available as contributed software on the \textit{Spitzer Science Center} website.  
Final image processing and mosaicing was performed using the MOPEX/APEX pipeline \citep{Mak05b}.
Prior to mosaicing, we applied the overlap corrector with bright source masking and 3$\sigma$ 
outlier clipping to match the background levels between BCD frames.  
For cosmic-ray rejection, we used the r-masks for each BCD frame supplemented by 
the \textit{mosaic outlier} and \textit{dual outlier} modules, which removed 
nearly all cosmic rays.  Finally, both the IRAC and MIPS data were mosaiced together 
using a bicubic interpolation with outlier rejection.
  The final mosaics were inspected and cleared of  
 image artifacts or pattern noise that could compromise the photometry.  

For the IRAC photometry, we perform source finding and aperture photometry using the IDLPHOT 
package and SExtractor \citep{Be96}.  For source identification, we determined the 
background rms level ($\sigma_{bkgd, rms}$) for each IRAC mosaic using SExtractor.  Sources were identified 
as groups of pixels with fluxes $\ge$ 3$\sigma_{bkgd, rms}$ in all filters using the 
\textit{find.pro} routine.  These sources lists were used as input for aperture photometry 
with the \textit{aper.pro} routine.  The source flux was computed in a 2 pixel ($\sim$ 2.44") 
and 3 pixel ($\sim$ 3.66") aperture radius with the local background calculated from 
a 4 pixel-thick annulus surrounding each source extending from 2--6 and 3--7 pixels.  
We multiplied the source fluxes by the appropriate aperture corrections given in the IRAC 
data handbook, version 3.0\footnote{http://ssc.spitzer.caltech.edu/irac/dh/iracdatahandbook3.0.pdf}.  
We fine-tuned the aperture correction by first comparing photometry for bright, unsaturated ($\sim$ 10th--12th magnitude) 
stars derived above with photometry derived using a 10 pixel aperture and background annulus of 
12 to 20 pixels, which requires no aperture correction.  Additionally, we then compared our results  
with photometry for bright stars measured using a 5 pixel aperture with a 12--20 pixel background annulus.

Photometry from the 2 and 3 pixel apertures show excellent agreement
through $\sim$ 13th (14th) magnitude for the [5.8] and [8] ([3.6] and [4.5]) 
channels with dispersions of $<<$ 0.025 magnitudes.  The dispersion
increases to $\sim$ 0.1--0.2 mags beyond these limits.  The measured pixel 
area of $\ge$ 13th magnitude stars is enclosed by the 2-pixel 
aperture radius.  Using the 2-pixel aperture may introduce $\sim$ 1--2\%
uncertainties in the spatial variation of the aperture correction but results in 
a higher signal-to-noise, which is important for obtaining accurate 
photometry of stars whose fluxes are $<$ 10$\sigma_{bkgd, rms}$.  
Furthermore, NGC 2362 is a massive cluster projected against the densely populated 
galactic plane.  Source crowding in the sensitive [3.6] and [4.5] filters 
can be severe.  Several stars in the cluster center appear partially blended 
with the three-pixel aperture used.
Therefore, we chose photometry from the 2-pixel aperture in all filters for all stars.
The final IRAC catalogs were trimmed of sources lying within 5 pixels of the 
image edges.  We combine the IRAC catalogs with the 2MASS JHK$_{s}$ catalog 
using a 2" matching radius, resulting in 6,916, 7,075, 4,875, and 3,855 matches 
in the [3.6], [4.5], [5.8], and [8] channels

\begin{figure}
\centering
\plotone{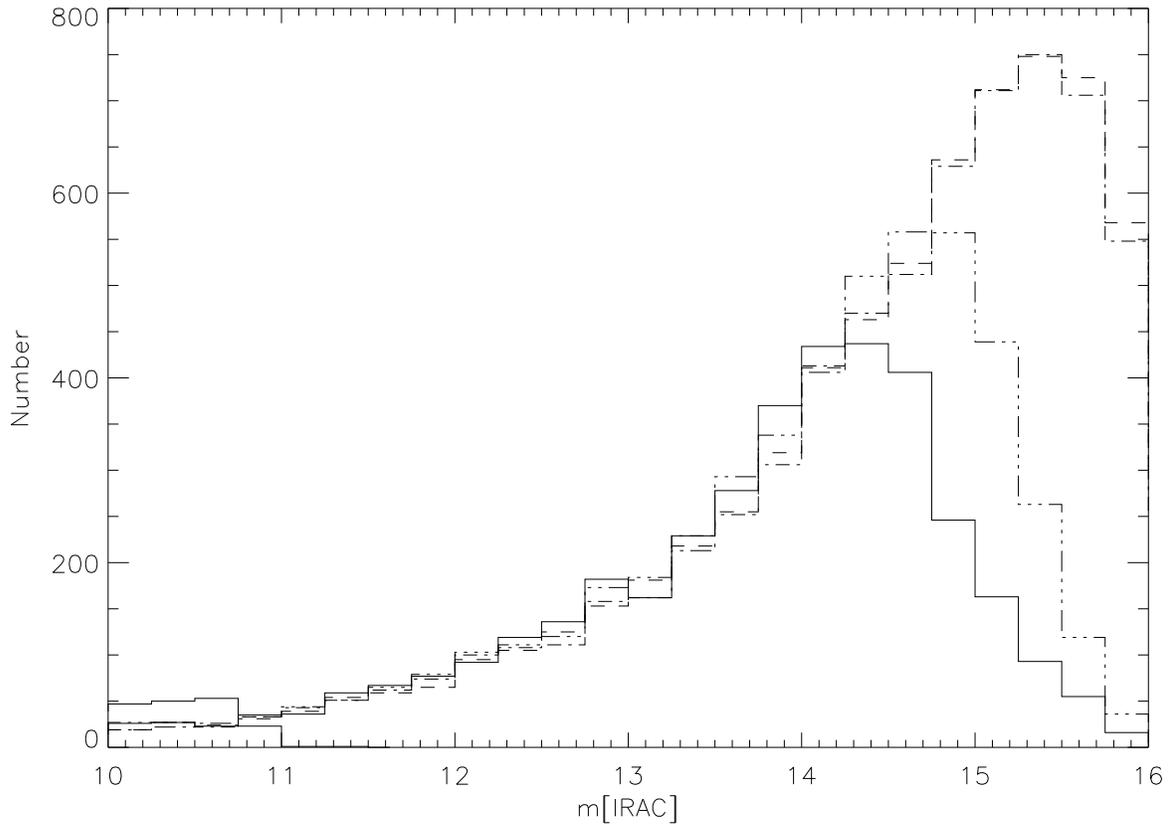}
\caption{Histogram of IRAC detections with 2MASS counterparts.  The 5.8 $\mu m$ (dash-three dots) 
and 8 $\mu m$ (solid) source counts peak at m[IRAC] $\sim$  15 and 14.25, respectively.  
The 3.6 $\mu m$ (dash-dot) and 4.5 $\mu m$ (dashed line) both peak at m[IRAC] $\sim$ 15.5. }
\label{iracdist}
\end{figure}

The photometric errors are calculated from the Poisson error in the 
source counts, the read noise, the Poisson error in the background 
level, and the uncertainty in the 
background\footnote{See https://lists.ipac.caltech.edu/pipermail/irac-ig/2007-February/000037.html}.  
Figure \ref{iracdist} shows the number 
counts of IRAC-detected sources with 2MASS counterparts as a function of magnitude.  
The source counts peak at [3.6] $\sim$ 15.5, [4.5] $\sim$ 15.5, [5.8] $\sim$ 15.0, 
and [8] $\sim$ 14.25--14.5.  IRAC approaches the 10$\sigma$ limits at [3.6] $\sim$ 15.75, 
[4.5] $\sim$ 15.75, [5.8] $\sim$ 13.5, and [8] $\sim$ 13.25.  The 5$\sigma$ limits 
for [5.8] and [8] are [5.8] $\sim$ 14.8 and [8] $\sim$ 14.4.  Source counts 
for [3.6] and [4.5] peak before the 10$\sigma$ limit is reached because of 
incompleteness in the 2MASS catalog.
\subsubsection{MIPS 24 $\mu m$ Data}
NGC 2362 was imaged with the Multiband Imaging Photometer for Spitzer
\citep[MIPS;][]{Rie04} on March 17, 2004 also as a part of 
the GTO program (ID 58).  MIPS data cover
$\sim$ 0.5 square degrees on the sky with boundaries of $\alpha_{2000}$ $\sim$
07$^{h}$17$^{m}$45$^{s}$ to 07$^{h}$20$^{m}$00$^{s}$ and $\delta_{2000}$ $\sim$
-25$^{o}$26$\prime$00" to -24$^{o}$33$\prime$00".
MIPS 24 $\mu m$ observations were taken in scan mode with an 
integration time per pixel of 80s.  Background cirrus levels at 24 $\mu m$
 are low ($\approx$ 19 MJy/sr) and vary by less than $\sim$ 
0.1 MJy/sr for most of the field.  Two regions centered on 
$\alpha_{2000}$ $\sim$ 07$^{h}$18$^{m}$8$^{s}$, $\delta_{2000}$ $\sim$ -25$^{o}$16$\prime$00" and 
$\alpha_{2000}$ $\sim$ 07$^{h}$19$^{m}$30$^{s}$, $\delta_{2000}$ $\sim$ -24$^{o}$51$\prime$30" 
have slightly higher background levels of $\approx$ 19.5 MJy/sr.  
The MIPS BCD data are largly free of artifacts.  

\begin{figure}
\epsscale{0.9}
\centering
\plotone{f2a.eps}
\plotone{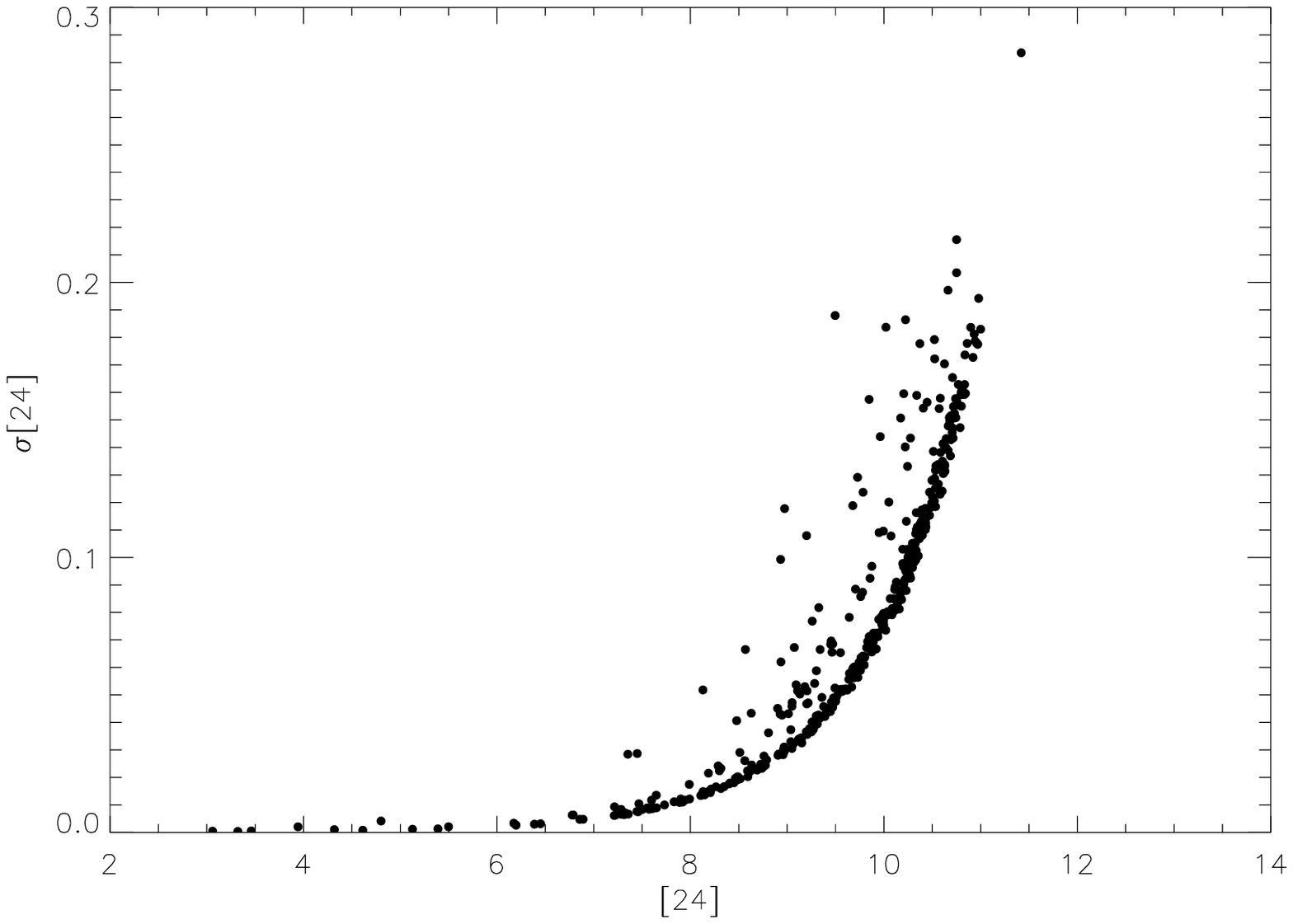}
\caption{(top) Histogram of MIPS 24 $\mu m$ detections (solid line) and 
sources with 2MASS counterparts (filled region). (bottom) Error distribution of 24 $\mu m$ detections 
with 2MASS counterparts.  Sources with photometric errors slightly greater than the main distribution 
typically lie closer to the image edges.}
\label{24dist}
\end{figure}

\begin{figure}
\centering
\plotone{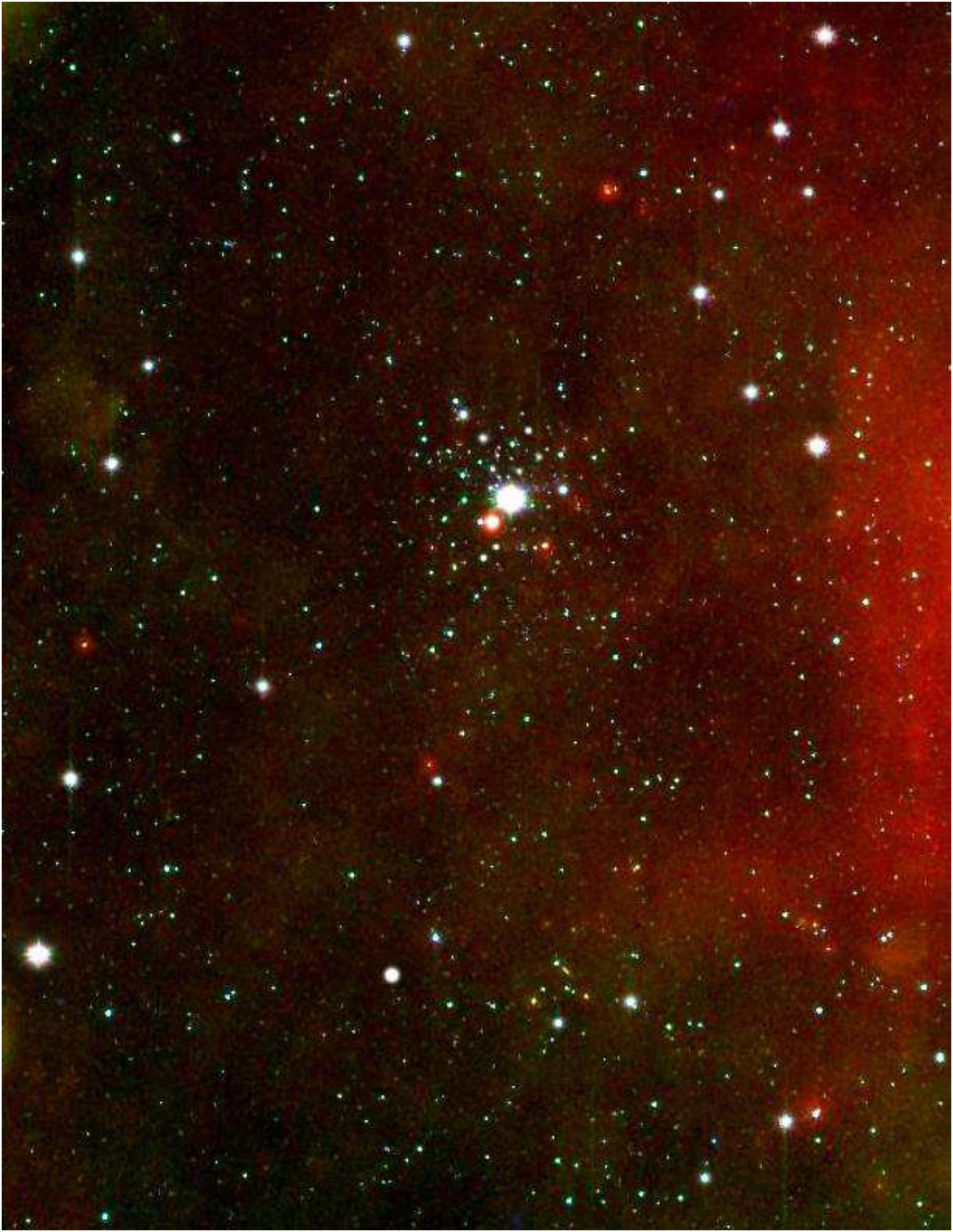}
\caption{False-color image of NGC 2362 made with the PBCD mosaic images at 
4.5 $\mu m$ (blue), 8 $\mu m$ (green), and 24 $\mu m$ (red).  The bright star 
$\tau$ CMa is shown near the image center.}
\label{3color}
\end{figure}

 Because the MIPS point-spread function is well characterized, we perform photometry on the 24$\mu m$ MIPS
data using pixel response function (PRF) fitting with APEX \citep{Mak05}.  We detect 1,761 sources at [24] with 
S/N $>$ 3.  Using a 2" matching radius, we identify 447 MIPS 24 $\mu m$ sources with 
2MASS counterparts.  Figure \ref{24dist} shows that the source counts peak at [24] $\sim$ 10.5 ($\sim$ 460 $\mu$Jy); the 
 (5$\sigma$) 10$\sigma$ limits are [24] $\sim$ (10.5) 9.75\footnote{The uncertainties quoted here
do not include the zero-point uncertainty, which is $\approx$ 4\%.}.
Table \ref{allphot} lists the full photometric catalog of 8,642 sources on the IRAC/MIPS fields with 
2MASS counterparts and a detection in at least one IRAC and/or MIPS channel.
Figure \ref{3color} shows a three-color image of NGC 2362 made with the 4.5 $\mu m$ and 8 $\mu m$ 
IRAC filters (blue and green) and MIPS 24 $\mu m$ filter (red).  Most 
sources discussed in this paper lie within $\sim$ 7$\prime$ of $\tau$ CMa, near the center of the mosaic 
and interior to the region with slightly higher 24 $\mu m$ background (righthand side of the figure).
%

\subsection{Ancillary Data: Identifying Cluster Members}
In the absence of proper motion data, we establish cluster membership by 
identifying stars with indicators of youth.  Most intermediate and late-type 
young stars are chromospherically active and thus are luminous 
X-ray sources \citep[e.g.][]{Pr05}.   Chromospheric activity as well as 
circumstellar gas accretion onto a star can result in H$_{\alpha}$ emission, a 
common feature in the optical spectra of stars with ages $\lesssim$ 10 Myr 
\citep[e.g.][]{Kh95, Da05, Cu07c}.  Young low-mass stars are also identified 
as having strong Li 6708 \AA\ absorption.  Finally, optical color-magnitude 
diagrams (e.g. V/V-I) can determine the cluster locus and thus 
identify candidate members, especially for massive clusters \citep[e.g.][]{Ly06, Cu08c}. 

In this section, we describe ancillary data for NGC 2362 that identifies confirmed and candidate cluster members.  
We draw our primary list of confirmed and candidate members from \citet{Dh07} identified by Li 6708 \AA\ absorption, 
X-ray activity, H$_{\alpha}$ emission and optical photometry.  To this list, we add a second list of candidate members 
identified by deep optical photometry from \citet{Ir08}.  
\subsubsection{Optical Photometry and Membership List from \citet{Dh07}}
\citet{Dh07} list 232 members from a sample of 
stars with H$_{\alpha}$ emission, Li 6707 \AA\ absorption, 
and X-ray activity.
About 20 $\%$ of these stars show evidence of youth from 
all three indicators, $\approx$ 36 \% have at least two indicators 
of youth, and nearly half ($\approx$ 44 \%) have just one 
indicator.
To this list, \citet{Dh07} add 105 sources whose optical colors/magnitudes 
are consistent with colors of $\sim$ 5 Myr-old stars and are thus 'candidate' members.  While none of these stars 
have detected H$_{\alpha}$ emission, Li absorption, or X-ray activity, many of them 
are too faint (V $\gtrsim$ 21) to be studied with spectroscopy or X-ray surveys.  Other stars are 
 too early in spectral type for the nondetections 
of H$_{\alpha}$ emission, Li absorption, or X-ray activity to rule out 
membership.  


Using a 2" matching radius, we identify all cluster members in at 
least one IRAC channel and 39 cluster members with MIPS.  
Increasing the radius to 3" does not yield any new MIPS matches and increases 
the likelihood of contamination by background galaxies. 
Thus, about 12 \% of the known cluster members are detected by MIPS: 
analysis of the mid-IR properties of cluster members (\S 3) is limited by many 
upper limits.  Table \ref{dhmemphot} lists the combined 
optical/IR (IRAC and MIPS) photometry of members from the \citet{Dh07} catalog, using 
IRAC photometry from this paper.

While the membership list from \citet{Dh07} is robust, it is 
spatially limited and does not include the lowest-mass stars.  The grism spectra used to identify 
H$_{\alpha}$ emitters are restricted to stars within a 11'x11' region surrounding 
 the cluster center (near $\tau$ CMa; $\alpha$ = 07$^{h}$ 18$^{m}$ 40.4$^{s}$, 
$\delta$ = -24$^{o}$ 33' 31.3").
Chandra/ACIS observations which detect identify X-ray active members 
are limited to a 16.9' x 16.9' area surrounding the cluster center.  
Finally, \citet{Dahm05} identifies candidate cluster members from photometry 
for stars with V $\lesssim$ 21 within 7' of the cluster center.   

According to \citet{Dahm05}, cluster members with H$_{\alpha}$ emission, 
 X-ray activity, and/or Li absorption are very centrally concentrated.  
The surface density of members drops by a factor of two at a cluster radius of $\approx$ 2' and merges 
with background by $\approx$ 6'--7'.  
However, many massive clusters 
\citep[e.g. h and $\chi$ Persei;][]{Cu07a} contain a low-density halo/distributed population of 
young stars located many core radii away from the cluster centers.   
The halo population may contain as much mass as the bound cluster population \citep{Cu08c}.
\subsubsection{Optical Photometry and Expanded List of 
Candidate Members from \citet{Ir08}}
To identify candidates for fainter members of the cluster 
and members of a surrounding halo-population, we rely on
 the deep optical photometry of NGC 2362 from \citet{Ir08}.  As a part of the 
MONITOR program \citep{Ai07}, \citet{Ir08} imaged a 36' x 36' field 
centered on $\tau$ CMa with the MOSAIC Imager on the 4m Blanco telescope 
at CTIO.  The \citeauthor{Ir08} catalog includes $\approx$ 56,000 stars 
with 5$\sigma$ detections at a magnitude limit of I $\sim$ 23.6.

\citet{Ir08} identify $\sim$ 1,826 stars brighter than V $\sim$ 25
whose positions on the V/V-I color-magnitude diagram are consistent with 
cluster membership.  About 1,465 of these stars lie outside the \citet{Dahm05} survey 
limits, greater than 7' away from the NGC 2362 cluster center.  All 
of these stars are fainter than V $\sim$ 16.5\footnote{Brighter stars are saturated.}.  
Using a simulated catalog of field objects from the Besacon Galactic 
models \citep{Rob03}, \citet{Ir08} estimate a contamination level of 
$\sim$ 65 \%, which implies that NGC 2362 contains a population 
of $\sim$ 600-650 faint stars, $\sim$ 500 of which lie in a halo population
 surrounding the cluster.  Thus, the candidate member list from \citet{Ir08} may 
identify new members.

Because the \citet{Ir08} catalog may include cluster members, combining their 
source list with the MIPS detections may reveal additional cluster stars with 
mid-IR excess emission from disks. In total, 1,143 stars from \citet{Ir08}  
have IRAC and/or MIPS counterparts and are not in the \citet{Dh07} 
membership sample.  Nearly all (1,120) of these stars are located beyond the bound 
cluster region ($\le$ 7' from the cluster core).
Table \ref{irmemphot} lists the optical, 2MASS, and Spitzer photometry of these 
candidate members.
\section{2MASS/IRAC and MIPS Colors of NGC 2362 Cluster Members and Candidate Members}
To identify disk-bearing stars in NGC 2362, 
we analyze the mid-IR colors of confirmed and new candidate cluster members
\footnote{No source has a galaxy contamination flag in the 
2MASS Point Source Catalog, indicating that its emission overlaps with that of an extended source.}.  
We use IRAC colors to identify stars with warm dust emission 
from the inner disk.  Then we analyze MIPS colors to identify stars with colder emission. 
Comparing the IRAC and MIPS colors provides a first-order investigation 
of the NGC 2362 disk population.
Comparisons between our photometry and that from \citet{Dh07} are discussed in the Appendix.

\subsection{IRAC Colors}
Figures \ref{jkk34dh} and \ref{jkk34ir} show the distributions
 of K$_{s}$-[5.8] (left panels) and 
K$_{s}$-[8] (right panels) colors for NGC 2362 stars from the \citet{Dh07} 
sample (Figure \ref{jkk34dh}) and candidates from 
\citet[][Figure \ref{jkk34ir}]{Ir08}.  
To show how these colors vary with stellar properties, 
we plot them against the 2MASS J--K$_{s}$ color.
For cluster stars with similar extinction, the 
observed J--K$_{s}$ serves as a good tracer of spectral type.  Late-type K and M
stars have red J--K$_{s}$ colors ($\gtrsim$ 0.6); early-type B and A 
stars have J--K$_{s}$ $\approx$ 0.  The positions of these stars in 
optical color-magnitude diagrams define a tight locus \citep{Dh07,Moi01}, 
which suggests that the cluster lacks a large age spread.  
Therefore, the J--K$_{s}$ color also correlates well with stellar mass.  

To compare the observed K$_{s}$-[5.8, 8] colors to predicted photospheric colors, 
we overplot the locus of photospheric colors (dotted lines) from the STAR-PET interactive tool available
from the \textit{Spitzer Science Center} website\footnote{STAR-PET computes the 2MASS-Spitzer 
colors based on the Kurucz-Lejuene stellar atmosphere models \citep[e.g.][]{Ku93, Le97}.  
\citet{Rie08} shows that differences in IRAC and MIPS colors between synthetic and empirical 
A type and solar-type photospheres are $\lesssim$ 0.05 mags.}.  All sources with 
 photometric uncertainties of $\le$ 0.5 mags are shown.  Sources with (without) MIPS 
detections are shown as black dots (triangles).

In both panels of Figure \ref{jkk34dh}, the distribution of the IRAC colors 
follows the trends expected for a young cluster.  Most sources lie within $\sim$ 0.2 mags 
of the predicted photospheric colors, which range from K$_{s}$-[8] $\sim$ 0 for early-type 
stars (J--K$_{s}$ $\sim$ 0) to $\sim$ 0.5--0.6 for late-type stars (J--K$_{s}$ $\sim$ 0.8).  
Very few stars have K$_{s}$-[5.8, 8] $\lesssim$ 0.  The \citet{Dh07} sample lacks 
 high-mass stars (J--K$_{s}$ $\le$ 0.4) with red IRAC colors that are indicative of 
excess emission from warm dust.  In contrast, many late-type, lower-mass stars 
(J--K$_{s}$ $\gtrsim$ 0.6 -- M$_{\star}$ $\lesssim$ 1.4 M$_{\odot}$ at 
5 Myr \citep{Ba98}) show evidence for IR excess emission.  

Most stars with clear IRAC excesses have MIPS detections.
The MIPS-detected population (black dots)
has a narrower range in J--K$_{s}$ ($\sim$ 0.8--1.4) than the IRAC-detected 
population.  The IRAC colors of MIPS-detected stars also show evidence for 
a wavelength-dependent luminosity of disk emission.  MIPS-detected stars characteristically 
have K$_{s}$-[5.8] $\sim$ 1.  The K$_{s}$-[8] colors for most of these stars 
are slightly redder (K$_{s}$-[8] $\sim$ 1.25--1.75).

The distribution of IRAC colors for the \citet{Ir08} candidates 
 shows similar trends (Figure \ref{jkk34ir}).  The dispersions in K$_{s}$--[5.8] and K$_{s}$--[8] 
colors are $\sim$ 0.2 mags.  The \citet{Ir08} sample shows evidence 
for a clear IR-excess population (K$_{s}$-[5.8, 8] $\gtrsim$ 0.5--0.75), which 
is smaller than that for the \citet{Dh07} sample.
MIPS-detected stars in this sample have 
K$_{s}$--[5.8] $\sim$ 1 and K$_{s}$--[8] $\sim$ 1--2.  

\subsection{K$_{s}$-[24] Colors and Upper Limits}
\begin{figure}
\centering
\plottwo{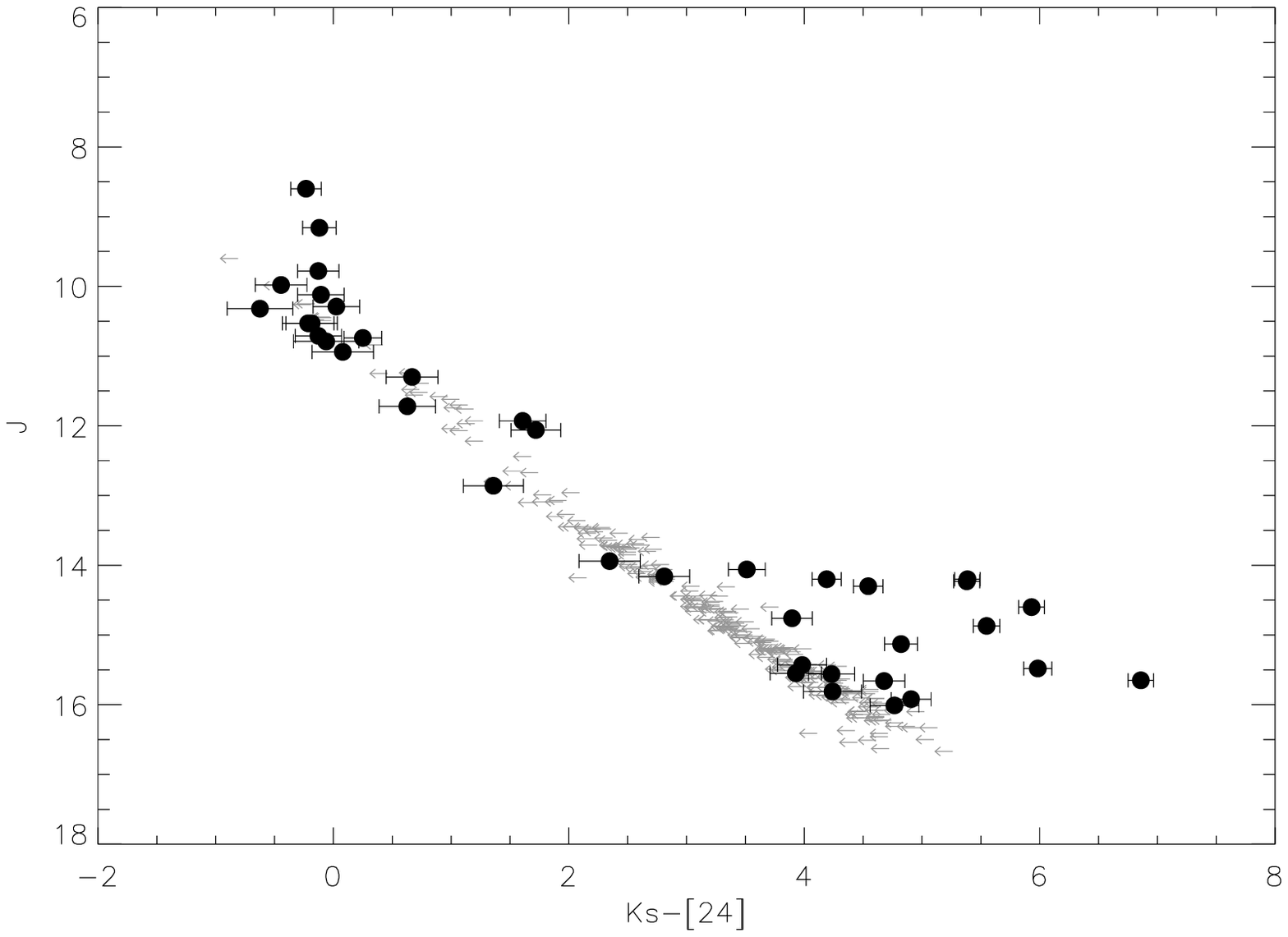}{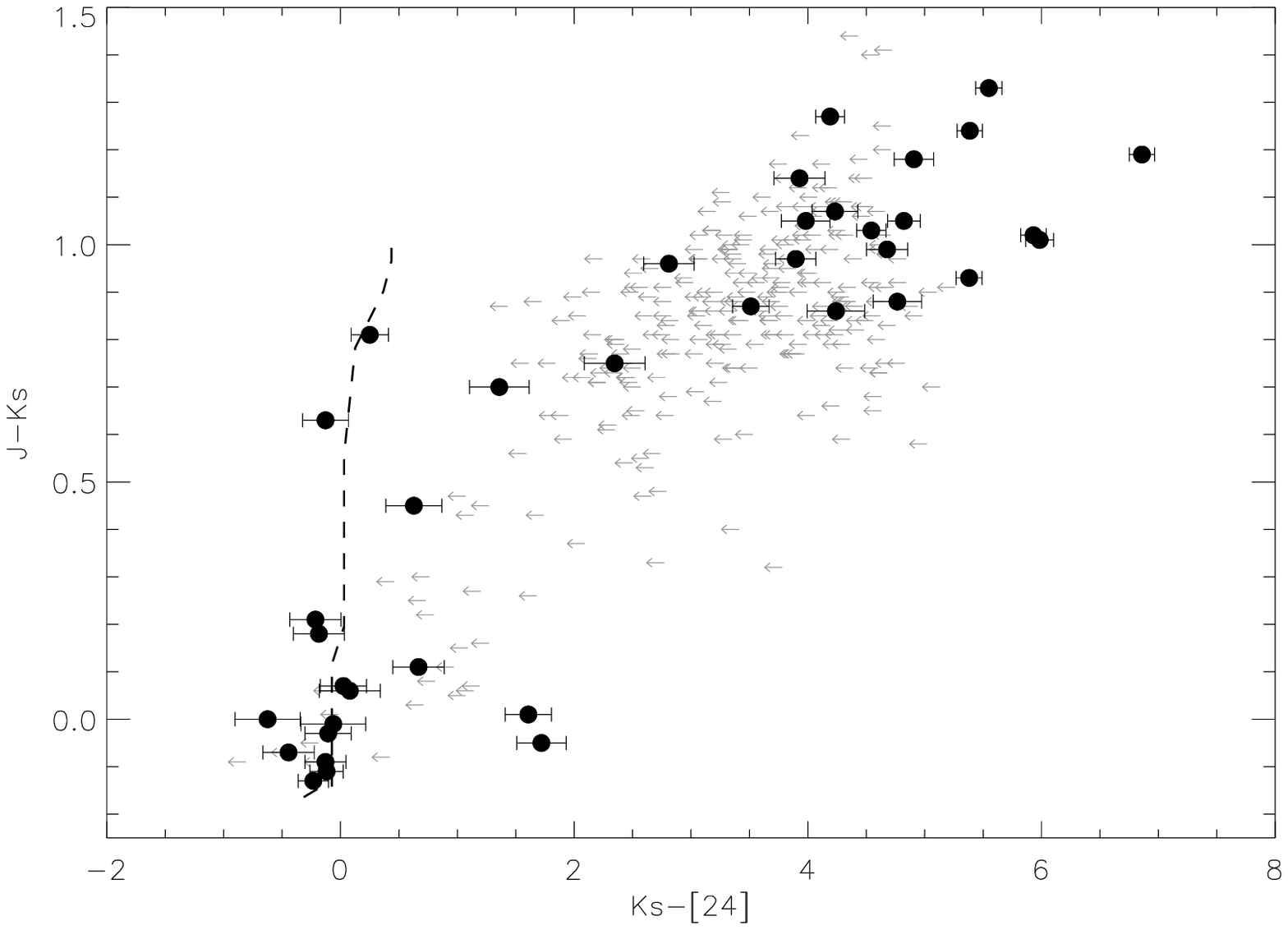}
\plottwo{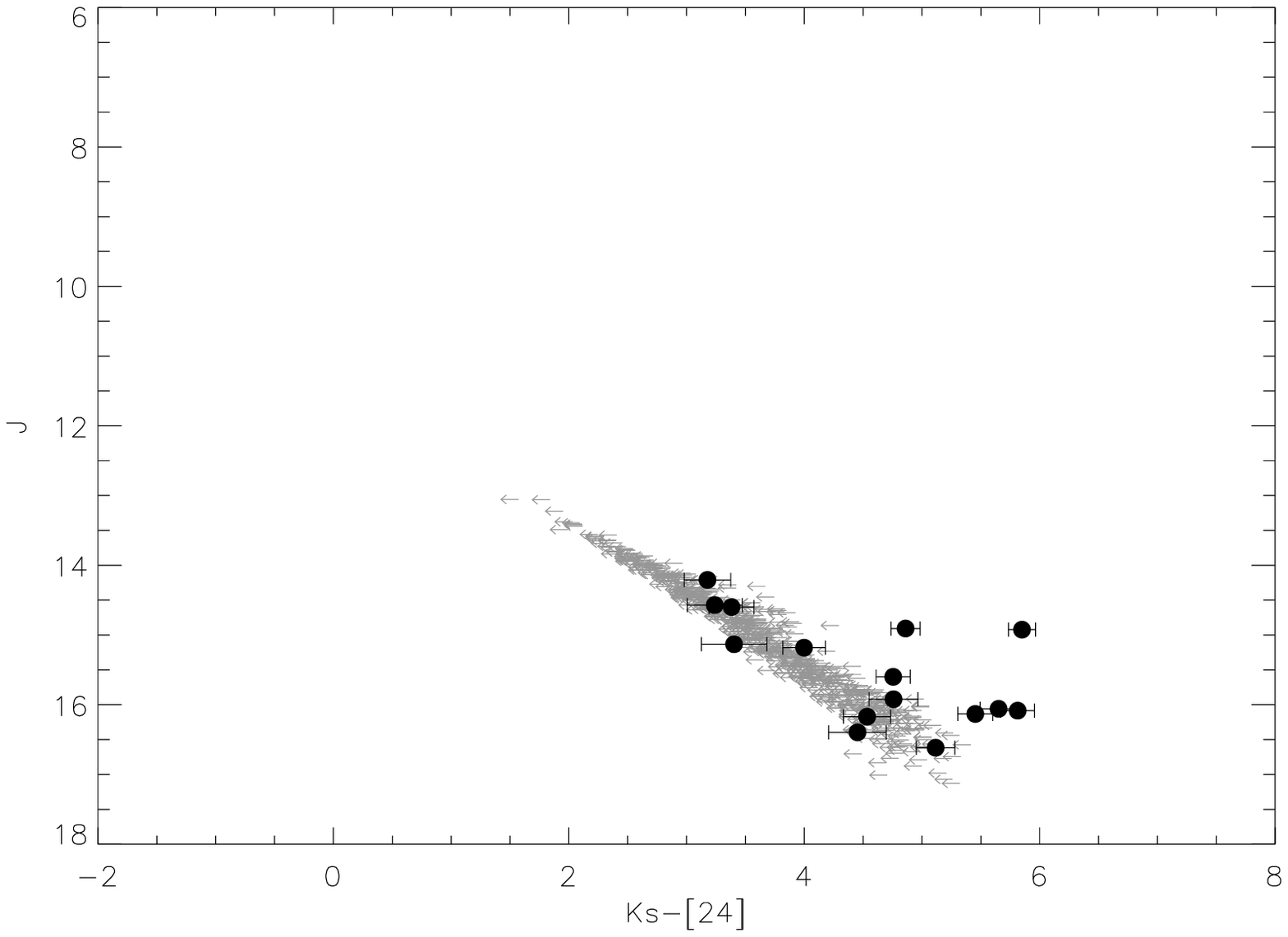}{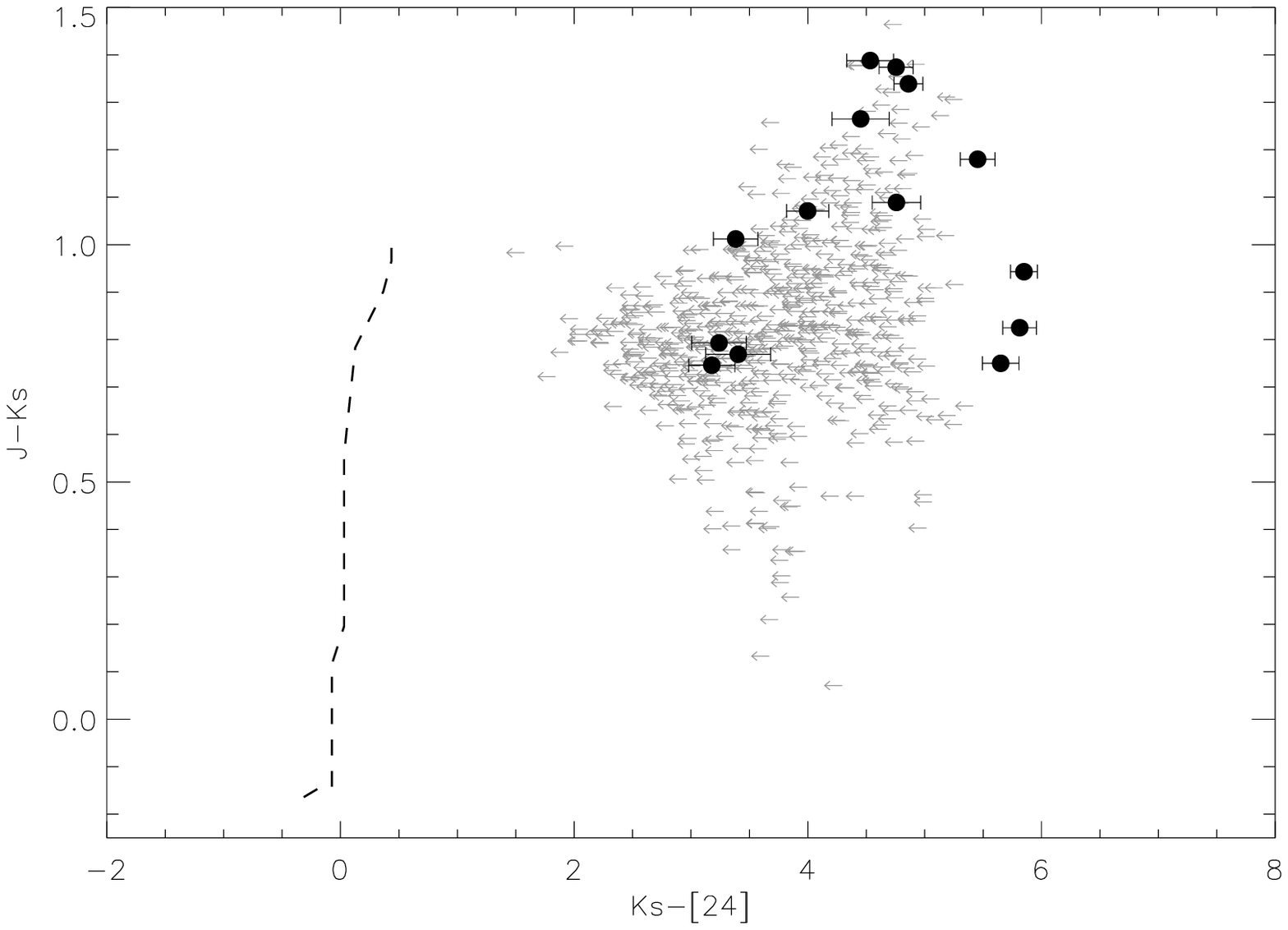}
\caption{J vs. K$_{s}$-[24] (left) and J-K$_{s}$ vs. K$_{s}$-[24] (right) color-color diagrams for 
confirmed/candidate cluster members from \citet[][top]{Dh07} and candidates from \citet[][bottom]{Ir08}.  
The leftward arrows identify K$_{s}$-[24] upper limits for sources lacking MIPS detections.}
\label{jvk24}
\end{figure}

Figure \ref{jvk24} shows the J vs. K$_{s}$-[24] and J-K$_{s}$ vs. K$_{s}$-[24] 
diagrams for confirmed/candidate cluster members from
 \citet[][top panels]{Dh07} and candidates from 
\citet[][bottom panels]{Ir08}.  
Extinction to the cluster is low \citep[avg. E(B-V) $\approx$ 0.1;][]{Bal96, Moi01} and affects 
optical fluxes far more than infrared fluxes \citep[e.g.][]{Ma90}.  Therefore, potential 24 $\mu m$ excess sources 
are not affected by small dispersions in the reddening of cluster stars.
While the \citet{Dh07} sample contains some stars with K$_{s}$-[24] colors consistent 
with stellar photospheres, the \citet{Ir08} sample lacks photospheric sources 
because it contains only faint stars.

The K$_{s}$-[24] colors of confirmed/candidate members indicate 
 that NGC 2362 harbors many stars with 24 $\mu m$ excess emission 
plausibly due to circumstellar disks.  
MIPS detects 13/22 of the known cluster members brighter than J=11 (Figure \ref{jvk24}, top-left panel).
These bright sources have J--K$_{s}$ $\approx$ 0,  
indicative of early spectral-type stars (Figure \ref{jvk24}, right panels).  
Thus, the MIPS observations probably detect photospheric emission from
 many early-type (B--A) cluster members.  The locus of 
sources with 24 $\mu m$ detections and upper limits in Figure \ref{jvk24} (left panels) shows 
that MIPS detects both weak (K$_{s}$-[24] $\sim$ 1--3) and strong (K$_{s}$-[24] $\gtrsim$ 3--4)
 excesses around stars with J $\sim$ 12--14.5 and only strong excesses around fainter stars.  
By comparing the locus of colors as a function of J magnitude and J--K$_{s}$ color, 
we rule out the presence of strong 24 $\mu m$ emission from cluster stars with J $\lesssim$ 14, 
J--K$_{s}$ $\lesssim$ 0.6.  For a 5 Myr-old cluster with negligible reddening, stars with 
J-K$_{s}$ $\lesssim$ 0.6 have spectral types earlier than K3 and masses greater than $\sim$ 
1.4 M$_{\odot}$ \citep{Ba98}.

To summarize, the population of stars with 24 $\mu m$ excesses reveals a wide range of 
mid-IR disk luminosities.  Most bright, early-type stars have 
photospheric MIPS emission and thus may lack circumstellar dust.  
Early-type stars with MIPS excesses have weak excesses, $\approx$ 0.5--2 magnitudes.
Later-type stars with MIPS detections typically have strong excesses. 
However, sensitivity limits preclude MIPS from detecting
 many late-type stars with weak excesses.

The positions of several stars in Figure \ref{jvk24} call into question their membership 
status.  Specifically, IDs 28 and 166, listed as members in \citet{Dh07} based on X-ray activity, have J--K$_{s}$ colors 
indicative of late-type stars (J--K$_{s}$ $\sim$ 0.6 and 0.8) but apparently have photospheric 
K$_{s}$-[24] colors.  These stars have optical and 2MASS fluxes comparable 
to the earliest-type stars in the cluster and are brighter than other late-type cluster members by 
$\sim$ 3 magnitudes.  If these stars are true pre-main sequence members, the huge implied age spread is 
at odds with recent work showing that NGC 2362 likely had a single, explosive burst of star formation 
$\sim$ 4--5 Myr ago \citep[e.g.][]{Mn08}.  These stars are probably either X-ray active foreground M stars, 
extremely high-mass, post-main sequence cluster members, or extragalactic sources.

\subsection{Combined IRAC and MIPS Colors of Cluster Members from \citet{Dh07} and 
Candidates from \citet{Ir08}}

\begin{figure}
\epsscale{1.0}
\centering
\plottwo{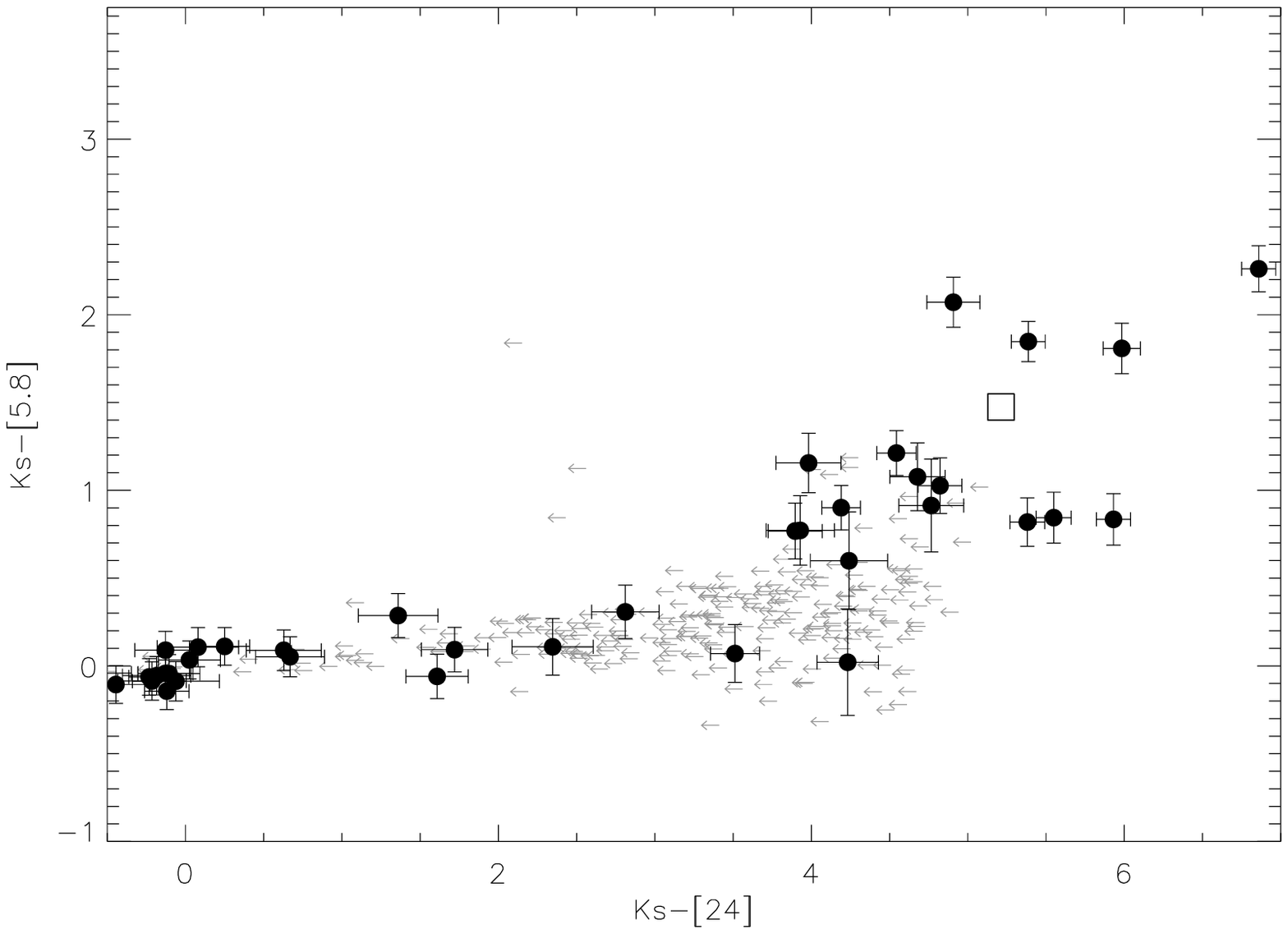}{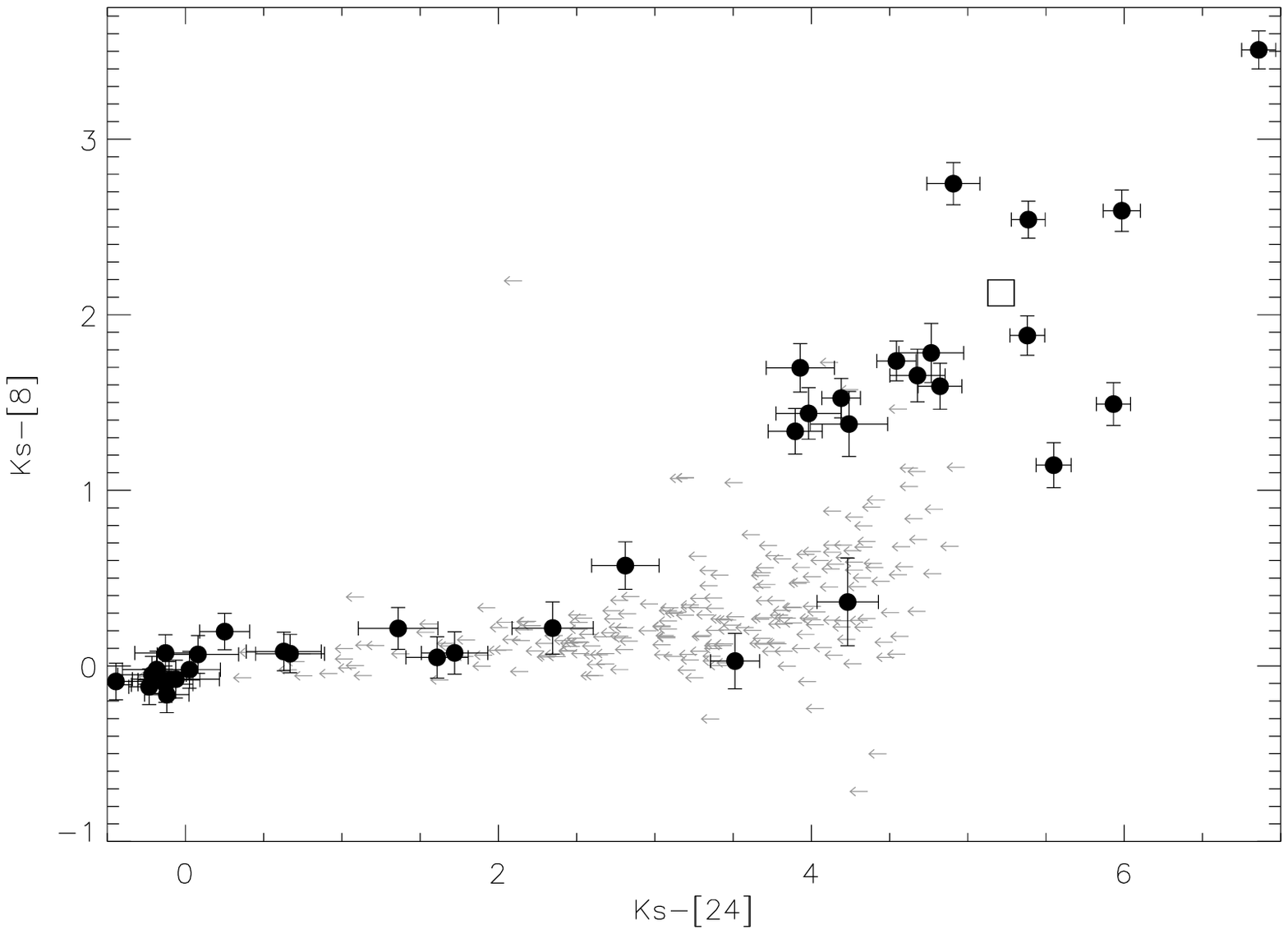}


\plottwo{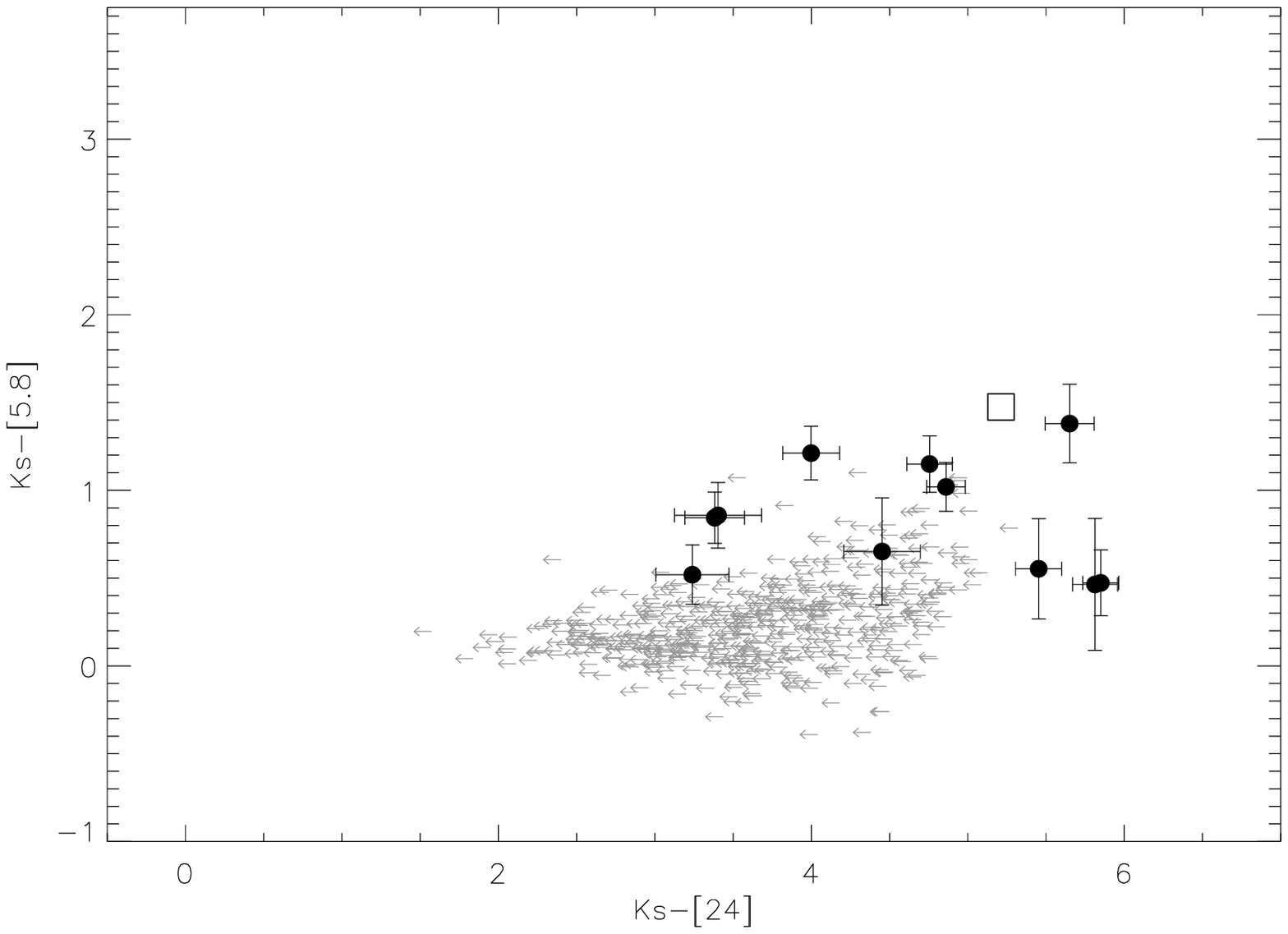}{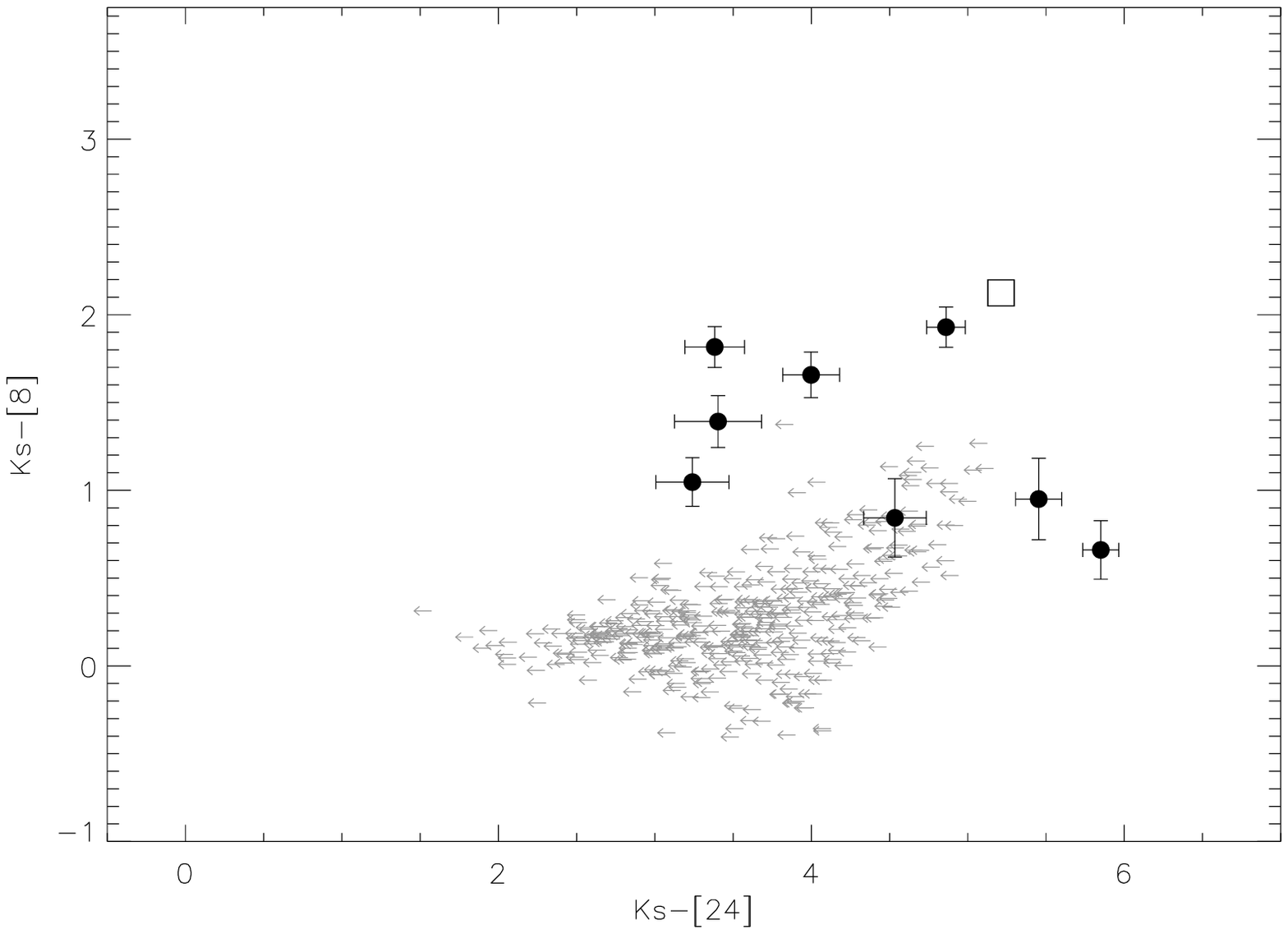}
\caption{The K$_{s}$-[5.8,8] vs. K$_{s}$-[24] color-color diagrams 
for the \citet{Dh07}sample (top).  Overplotted are the colors from the median Taurus SED \citep{Fu06} reddened 
by E(B-V) = 0.1 (square).  (top-left) The K$_{s}$-[5.8] vs. K$_{s}$-[24] color-color diagram.
  Sources with 24 $\mu m$ upper limits 
are shown as grey left-pointing arrows.  MIPS-detected sources (black dots) 
are shown overplotted with 1$\sigma$ photometric errors.  (top-right) 
The K$_{s}$-[8] vs. K$_{s}$-[24] color-color diagram for \citeauthor{Dh07} 
sources.  The symbols are the same as in the left-hand panel.
(bottom)K$_{s}$-[5.8,8] vs. K$_{s}$-[24] color-color diagrams for the \citet{Ir08} 
sample.}
\label{k4k24}
\end{figure}

When combined with the IRAC data, the MIPS data reveal a diversity of mid-IR disk emission.  
Figure \ref{k4k24} shows the K$_{s}$-[5.8, 8] vs. K$_{s}$-[24] 
color-color diagrams for stars in the \citet{Dh07} sample (top panels) and 
\citet{Ir08} sample (bottom panels).
To compare the observed colors with colors typical of 
primordial circumstellar disks, we overplot the colors from the 
median Taurus SED \citep{Fu06} reddened by E(B-V) = 0.1.
Most sources with MIPS upper limits (grey arrows) have 
K$_{s}$-[8] $\sim$ -0.2--0.5.  
About 18 stars have $\gtrsim$ 3.5 magnitude MIPS excesses.
Comparing their K$_{s}$-[24] colors in Figure \ref{k4k24} to the 
distribution of K$_{s}$-[24] colors vs. J--K$_{s}$ in Figure \ref{jkk34dh} 
shows that all stars with strong MIPS \textit{and} IRAC excesses are late type, 
low-mass stars.

Most sources with IRAC and MIPS excess emission have weaker 
emission than the Taurus sources with primordial disks.
Because many of the 85 Taurus sources used to construct the median 
SED have evidence for inner holes/gaps (e.g. DM Tau, UX Tau), 
stars surrounded by full primordial disks should have K$_{s}$--[5.8, 8] 
colors comparable to or greater than those for the median Taurus SED.
Only four NGC 2362 sources 
have IRAC excesses exceeding that of the 
median Taurus SED.  Most sources with excess emission have 
weaker emission than the median Taurus SED at both IRAC and MIPS wavelengths.
Sources with IRAC and/or MIPS excesses weaker than the median Taurus SED, 
show a further diversity.
Among the sources with 24 $\mu m$ excesses equal to or greater than the 
median Taurus SED, nearly half have IRAC fluxes in at least one filter 
that are reduced by a factor of $\sim$ 2 from the median Taurus SED.  

\section{Analysis of the NGC 2362 Disk Population}
The combined IRAC and MIPS photometry of confirmed/candidate cluster members 
reveals sources with a wide range of mid-IR excess emission consistent with a 
range of disk evolutionary states.  In general, stars with blue J--K$_{s}$ colors ($\le$ 0.4; early-type stars) 
either lack 24 $\mu m$ excess emission or have weak 24 $\mu m$ excess emission (K$_{s}$ -[24] $\sim$ 0.5-2).  
Stars with redder (J--K$_{s}$ $>$ 0.6; late-type stars) colors have strong 
24 $\mu m$ excess emission (K$_{s}$-[24] $\sim$ 3--7) 
presumably from optically-thick circumstellar disks.
Furthermore, the combined IRAC and MIPS colors show that stars with strong 24 $\mu m$ disk emission 
have a wide range of emission from inner disk regions probed by the IRAC 3.6--8 $\mu m$ filters.
The variety of mid-IR fluxes from disks surrounding NGC 2362 stars then warrants and enables 
detailed comparisons with predicted fluxes from disks spanning a range of evolutionary states.

In this section, we explore the NGC 2362 disk population in more detail by using models to constrain the 
evolutionary state of disks surrounding early type, high/intermediate-mass stars and late type, low-mass stars.
By comparing the observed SEDs of early-type stars to best-match SEDs from the \citet{Ro06} radiative 
transfer models, we estimate the mass of emitting dust needed to reproduce the observed disk emission.  
We then compare this dust mass to dust masses for disks in several evolutionary states and derive the dust removal timescales.

For late-type stars, our approach is twofold.  First, we compare the source SEDs to the median Taurus SED and 
to predictions for warm debris disk emission.  The median Taurus SED serves as an empirically-based fiducial primordial 
disk model.  In the debris disk model \citep{Kb04}, debris emission from terrestrial planet formation is calculated at 
0.7--2 AU from a 1 M$_{\odot}$ star and tracked for $\sim$ 100 Myr.  
  The disk luminosity as a function of wavelength at 5 Myr is converted into a flux and scaled to the stellar photosphere.
In these comparisons, we emphasize that \textit{no attempt is made to fit the observed SEDs}.  Rather, 
comparing the source SEDs to the median Taurus and debris disk models can constrain whether the 
SED is `primordial disk like', `debris disk like', or in between.  
For a second check on the disk evolutionary state, we 
fit the observed SEDs to the \citet{Ro06} models to determine the allowed ranges of dust masses, inner hole sizes, and 
disk flaring indices using the SED fitting tool of \citet{Rob07}.
 From this analysis, we determine the relative frequency of primordial disks, debris disks, 
and disks in an intermediate stage surrounding late-type stars.
\subsection{Mid-IR Disk Emission from $\gtrsim$ 1.4 M$_{\odot}$-Mass Stars}
\subsubsection{SED Modeling of $\gtrsim$ 1.4 M$_{\odot}$ Stars}
To model the mid-IR emission from early-type stars with 24 $\mu m$ excess emission, we compare 
their SEDs to the grid of 200,000 radiative transfer models 
from \citet{Ro06}.  The \citet{Ro06} models predict the thermal emission of sources 
including contributions from stellar photospheres, infalling envelopes, bipolar cavities, and 
accreting circumstellar disks.  The model parameters are varied for stars with 
a wide range of stellar temperatures.  Dust masses are derived assuming a standard 
ISM size distribution (n(a) $\propto$ a$^{-3.5}$).  By selecting the best-matching SEDs, we estimate 
the masses of small ($\lesssim$ 100 $\mu m$--1 mm) dust grains that are required to reproduce the observed level of 
emission.  We then compare these masses to typical dust masses for debris disks and primordial disks.  
 Typical dust masses for debris disks range from 10$^{-11}$ M$_{\odot}$ to
5$\times$10$^{-9}$ M$_{\odot}$ \citep[cf.][]{Ch05, Lo05}.  Typical dust masses
for primordial disks range from 10$^{-5}$--10$^{-3}$ M$_{\odot}$
 \citep[cf.][]{Aw05}.  To estimate spectral types of these stars, we use their observed 
J--K$_{s}$ colors.  We model the SEDs of the four sources -- IDs 31, 144, 303, and 314 -- 
with K$_{s}$-[24] = 0.5 -- 2.5.

\begin{figure}
\epsscale{0.85}
\centering
\plottwo{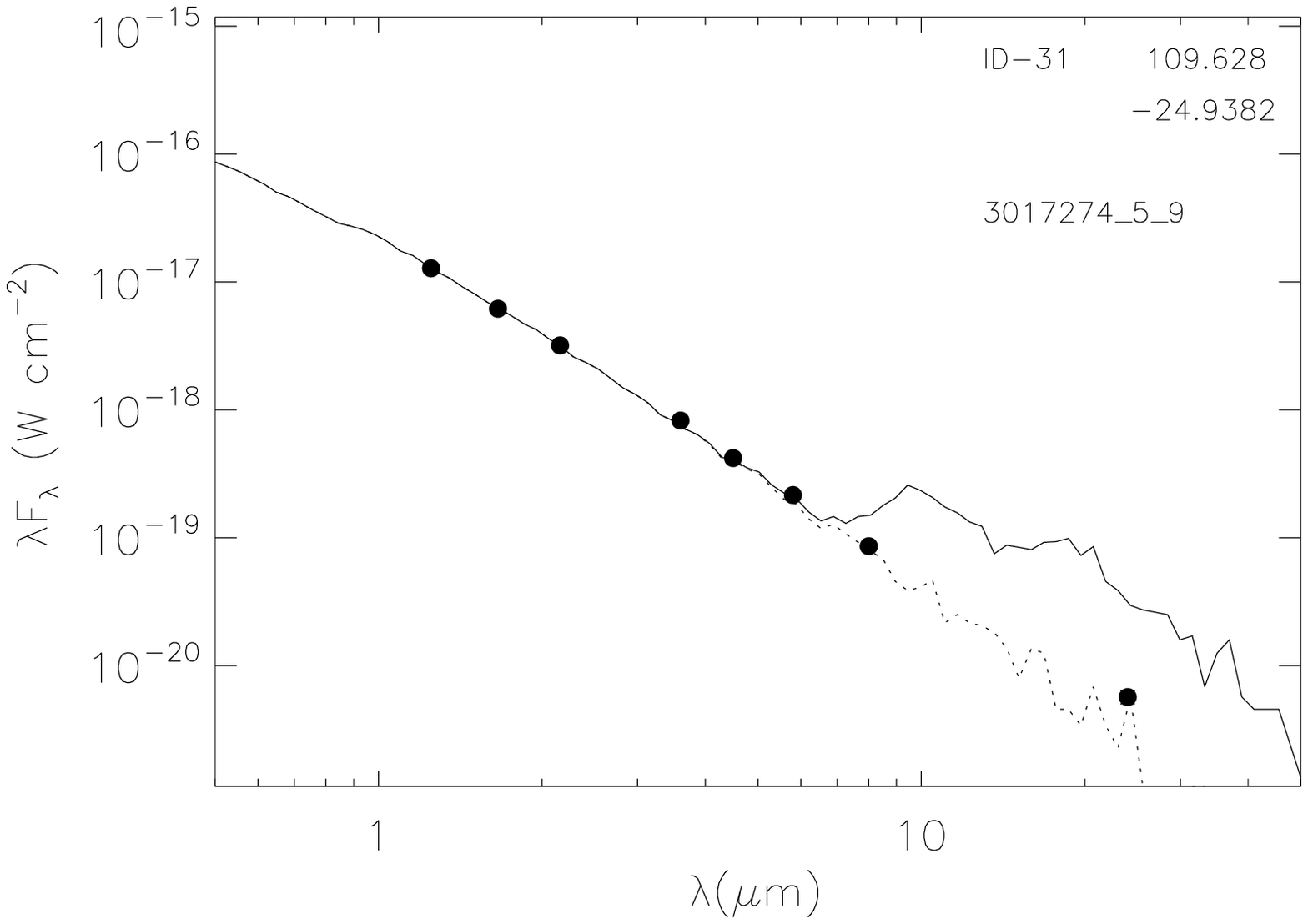}{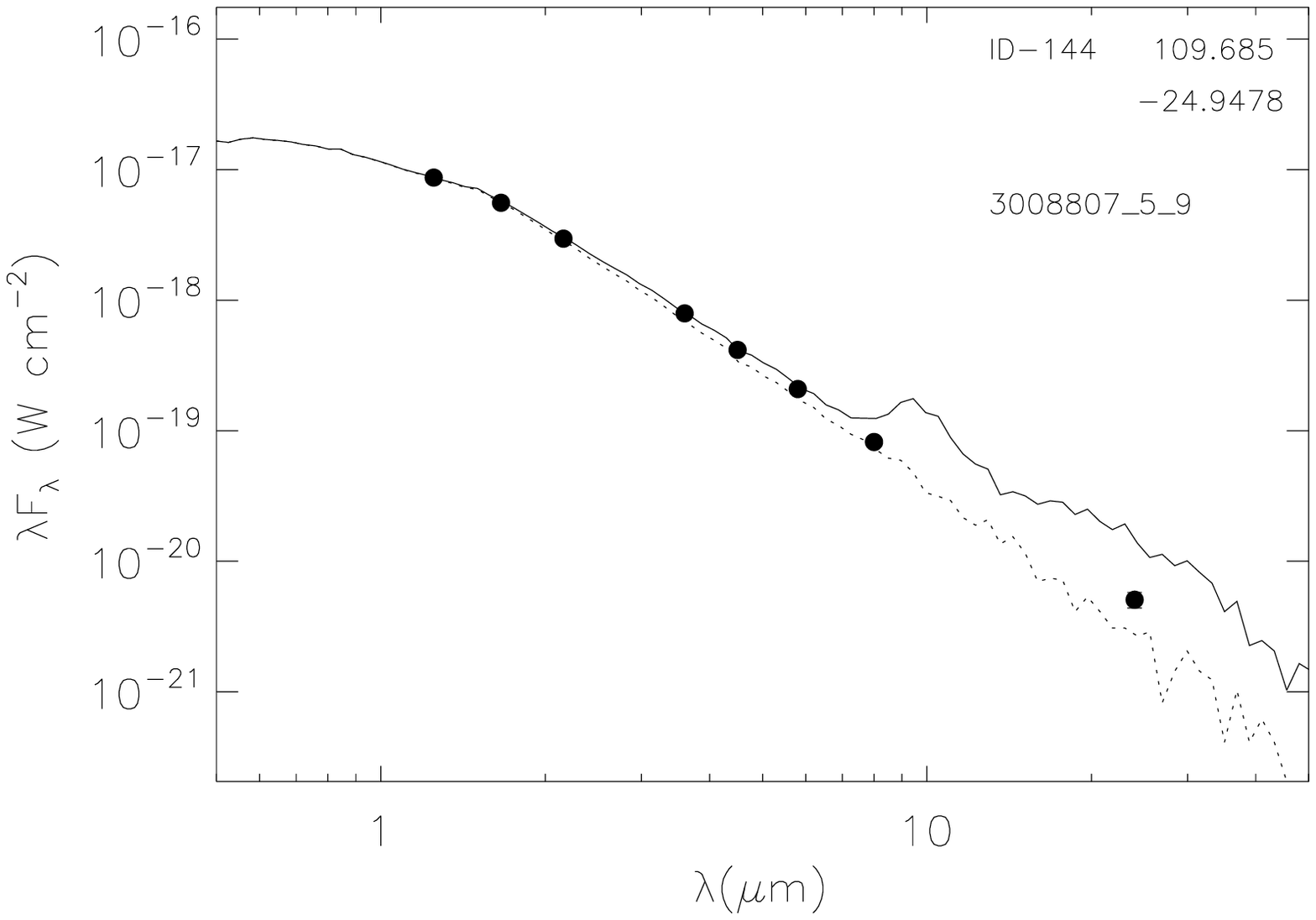}
\plottwo{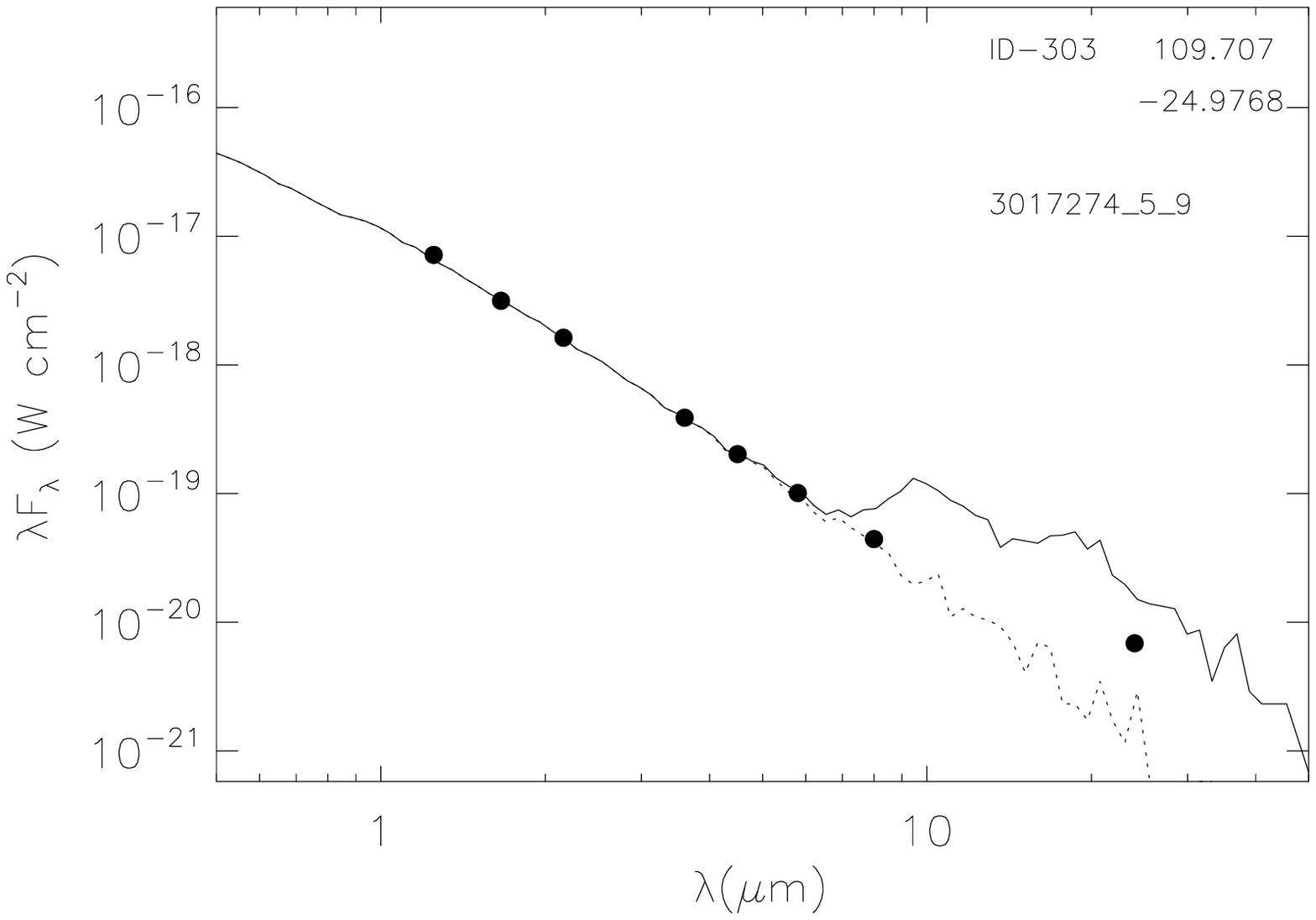}{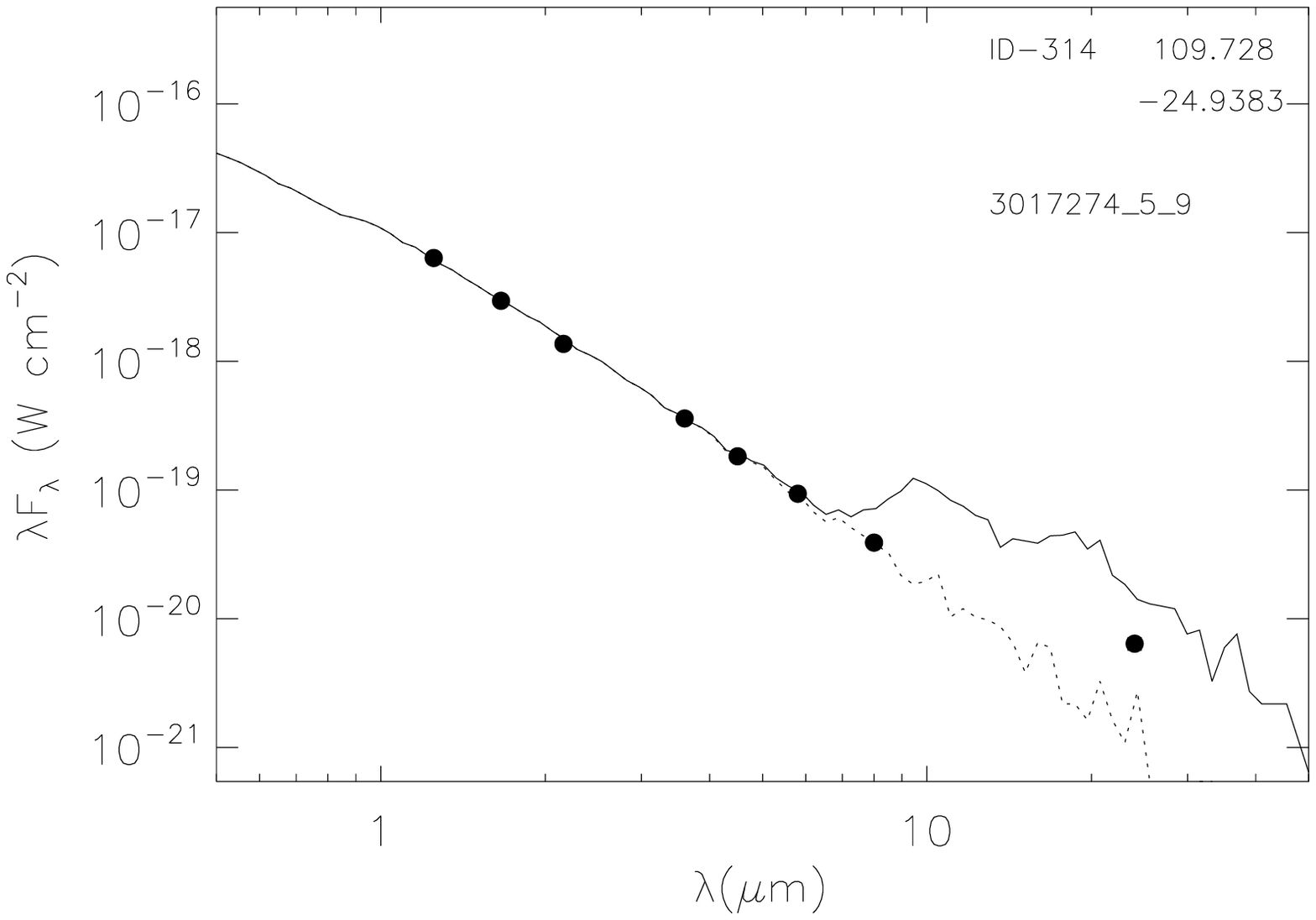}
\caption{SEDs of early/intermediate-type stars with masses $\gtrsim$ 1.4 M$_{\odot}$.  
Overplotted are the best-fitting SEDs from \citet[][solid line]{Ro06} along with
the stellar photosphere (dotted line).  
Also listed is the J2000 source position (in degrees).  
Photometric uncertainties are smaller than the symbol sizes.  Inferred dust masses are 
10$^{-10}$ M$_{\odot}$ to 10$^{-9}$ M$_{\odot}$, consistent with debris disks.  
}
\label{sedstr}
\end{figure}

 Figure \ref{sedstr} shows SED model fits to the four early/intermediate-type stars with 
MIPS excesses, the results of which are summarized in Table \ref{noremnant}.  
The best-matching \citet{Ro06} models overpredict the observed fluxes from these sources at 
all mid-IR wavelengths\footnote{Even though the "fits" are obviously poor and 
clearly overestimate the mass of dust required to reproduce the observed emission,
 the \citet{Ro06} grid lacks SEDs for disks with lower dust masses.  Therefore, these are technically 
the "best-fit" models.}, which indicates that the dust masses required 
to reproduce the observed disk emission through 24 $\mu m$ are considerably less than the fitted masses.  
Three of the four sources -- IDs 144, 303, and 314 -- have significant 24 $\mu m$ excesses above the 
model stellar photosphere (dotted line).   The 24 $\mu m$ flux from ID 31 lies just barely 
above the photosphere ((K$_{s}$-[24])/$\sigma([24])$ $\approx$ 4) so this source has marginal evidence for a disk.
The dust masses for the best-fit \citeauthor{Ro06} models are $\sim$ 1--10 $\times$ 10$^{-10}$ M$_{\odot}$, or 
$\sim$ seven orders of magnitude less than typical masses of dust in primordial disks.  
Unless these sources have enormous reservoirs of 
extremely cold dust, which would not contribute significantly to the observed 24 $\mu m$ emission, 
the total mass of dust in small grains is likely small compared to primordial dust masses as well.
The extremely low required dust masses required imply that the disks are optically thin through 24 $\mu m$.  
These dust masses are inconsistent with masses typical of primordial disks but 
are consistent with debris disk masses \citep[e.g.][]{Ch06}.
\subsubsection{Dust Dynamics in Disks Surrounding $\gtrsim$ 1.4 M$_{\odot}$ Stars}
Next, we use dynamical arguments to show that,
that dust surrounding high/intermediate-mass stars is most likely from 
debris disks (Table \ref{noremnant}).  In optically-thin conditions, 
small dust grains are removed from the circumstellar environment
by Poynting-Robertson drag and radiation pressure.  
The timescale for dust removal by Poynting-Robertson
drag is \citep[e.g.][]{Bp93, Au99}:
\begin{equation}
t_{P-R} (\mbox{yr}) \sim 705 \frac{\rho_{s}L_{\odot}}{<Q_{abs}> L_{\star}}s_{\mu m} r^{2}_{AU},
\end{equation}
where r$_{AU}$ is the distance from the central star, s is the radius of
the grain, $\rho_{s}$ is the grain volume density and $<Q_{abs}>$ is the
absorption coefficient.

 To derive the drag timescale for grains in the NGC 2362 sources, we consider typical
parameters for r$_{AU}$, $<Q_{abs}>$, and $\rho_{s}$.  These sources have  
negligible 8 $\mu m$ excesses but clear 24 $\mu m$ excesses.  Thus, the
typical grain temperature is likely $\approx$ 100-200 K.  For grains in radiative equilibrium,
r$_{AU}$ $\sim$ (L$_{\star}$/L$_{\odot}$)$^{0.5}$ (T/280 K)$^{-2}$.  
Assuming an age of 5 Myr, early and intermediate-type stars in NGC 2362 
have luminosities $\sim$ 10--56 L$_{\odot}$ \citep{Si00}.  Therefore,  r $\sim$ 6--40
AU for T $\sim$ 100-200 K.  We adopt r $\sim$ 10--30 AU, roughly
the mean distances for 200 K and 120 K dust in our sample.  For grains with
$\rho_{s}$ $\sim$ 1 g cm$^{-3}$ and $<Q_{abs}>$ $\sim$ 1 \citep{Bu79},
\begin{equation}
t_{P-R} (\mbox{yr}) \sim 7\times10^{4}~(\frac{L_{\odot}}{L_{\star}})~(\frac{s}{1 \mu m})~(\frac{r}{10 AU})^{2}.
\end{equation}

For these massive stars, P-R drag removes small grains on timescales shorter
than the $\sim$ 5 Myr age of sources in NGC 2362.  For 10 $\mu m$ grains
orbiting stars with L$_{\star}$ $\sim$ 10--56 L$_{\odot}$,
the drag time is t$_{P-R}$ $\lesssim$ 6.3 $\times$ 10$^{5}$ yr.
Grains with s $\lesssim$ 10 $\mu m$ are probably porous, with
$\rho_{s}$ $\sim$ 0.1 g cm$^{-3}$ instead of our adopted
$\rho_{s}$ $\sim$ 1 g cm$^{-3}$ \citep[e.g.][]{Au99}.
Porous grains have very short lifetimes, t$_{P-R}$ $\lesssim$ 6.3 $\times$ 10$^{4}$ yr.
Thus, our derived upper limit for t$_{P-R}$ is conservative.

If the ratio of the force from radiation pressure to gravity ($\beta$) exceeds 1/2\footnote{
\citet{StrCh06} argue that grains more easily become unbound, on 
a dynamical timescale once $\beta$ $>$ 1/2, not 1.},
radiation pressure can remove smaller grains on faster, dynamical timescales
 \citep[e.g.][]{Bu79, StrCh06}:
\begin{equation}
\beta_{pr} = \frac{3L_{\star}<Q_{pr}(a)>}{16\pi GM_{\star}cs\rho_{s}} > 1/2.
\end{equation}
This equation can be rearranged to solve for the grain
size below which radiation pressure can remove dust from the system \citep{Bu79, Bp93}:
\begin{equation}
s_{max, \mu m} < 1.14 \mu m~(\frac{L_{\star}}{L_{\odot}})~(\frac{M_{\odot}}{M_{\star}})~(\frac{1 g cm^{-3}}{\rho_{s}}) <Q_{pr}>.
\end{equation}
 
 For stars listed in Table \ref{noremnant}, s$_{max}$ ranges from 5.8 to 23.8 $\mu m$, assuming $<Q_{pr}>$ $\approx$ 1
and $\rho_{s}$ $\approx$ 1 g cm$^{-3}$.  Using a more realistic dust grain porosity makes even more grains
subject to blowout by radiation pressure.  
Therefore, grains responsible for producing the observed 24 $\mu m$ emission
should be blown out unless they are replenished by collisions between larger bodies.

As long as the gas density is low, radiation pressure can remove small
grains on short timescales.  When the grains are poorly coupled to the gas,
radiation removes grains on timescales comparable to 
the local dynamical timescale, $\lesssim$ 10$^{3}$ yr.
In gas disks with M$_{gas}$ $\sim$ 1--10 M$_{\oplus}$, small ($\lesssim$ 10-50 $\mu m$)
dust grains are well coupled to the gas.  Even in these circumstances, they are pushed by
radiation pressure from $\approx$ 10 AU to hundreds of AU in $\approx$ 10$^{4}$ years 
if the disk is optically thin \citep{Tak01}.  Disk emission originating at tens of AU 
then requires an active replenishment source and thus is likely second generation.

The properties of disks surrounding early-type NGC 2362 stars are highly suggestive 
of debris disks \citep{Bp93}.  The small IR excesses in dusts imply that the disks have very low luminosities 
compared to their stars and have a low mass in dust.  The rapid removal timescales for 
this dust, regardless of whether there exists residual gas, indicates that the dust 
is likely second generation.   Disks with copious amounts of circumstellar gas almost 
without exception have much stronger broadband near-to-mid infrared disk emission than 
that observed here.  Thus, all known observed and inferred disk properties fit standard 
debris disk characteristics and not primordial disk characteristics.  
More sensitive measures of circumstellar gas (e.g. optical echelle spectroscopy) would 
provide an even better test of our conclusion that these disks are debris disks.
\subsection{SEDs of Solar and Subsolar-Mass Stars}
The Appendix shows the
atlas of SEDs for all late-type (candidate) cluster members from \citet{Dh07} and \citet{Ir08}
with K$_{s}$-[24] $\gtrsim$ 1.  The two late-type stars in \citet{Dh07} with 
questionable membership status lack disk emission and are not shown.
In each panel, we overplot 
the 1$\sigma$ error bars for IRAC and MIPS data.  
For the stars with spectral types from \citet{Dahm05}, 
we overplot the SED of the stellar photosphere (solid 
line/squares) using a model specific to the spectral type.  
We calculate the range in reddening and extinction from comparing 
the observed V-I colors to intrinsic colors from \citet{Kh95}, assuming 
a one-subclass uncertainty in the spectral type.  All but five sources 
have a range of extinctions consistent with E(B-V) $\sim$ 0.1 \citep{Moi01}.
For these five sources, we use a value for reddening halfway in between 
the minimum and maximum A$_{V}$ values allowed.  For the other stars with spectral types, 
we assume A$_{v}$ = 0.31, which is the mean extinction to the cluster \citep{Moi01}.  
For the stars without known spectral types, we fit the source SEDs 
from I through K-band with stellar photospheres to determine the best-fit 
spectral type, using colors from 
\citet{Kh95} and assuming a reddening of E(B-V) = 0.1 (A$_{v}$ $\sim$ 0.31).  

Because most Taurus sources lack clear J-band excess emission \citep{Fu06}, we scale 
the median Taurus SED (dotted line) by its J-band flux, setting it equal to the stellar photosphere.
The median Taurus SED clearly has the strongest disk emission.  
The warm debris disk models (dashed line/crosses) have weaker disk emission 
at all mid-IR wavelengths, slightly in excess of the stellar photosphere.
  The nominal MIPS 24 $\mu m$ detection limit in $\lambda$F$_{\lambda}$ 
space is overplotted as a grey dashed line ($\lambda$F$_{\lambda}$ $\sim$ 
5.75 $\times$ 10$^{-21}$ W cm$^{-2}$).  

To compare the inferred physical states of disks with more simple empirical classifications, 
we measure the IRAC [3.6]-[8] flux slope for each star.  
We divide the sample according to the three empirical divisions from \citet{La06}.  We label sources 
with $\alpha$ $\ge$ -1.8 as those with "strong IRAC" emission, those with $\alpha$ = -2.56 -- -1.8 
as having "weak IRAC" emission, and those with $\alpha$ $<$ -2.56 as those with "photospheric" emission.
\citet{Dh07} use this flux slope to infer the physical structure of the disk, specifically identifying 
sources with $\alpha$ $\ge$ -1.8 as having primordial disks.  Comparing the empirical disk classifications 
with evolutionary states determined from SED modeling allows us to compare our results with 
\citeauthor{Dh07}'s and test the utility of using a single IRAC slope to probe disk states.

Table \ref{diskevostate} summarizes our results and is divided into three samples: 
MIPS-detected sources from the \citet{Dh07}, MIPS-detected sources from the \citet{Ir08} sample, 
and sources lacking clear MIPS detections that have disk emission according to \citet{Dh07}.
We list basic sources parameters (spectral type, extinction) as well as 
derived disk parameters -- disk mass, minimum inner disk radius, and flaring index ($\beta$) -- 
using the \citet{Ro06} models which, along with comparisons to the Taurus SED and debris disk SED, determine 
the disk evolutionary state \citep[e.g.][]{Rob07}.
For each source, we consider all models which satisfy the goodness-of- 
fit criterion $\chi^2-\chi^2_{\rm best} < n_{\rm data}$ rather than  
simply the best-fit model.
The derived disk evolutionary states for each sample are discussed in the next two sections.  
For the simple comparisons to the Taurus and debris disk SED, we assume a single extinction.  
The extinction is treated as a free parameter in the \citet{Ro06} models.  
However, the derived evolutionary states are insensitive to changes in the 
assumed extinction over the range of plausible values.
\subsubsection{SEDs of Confirmed/Candidate Cluster Members with MIPS detections.}
\begin{figure}
\epsscale{0.95}
\centering
\plottwo{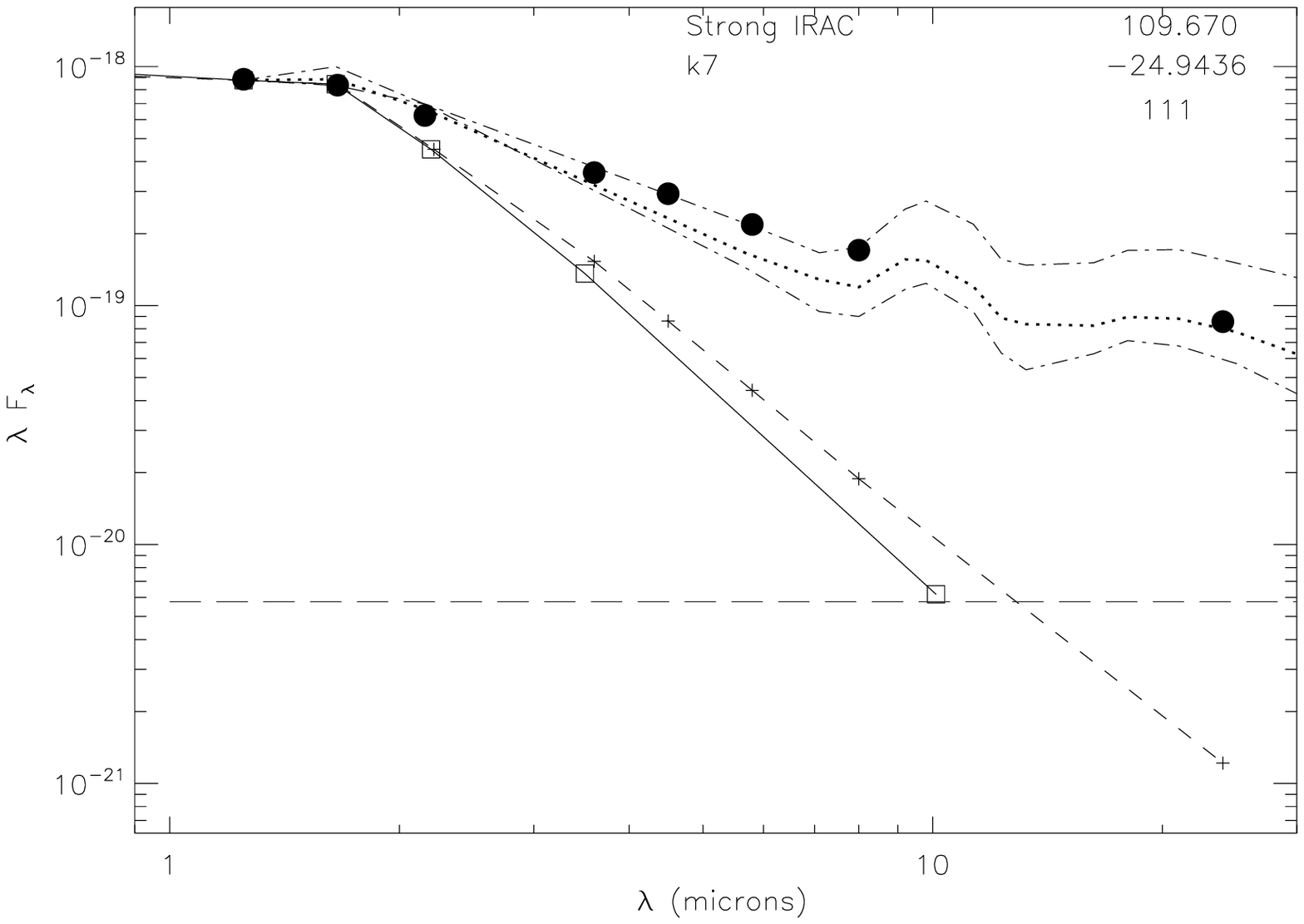}{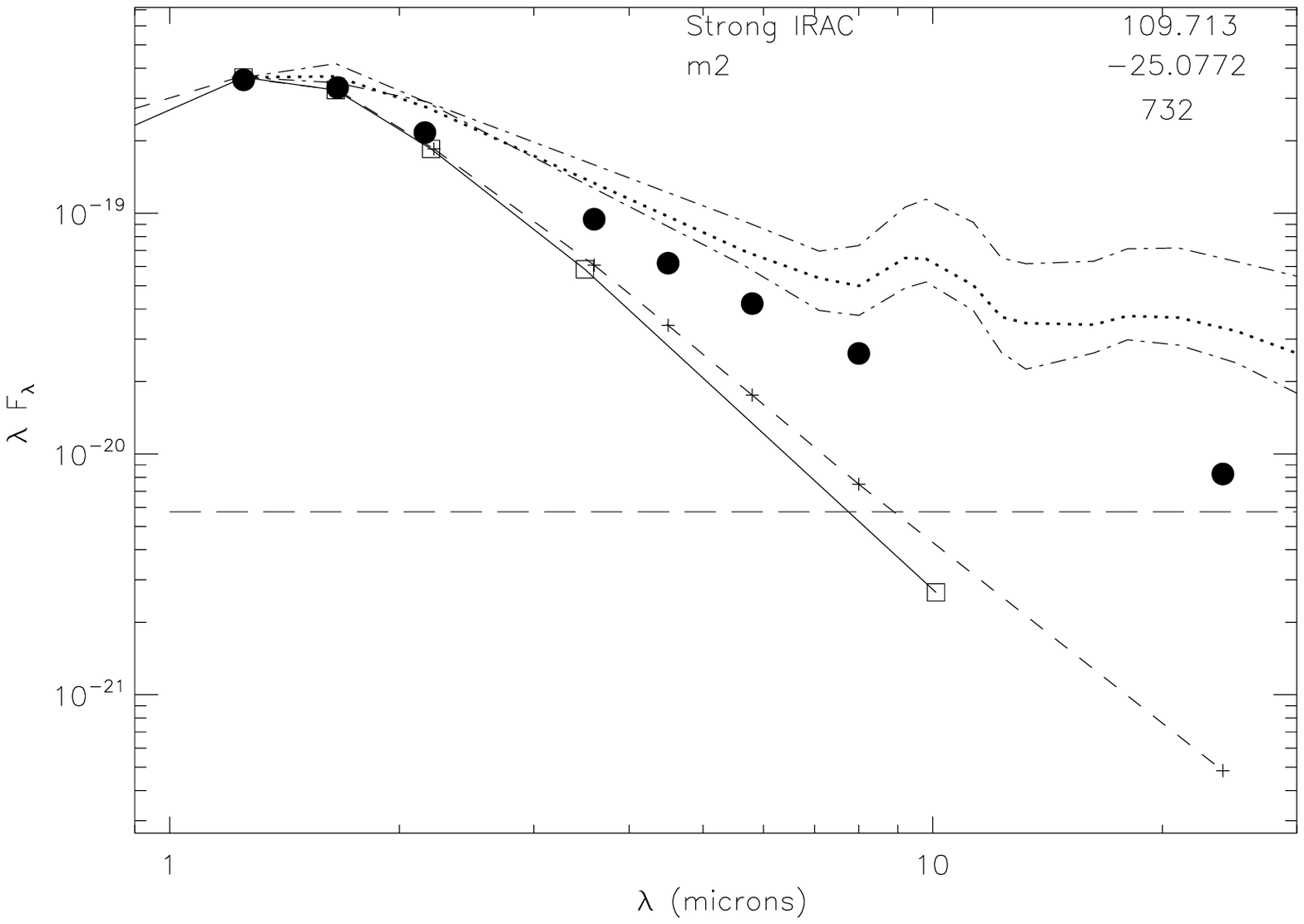}
\plottwo{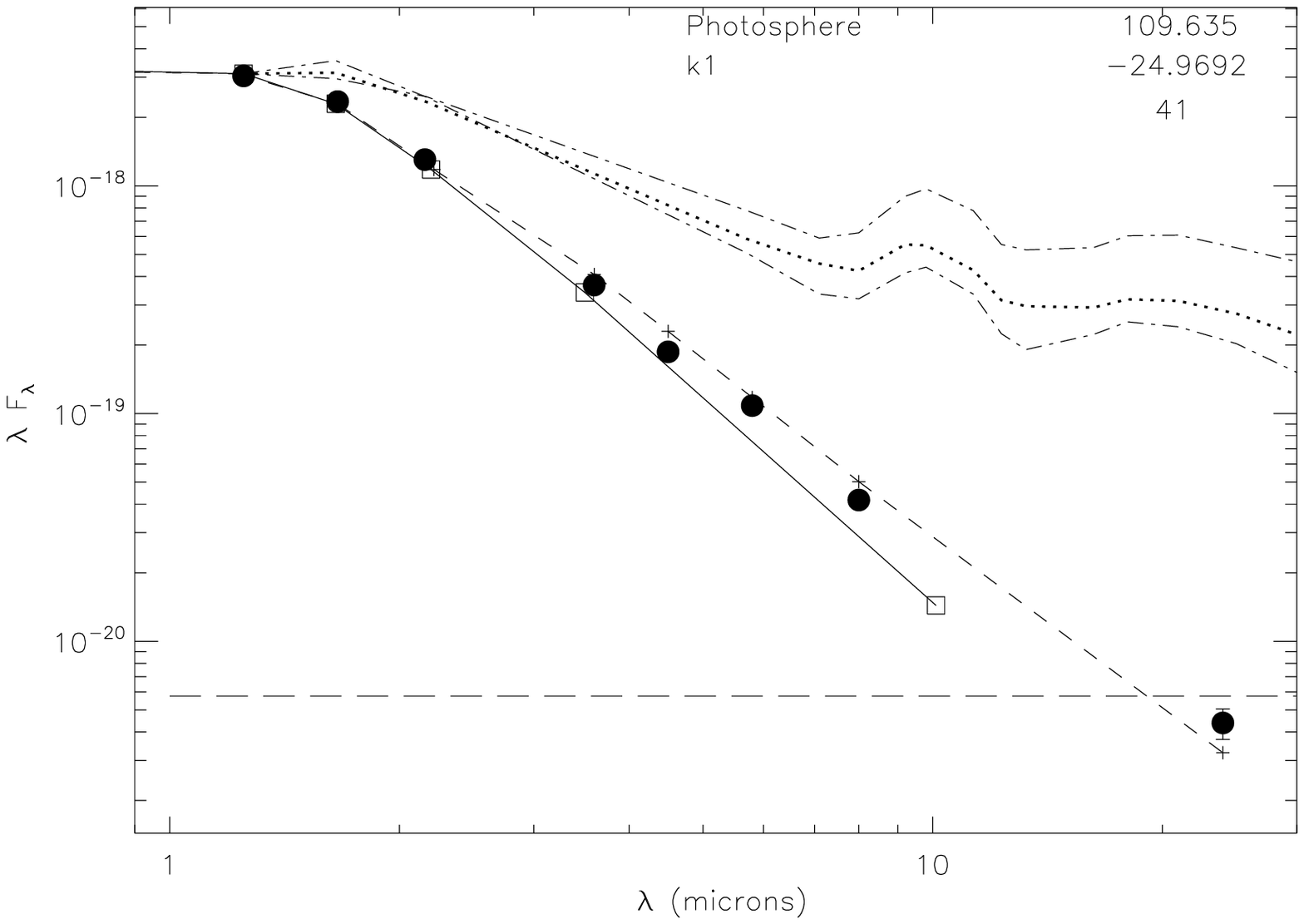}{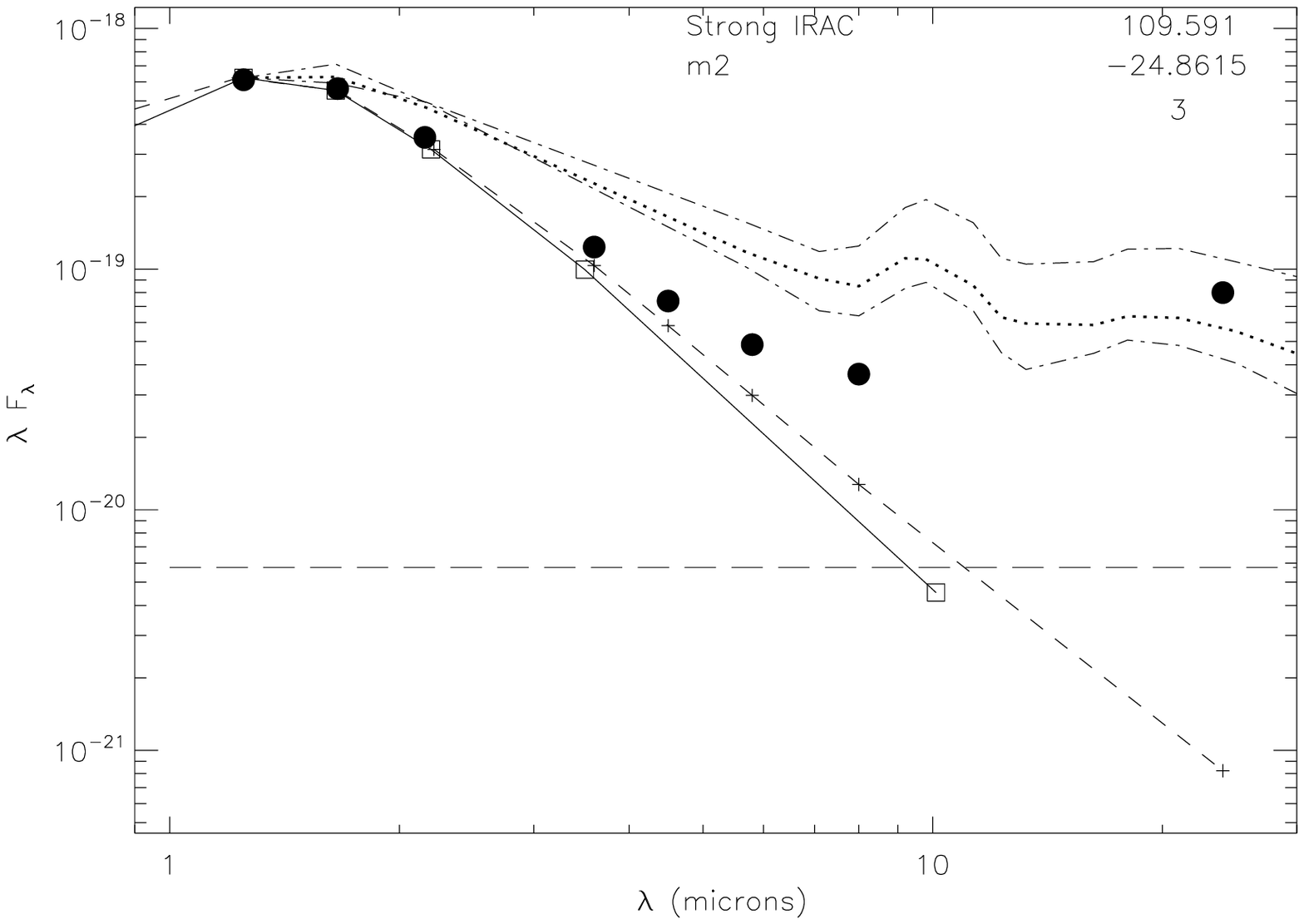}
\caption{Representative spectral energy distributions of late-type NGC 2362 members and candidate 
members with MIPS 24 $\mu m$ excesses illustrating the range of disk evolutionary states.  
Overplotted are the median Taurus SED (dotted line) with upper and lower quartiles (dot-dashed line).
A terrestrial zone debris disk model is shown as a dashed line with mid-IR flux slightly greater 
than the photosphere (solid line, connected by open squares).
The MIPS 5$\sigma$ detection limit is shown as a horizontal grey dashed line.  
The derived disk evolutionary states are (clockwise from the top left) primordial, homologously 
depleted, transition, and debris.}
\label{sedrep}
\end{figure}

\begin{figure}
\centering
\plotone{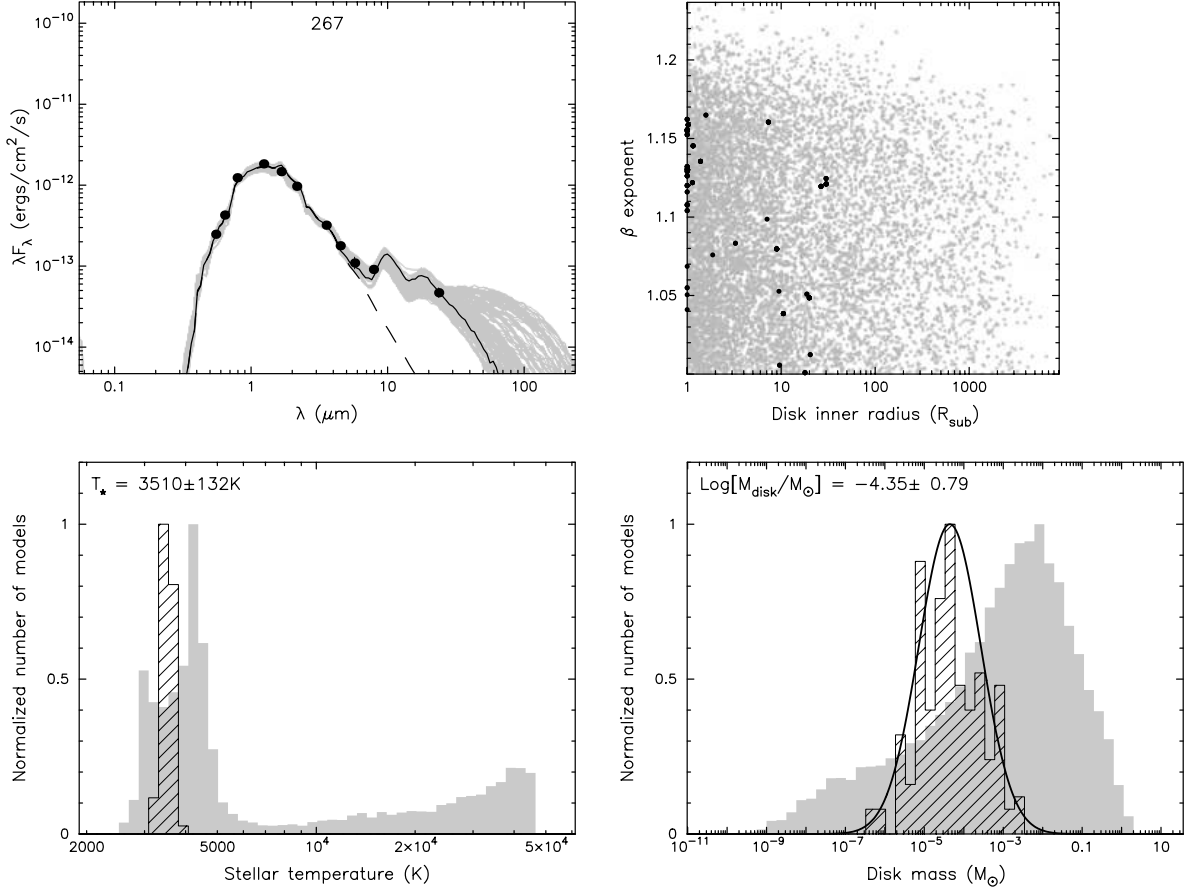}
\caption{SED analysis of ID-267 using the \citet{Ro06} models to fit the observed optical/infrared 
flux (top-left), constrain the inner disk radius and flaring index $\beta$ (top-right), constrain 
the stellar temperature (bottom left), and constrain the disk mass (bottom right).  The solid line 
in the top-left plots shows the best-fitting SEDs from the \citeauthor{Ro06} grid with other possible 
models overplotted as faint, grey lines.  In the plot of $\beta$ vs. R$_{in}$ (top-right panel), black dots denote 
the flaring index and inner disk size (in units of the sublimation radius) for the models shown in 
the top-left panel.  The grey dots show values for the entire grid.  The black curves in 
the bottom panels define the mean and standard deviation of the stellar temperature and disk masses from 
the best-fitting models, the latter assuming a gas-to-dust ratio of 100.  The hatched regions show the actual disk masses for each SED 
model shown the top-left panel.  The distributions of temperatures and disk masses for the entire grid are 
shown in grey.}
\label{trob_full}
\end{figure}

The SEDs of MIPS-detected sources shown in the Appendix reveal disks with a 
variety of morphologies.  Figure \ref{sedrep} shows comparisons between the Taurus and debris disk SEDs with 
four representative SEDs from the atlas illustrating 
the range of evolutionary states listed in Table \ref{diskevostate}.  
Figure \ref{trob_full} shows modeling results for one SED (ID 267) using the \citet{Ro06} models.
Below, we describe the characteristics for each evolutionary state 
and discuss their relative frequencies:
\begin{enumerate}
\item \textbf{Primordial Disks} -- Four stars in the \citet{Dh07} sample (IDs 111, 137, 139, 202) and 
two stars in the \citet{Ir08} sample (ID-IR526 and ID-IR1046) likely harbor primordial disks.
These sources typically have near-to-mid IR fluxes comparable to the 
median Taurus SED, though ID 202 appears to have a slight near-IR flux deficit.
  Only one of the six sources has an SED with mid-IR clearly stronger than the median Taurus SED (ID 137).
Its stronger emission may be due to a greater level of disk flaring or inclination.
Comparisons with the \citet{Ro06} models show that these disks probably do not have inner holes 
and most have masses of M$_{disk}$/M$_{\odot}$ $\sim$ 10$^{-1}$--10$^{-2.5}$ for a solar gas-to-dust ratio, 
comparable to both the Minimum Mass Solar Nebula and the typical masses of primordial disks 
in Taurus as derived from submillimeter data \citep{Aw05}.

\item \textbf{Homologously Depleted Disks} -- Many stars in both the 
\citet{Dh07} sample (12 stars) and \citet{Ir08} sample (5 stars) show evidence 
for a disk that is significantly depleted in small dust grains through 24 $\mu m$ 
relative to typical primordial disks: a `homologously depleted' disk.
Like the median Taurus SED, the SED slopes 
for these sources typically follow a nearly power-law decline with wavelength through 24 $\mu m$.  
However, these sources typically show a flux deficit in the IRAC and MIPS bands 
relative to the median Taurus SED.  Indeed, they are underluminous by a factor of $>$ 2--3 
from 5.8 $\mu m$ to 24 $\mu m$.
Thus, they have a steeper power-law decline of disk emission with wavelength compared to 
primordial disks: $\lambda$F$_{\lambda}$ $\propto$ $\lambda^{-n}$ where 
n = -7/4 to -5/4.  
These disks also lack inner holes but have dust masses in small grains that 
are $\sim$ 100-1000 times less massive than their primordial disk counterparts 
The dust masses in small grains required to reproduce the observed SEDs are
 100-1000 times less than those of their primordial disk counterparts.
They also often have little flaring (with a flaring parameter, $\beta$, of $\sim$ 1--1.1), 
which is consistent with a significant amount of dust settling.  
These features are consistent with a disk that loses a significant amount of mass 
of small dust grains at all disk radii simultaneously \citep[e.g.][]{Wo02}, which 
could plausibly be due to the growth of dust to larger, 1 mm--1 km-sized 
bodies and/or accretion onto the star.

\item \textbf{Transition Disks} -- Five stars in the \citet{Dh07} sample and 
seven stars in the \citet{Ir08} sample have SEDs consistent with the 
presence of inner holes/gaps indicative of so-called transition disks.
Compared to the median Taurus SED,
these sources have weaker IRAC fluxes but comparable 
MIPS 24 $\mu m$ fluxes, a feature typical of
 disks with inner holes/gaps such as 
DM Tau, Hen 3-600A, and TW Hya \citep{Cal05, Lo05}.  
The \citet{Ro06} models that provide a good fit have minimum 
inner holes ranging in size between 2.6 AU and 7.1 AU.
Their dust masses range between $\sim$ 10$^{-5}$ M$_{\odot}$ and 10$^{-7}$ M$_{\odot}$.  For a 
solar gas-to-dust ratio, the inferred disk masses range between $\sim$ 10$^{-3}$ M$_{\odot}$ 
and $\sim$ 10$^{-5}$ M$_{\odot}$.  These values indicate that the disks are 
 slightly depleted relative to primordial disks. 
One source (ID 194) has strong K-band and 24 $\mu m$ excesses but weak IRAC excesses.  
Its SED may be suggestive of a disk gap instead of an inner hole.
\end{enumerate}

While most sources have SEDs that are straightforward to classify, several 
have characteristics that make them slightly more ambiguous.  Several sources (IDs 41, 63, and 196) 
have photospheric emission shortwards of 5.8 $\mu m$ or 8 $\mu m$; their
24 $\mu m$ excess emission is much weaker than other late-type stars detected by 
MIPS (K$_{s}$-[24] $\sim$ 0.75--2.5).   In this sense, their disks are 
consistent with being debris disks. 
However, dust grain removal processes (e.g. Poynting-Robertson drag) likely operate 
on far longer ($\sim$ 1 Myr) timescales in disks surrounding these stars than disks surrounding 
high and intermediate-mass stars examined in \S 4.1.1.  Therefore,  
that dust in disks surrounding these stars need not be be second generation debris dust.  
Based on their dust masses and presence/absence of inner holes, 
we classify these sources as either homologously depleted disks or transition disks instead. 
Identifying these sources as having debris disks changes the disks statistics we discuss 
later but will not qualitatively change our results.

Most (15/21) of the \citet{Dh07} sample with large MIPS excesses have measured H$_{\alpha}$ equivalent 
widths.  Because H$_{\alpha}$ emission can be due to circumstellar gas accretion, 
its presence also provides some constraints on disk evolutionary states.
Using the results of \citet{Dh07} and the criteria of \citet{Wb03} to identify accretors,  
 two of the three sources with primordial disks and H$_{\alpha}$ measurements
 show evidence for strong H$_{\alpha}$ emission and thus 
are accreting.  None of the sources with extremely low disk masses (IDs 41, 63, and 196)
 show this accretion signature.
Stars surrounded by transition disks and homologously depleted disks include both accretors and 
those lacking evidence for accretion.  Not all sources 
 have optical spectra, so it is difficult to make definitive statements 
about the relative frequency of accretion for each disk state.  
However, the lack of evidence for accretion for the sources with the weakest disk emission is consistent 
with a depleted reservoir of circumstellar gas in these systems.  The 
low frequency of accretion for homologously depleted disks (20\%)
and their low mean H$_{\alpha}$ equivalent widths ( mean value is 9.2\AA) is 
expected if these systems have lower rates of accretion and are generally more evolved 
than their primordial disk counterparts.

\subsubsection{Disk-bearing Stars in NGC 2362 Lacking MIPS Detections}
\citet{Dh07} identify 14 stars with strong IRAC-excess emission and  
33 with weak IRAC-excess emission.  MIPS detects 16 stars 
included in the \citet{Dh07} list, leaving 31 without detections.  
Of the sources lacking MIPS detections, three have strong IRAC excesses and 
are classified as 'primordial' by \citet{Dh07}\footnote{\citet{Dh07} classify 
disks as 'primordial' or 'weak' according to their IRAC flux slopes.  
Thus, these classifications are empirical classifications.  
The division between classes follows that of \citet{La06}.  Usage of the term 
'primordial' in this paper always refers to the inferred \textit{physical} state 
of the disk from comparing its emission to disk models and the median 
Taurus SED.}; \citeauthor{Dh07} classify the remaining 28 sources as 
'weak'.  Disk candidates without MIPS detections 1) could have 24 $\mu m$ 
fluxes fainter than the MIPS detection limit, 2) could be close to 
a bright 24 $\mu m$ source, or 3) might not be true disk-bearing stars.

To infer the physical state of disks surrounding stars without MIPS detections, 
we first examine the processed MIPS mosaic image to identify any stars that are 
$\sim$ 2--3 $\sigma$ above the background or are blended with another star.  
Next, we identify stars that are accreting and thus either have 
primordial or more evolved primordial disks (transition/homologously depleted). 
Finally, we use the MIPS upper limits to constrain the possible range of 
24 $\mu m$ fluxes and thus constrain the disk evolutionary state.  

\begin{figure}
\epsscale{0.5}
\plotone{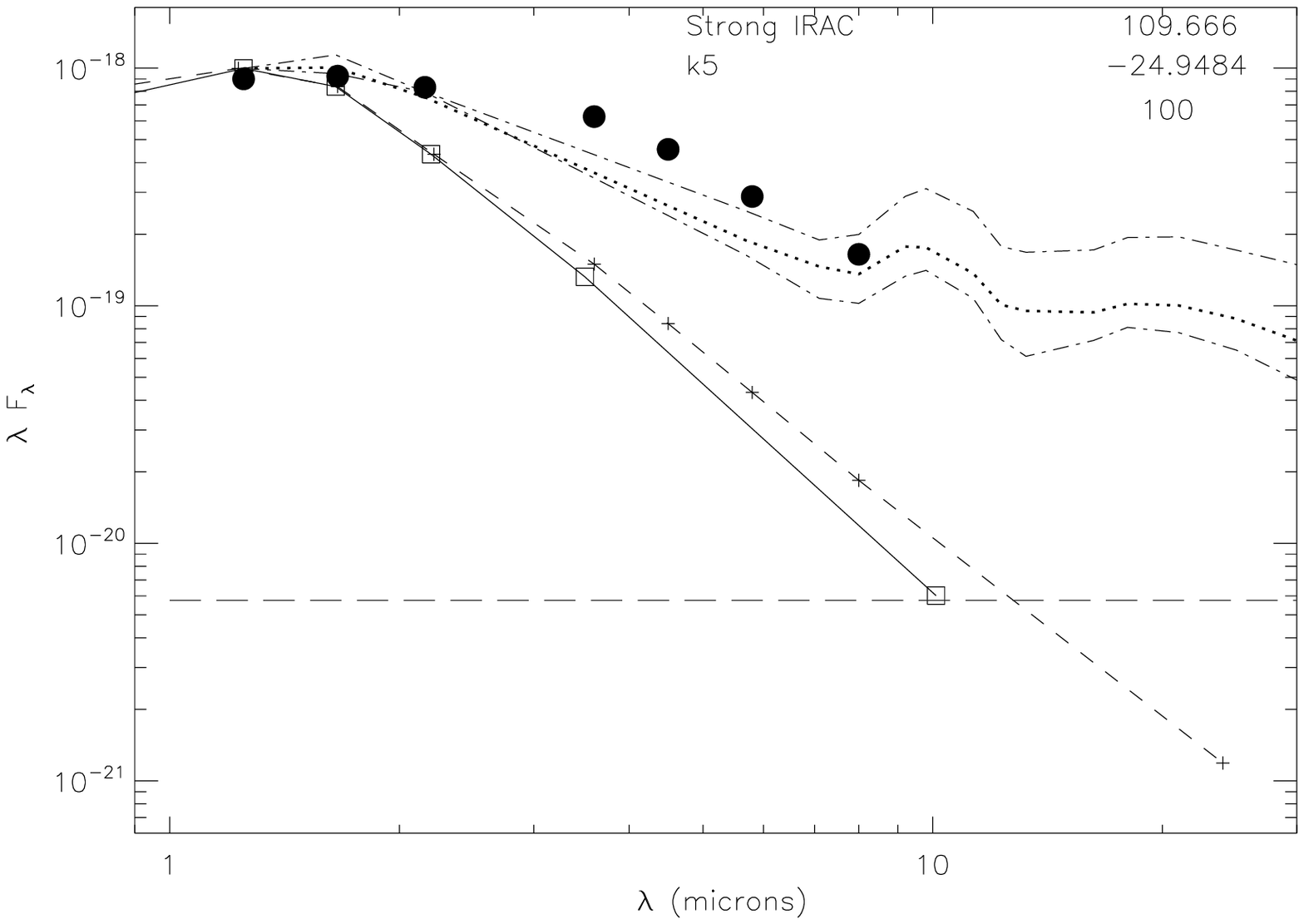}\\
\epsscale{0.9}
\plottwo{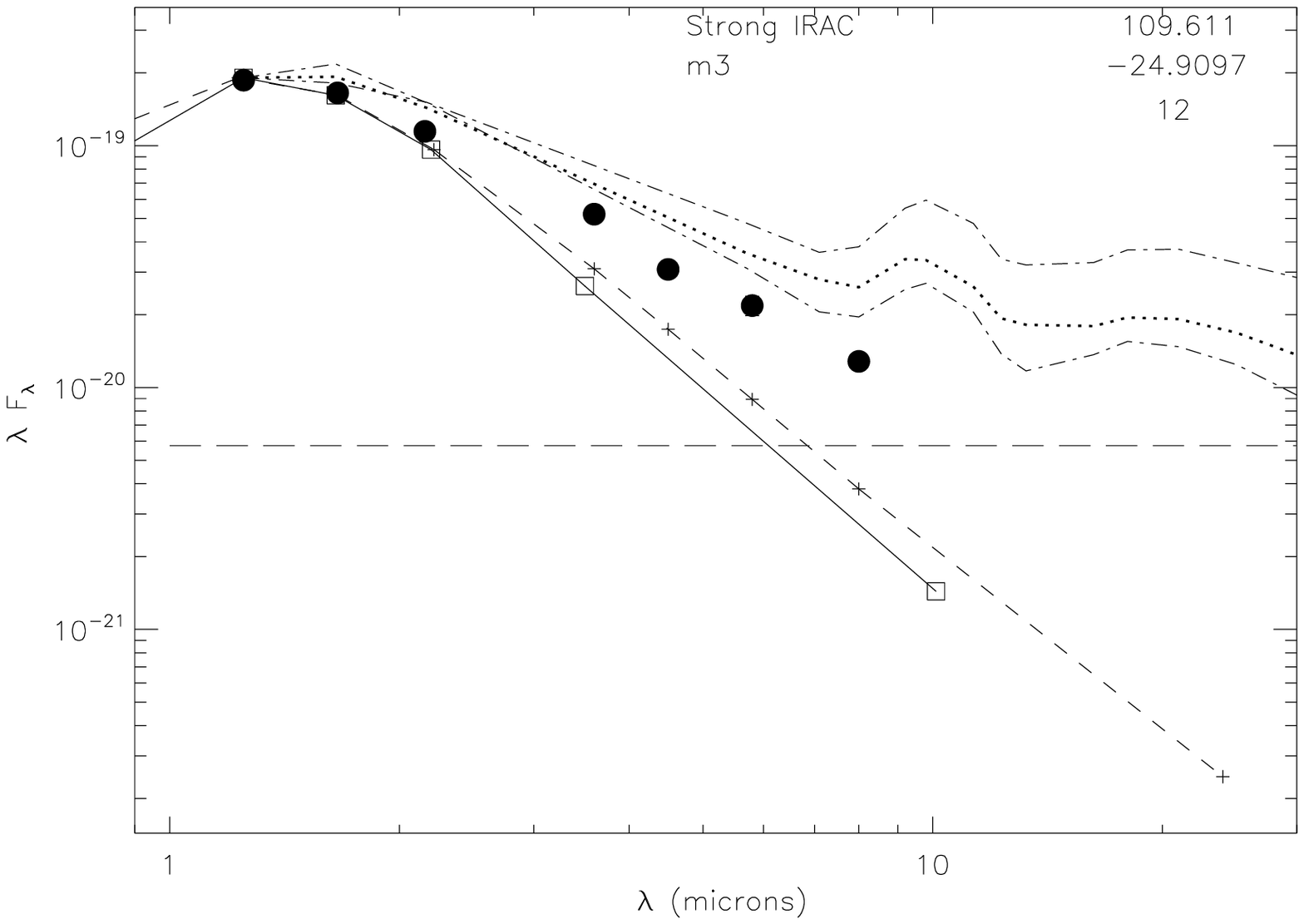}{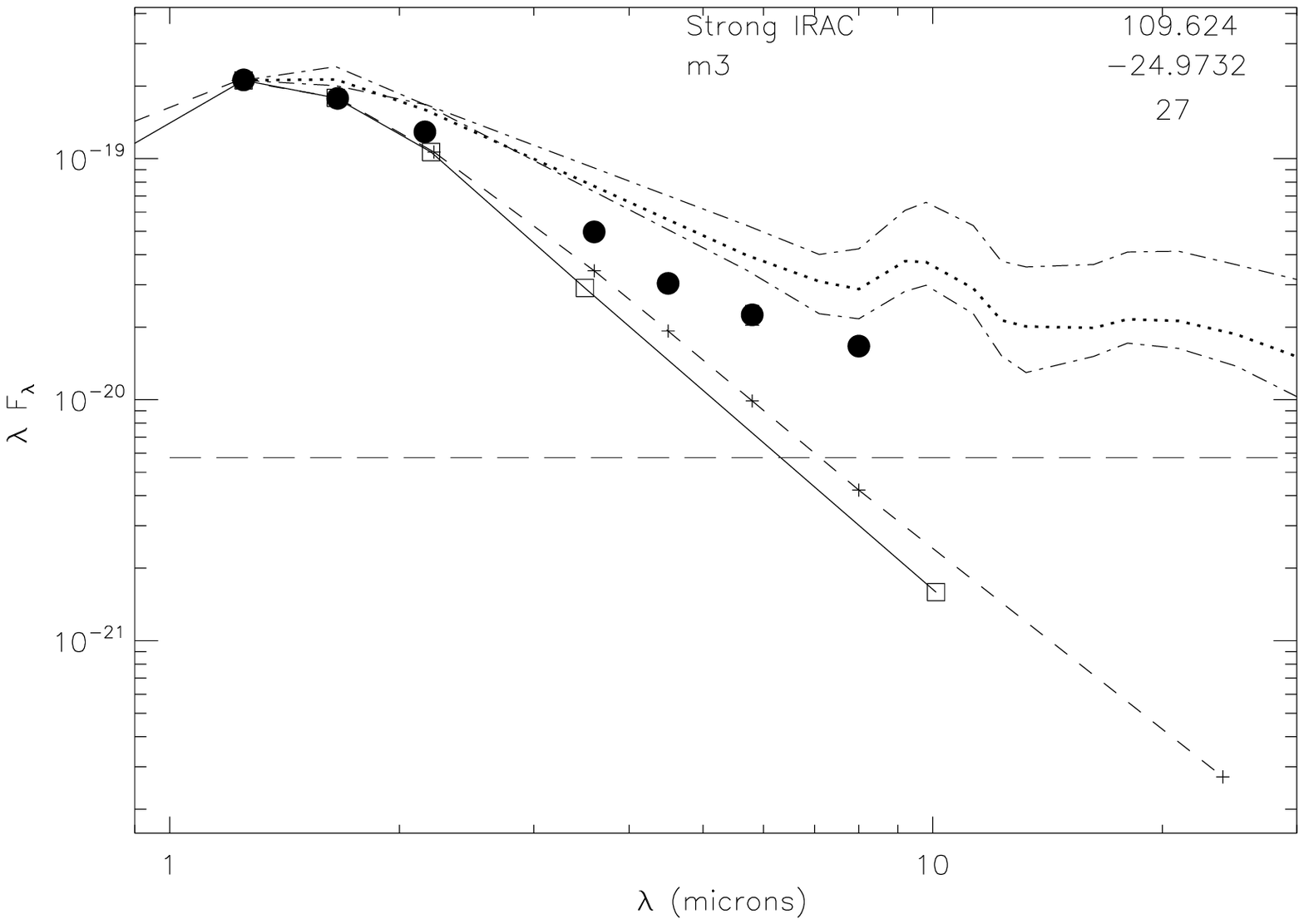}
\plottwo{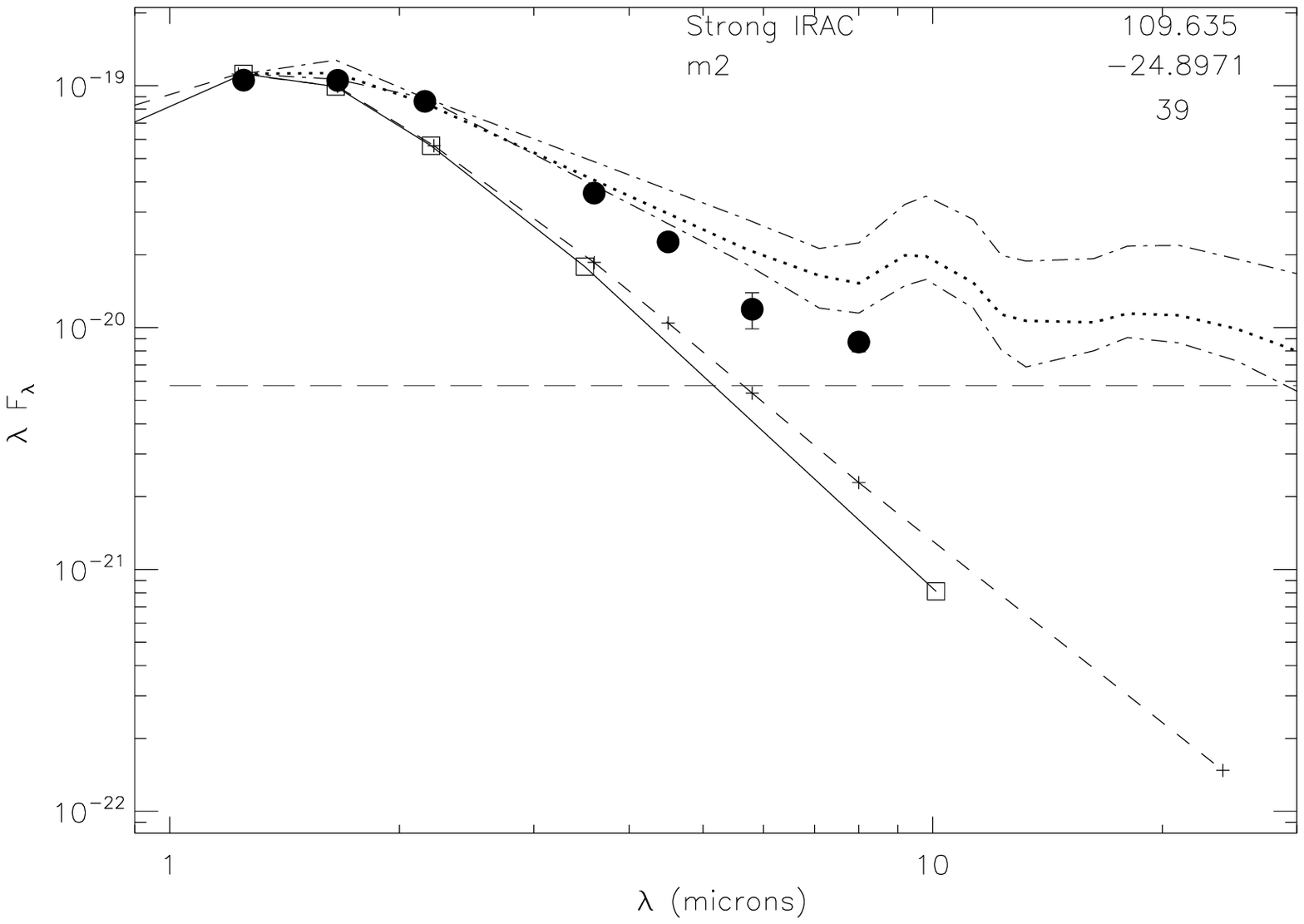}{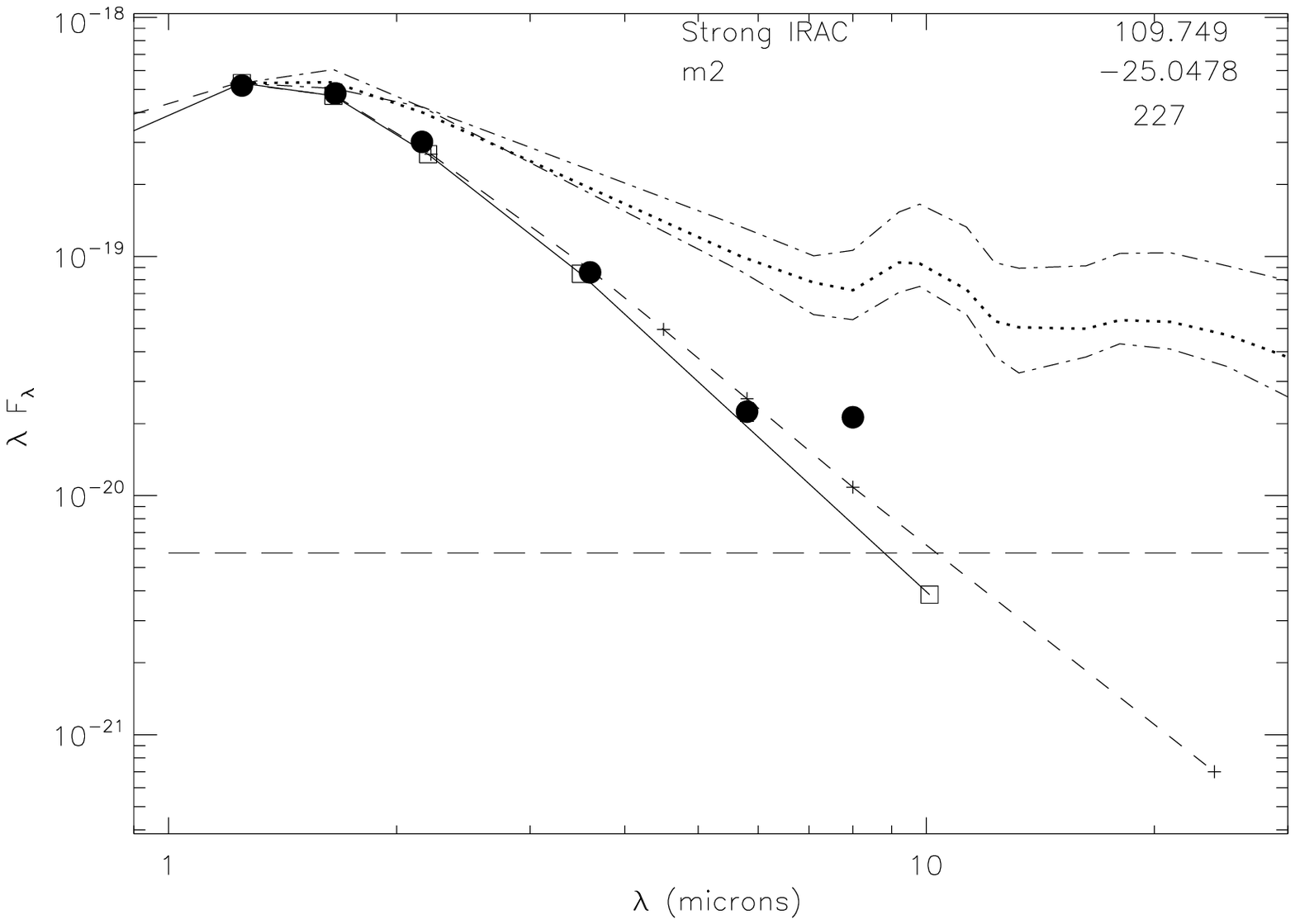}
\caption{Spectral energy distributions for primordial/evolved primordial 
disk candidates from the \citet{Dh07} sample lacking MIPS detections.  Symbols are the same as before.}
\label{sedupplim}
\end{figure}

The bottom section of Table \ref{diskevostate} lists the results of our analysis and Figure \ref{sedupplim} shows 
SEDs of five additional primordial and evolved primordial disk candidates.  One star, ID 100, 
is likely surrounded by a primordial disk given its strong 
IRAC excess emission.  In the MIPS 24 $\mu m$ filter, 
it appears to be blended with a brighter cluster member.
Four stars -- IDs 12, 27, 39, and 227 -- have clear IRAC excess emission and 
are barely discernable on the MIPS mosaic at very high contrast.  The signal-to-noise for these sources 
is likely $\sim$ 2--3, which would exclude them from the 
MIPS source list.  ID 12 shows evidence for accretion.  
The SEDs of these four stars lie below the lower quartile for the median Taurus 
SED, which indicates that they are probably most 
similar to homologously depleted disks.  The remaining sources 
lack evidence for accretion, are not discernable on the MIPS mosaic, and have 
 weak, $\lesssim$ 1 magnitude IRAC excesses.  These sources either not are true disk candidates, 
have very large inner holes, or harbor optically-thin disks, possibly debris disks.

\subsubsection{Summary of SED Analysis}
Analyzing the SEDs of late type, MIPS-detected 
NGC 2362 stars reveals disks with a wide range of evolutionary states.  
Primordial disks in NGC 2362 are very rare.  Only 
19\% of MIPS-detected confirmed/candidate cluster stars
from \citet{Dh07} have SEDs consistent with primordial disks.  
Combining the \citet{Dh07} sample with the \citet{Ir08} sample of 14 MIPS-detected stars 
indicates that the true fraction of 
primordial disks in NGC 2362 is even lower, $\sim$ 17\% (6/35).

Most disks surrounding late-type cluster stars are in an advanced evolutionary state.
  In the \citet{Dh07} sample, $\sim$ 57\% (12/21) of late type, MIPS-detected sources 
 have SEDs consistent with disks with a 
reduced mass at all disk radii (homologously depleted disks).  If candidates from \citet{Ir08} 
are included, the fraction of these disks is $\sim$  
49\% (17/35).  Disks with inner holes/gaps ('transition disks') also comprise a 
large fraction of MIPS-detected sources, representing $\sim$ 24\% (5/21) 
of MIPS-detected \citeauthor{Dh07} sources and $\sim$ 34\% (12/34) of all confirmed/candidate 
cluster members detected by MIPS.  In total, 81\% (83\%) of the MIPS-detected late-type stars in the 
\citeauthor{Dh07} sample (\citeauthor{Dh07}+\citeauthor{Ir08} samples combined) 
have disks in evolutionary states between primordial disks and debris disks.  
Any other disk-bearing stars lacking MIPS detections have 24 $\mu m$ emission lying 
below the detection limit.  Compared to the stars with MIPS detections, disks 
around these stars have even weaker emission suggestive of substantial evolution.
\textbf{Therefore, disks around most late type stars in NGC 2362 are leaving the 
primordial disk phase}. 

Our results stand in contrast to those from \citet{Dh07} who find that primordial disks 
comprise $\sim$ 34\% of the total disk population.
Our photometry identifies a comparable number of confirmed members with $\alpha$ $\ge$ -1.8 (16) and between 
-1.8 and -2.56 (30), so poorer photometry is unlikely to explain this discrepancy.  However, 
many sources (e.g. ID 3) listed in Table \ref{diskevostate} and identified as 'primordial' disks 
by \citeauthor{Dh07} clearly have SEDs that do not look anything like canonical 
primordial disks found in Taurus and other young star-forming regions.  
In part, some of these descrepancies result from the adopted definition of 
a primordial disk.  In this paper, we define a primordial disk to be a disk 
with an SED close to the median Taurus SED and a disk mass $\gtrsim$ 10$^{-3}$ M$_{\odot}$.
Modeling shows that only five of the 16 stars with $\alpha$ $\ge$ -1.8 have 
SEDs and disk masses typical of primordial disks.  
Using only a single parameter (the IRAC slope) then results in a larger number of 
primordial disks than identified from full SED modeling.  
\section{Implied Constraints on The Evolution of Circumstellar Disks and Planet Formation}

The disappearance of primordial disks and 
the emergence of debris disks sets an empirical limit for the formation 
timescale of gas giant planets.  The disappearance of IR excesses in disks 
implies the disappearance of the building blocks -- solid material -- for the 
cores of gas giant planets.  Similarly, the lack of gas accretion implies a lack of 
circumstellar gas in inner disk regions, which corresponds to a lack of material to build the gaseous 
envelopes of gas giant planets.  Therefore, in this section we use results from 
analysis presented in \S 4 to 
constrain the evolutionary states of disks and to investigate the formation timescale of 
gas giant planets as a function of stellar mass (\S 5.1).
We divide the sample into high/intermediate-mass stars (M$_{\star}$ $\gtrsim$ 1.4 M$_{\odot}$) 
and solar/subsolar-mass stars ($\lesssim$ 1.4 M$_{\odot}$).

From the population of NGC 2362 stars surrounded exclusively by debris disks, 
we compare their 24 $\mu m$ debris luminosity with that for $\gtrsim$ 5 Myr-old stars 
to investigate the evolution of debris emission and the connections 
this evolution has with the growth of icy planets (\S 5.2).  In 
\S 5.3, we compare the more evolved primordial disk population to disks in other 1--10 Myr-old clusters 
to examine the morphology of and timescale for the primordial-to-debris disk transition. 

The results presented in this section are based on analyzing the \textit{MIPS-detected} population 
of NGC 2362 confirmed/candidate members and other cluster stars with clear evidence 
for disks listed in Table \ref{diskevostate}.  Most faint confirmed/candidate cluster stars 
are not detected by MIPS because of its bright 5$\sigma$ limit ([24] $\sim$ 10.5).  
Therefore, completeness at 24 $\mu m$ may affect our results.  However, 
we show that any disk-bearing sources not detected by MIPS either harbor debris 
disks or homologously depleted disks with low masses.  MIPS is sensitive enough 
to detect levels of 24 $\mu m$ emission typical of primordial disks surrounding 
faint cluster stars and emission typical of all but low-luminosity debris disks 
around brighter cluster stars.  Thus, our ability to determine when primordial disks 
disappear is not undermined by sample incompleteness.  Because different dust masses are 
implied for detections as opposed to non detections, the detection 
limit may serve as a discriminator between disk types. 

Because not all of the MIPS-detected members and candidate members have published spectral types, 
this section uses the J--K$_{s}$ color as a tracer of spectral type and thus, assuming a 4--5 Myr age, stellar 
mass.  While the lack of published spectral types for some sources adds uncertainty to our analysis, 
known properties of NGC 2362 stars show that this uncertainty is not significant given the coarse 
division into stars with masses greater than or lesser than 1.4 M$_{\odot}$.  NGC 2362 has a small 
 dispersion in age \citep{Mn08}, which yields a reliable conversion from spectral type to stellar mass.  
The cluster also has a very low reddening (E(B-V) $\sim$ 0.1), which translates into E(J--K$_{s}$) $\sim$ 0.05--0.06.  
For early type, high/intermediate-mass stars (BAFG) to be misidentified as low-mass stars based on their 
observed J--K$_{s}$ colors, they must have a reddening $\sim$ 3--5 times the typical reddening of the cluster.  
There is little evidence for such a large spread in reddening.  Cluster stars with known spectral types have 
J--K$_{s}$ colors consistent with using J--K$_{s}$ = 0.6 as the division between high/intermediate-mass stars 
and low-mass stars.
\subsection{Constraints on the Formation Timescale for Gas Giant Planets}
Assuming an age of $\sim$ 4--5 Myr, NGC 2362 stars with masses $>$ 1.4 M$_{\odot}$
have G or earlier spectral types and J--K$_{s}$ $<$ 0.6 \citep{Ba98}.  Cluster 
stars with masses $\ge$ 2 M$_{\odot}$ have spectral types earlier than mid F and 
J--K$_{s}$ $\lesssim$ 0.3 \citep{Kh95, Si00}.
From Figure \ref{jvk24}, the K$_{s}$-[24] upper limits for stars with 
M$_{\star}$ $\sim$ 1.4--2 M$_{\odot}$ are $\sim$ 2.5; the upper limits for 
more massive, 2 M$_{\odot}$ stars are $\sim$ 1--1.5.  In comparison the 
predicted K$_{s}$-[24] excesses for primordial disks 
and transition disks discussed in \S 4 are $\sim$ 4--5.5 magnitudes 
over this spectral type range.  Thus, any primordial disks or 
transition disks around cluster stars with masses $>$ 1.4 M$_{\odot}$ 
are easily detectable.  Stars with 'homologously depleted' 
disks listed in Table \ref{diskevostate} have K$_{s}$-[24] $>$ 3 
and are also easily detectable.

Analysis from \S 4 shows that there are \textbf{no} primordial disks, 
transition disks, or homologously depleted disks surrounding 
NGC 2362 stars plausibly in this mass range. 
MIPS-detected stars with M$_{\star}$ $\ge$ 2 M$_{\odot}$ 
have photospheric emission or $\lesssim$ 2 magnitude excesses (cf. 
Figures \ref{jvk24} and \ref{k4k24}).
MIPS-detected stars with masses likely between 1.4 M$_{\odot}$ and 2 M$_{\odot}$
include only one star, ID 144, with nonphotospheric emission.
High/intermediate-mass cluster stars with optical spectra do not 
show any evidence for H$_{\alpha}$ emission indicative of gas accretion.

Based on SED modeling and dust removal timescales (\S 4), all disks surrounding 
high/intermediate-mass cluster stars are debris disks.  Furthermore, 
even though Spitzer observations only trace thermal emission from small
 circumstellar dust grains not gas, the observable signatures of 
disks in evolutionary states with gas (primordial, transition, homologously depleted) 
are clearly lacking from \textbf{any} high/intermediate-mass star 
studied here.  NGC 2362 then lacks any disks around high/intermediate-mass stars 
which have yet to reach the debris disk phase.  
Recent combined Spitzer and ground-based surveys of young clusters 
indicate that circumstellar gas should disperse on similar or faster timescales than 
dust \citep{La06}.  If this is the case, our 
results suggest that \textit{gas giant planet formation likely occurs 
prior to 4--5 Myr for $>$ 1.4 M$_{\odot}$ stars}.

MIPS-detected late-type stars in NGC 2362 have inferred masses between $\sim$ 0.3 M$_{\odot}$ and 
1.4 M$_{\odot}$ \citep[cf.][]{Ba98, Si00, Dh07}.  
Most late-type stars detected by MIPS have SEDs consistent with more evolved primordial 
disks or debris disks.  About 1/3 of MIPS-detected disks in the \citet{Dh07} sample are transition disks.
Nearly half of all MIPS-detected stars are surrounded by disks that are
  depleted in mass at a range of disk radii (homologously depleted).
There is no clear connection between disks with this morphology and active 
planet formation.  Only four stars have SEDs consistent with primordial disks.
Given the MIPS detection limit for this sample, we can rule out the presence of any 
$\gtrsim$ 4.5 magnitude excesses typical of primordial disks and 
transition disks.

  Disks surrounding late-type stars in NGC 2362 that are not detected 
by MIPS must have K$_{s}$-[24] $\le$ 4--4.5.  These disks
 are likely either debris disks or homologously depleted disks with 
levels of emission slightly below the MIPS detection limit.  The disk population of 
solar and subsolar-mass stars is dominated by the large number of 
 MIPS upper limits, making statistically robust conclusions about the 
bulk disk population difficult.  However, combined with the low frequency of 
accretion in NGC 2362 \citep{Dahm05, Dh07}, these results 
demonstrate that most disks surrounding late-type stars have significantly evolved 
and that most lack evidence for circumstellar gas needed to form gas giant planets.


While some primordial disks remain around 
$\lesssim$ 1.4 M$_{\odot}$ stars, the disk population for these stars is 
dominated by disks which have clearly undergone a significant amount of evolution 
(transition disks, homologously depleted disks).  The morphologies of 
transition disks and homologously depleted disks are consistent with 
disks actively leaving the primordial disk phase for the debris disk phase.
If these disks are actively making the primordial-to-debris disk transition, 
then gas giant planet formation also likely ceases by $\approx$ 5 Myr for many, perhaps most, 
solar/subsolar-mass stars.
\subsection{Debris disk evolution around High/Intermediate-Mass Stars}
Combining NGC 2362 data for high/intermediate-mass stars with data from 
other clusters probes the time evolution of debris emission from 
planet formation.  Specifically, we examine the time evolution of 24 $\mu m$ 
excess emission (see also \citealt{Ri05}, \citealt{Su06}, \citealt{He06}, \citealt{Cu08a}, 
and \citealt{Wy08}), focusing on 5--40 Myr-old
early-type stars.  We include all confirmed/candidate high/intermediate-mass 
members in NGC 2362 with MIPS detections.
In addition to our data, we include stars from clusters with debris disks:
 Orion OB1a and b (10 and 5 Myr; \citealt{He06}), $\eta$ Cha (8 Myr; \citealt{Gau08}),
the $\beta$ Pic Moving Group (12 Myr; \citealt{Re08}),
h and $\chi$ Persei ($\sim$ 13 Myr; \citealt{Cu08a}),
Sco-Cen, and NGC 2547 (38 Myr; \citealt{Ri05}).
For the Sco-Cen, we include data for Upper Scorpius (5 Myr),
Upper Centaurus Lupus (16 Myr), and Lower Centaurus Crux (17 Myr)
from \citet{Ch05}.  We limit our literature sample to
BAF stars.

\begin{figure}
\plotone{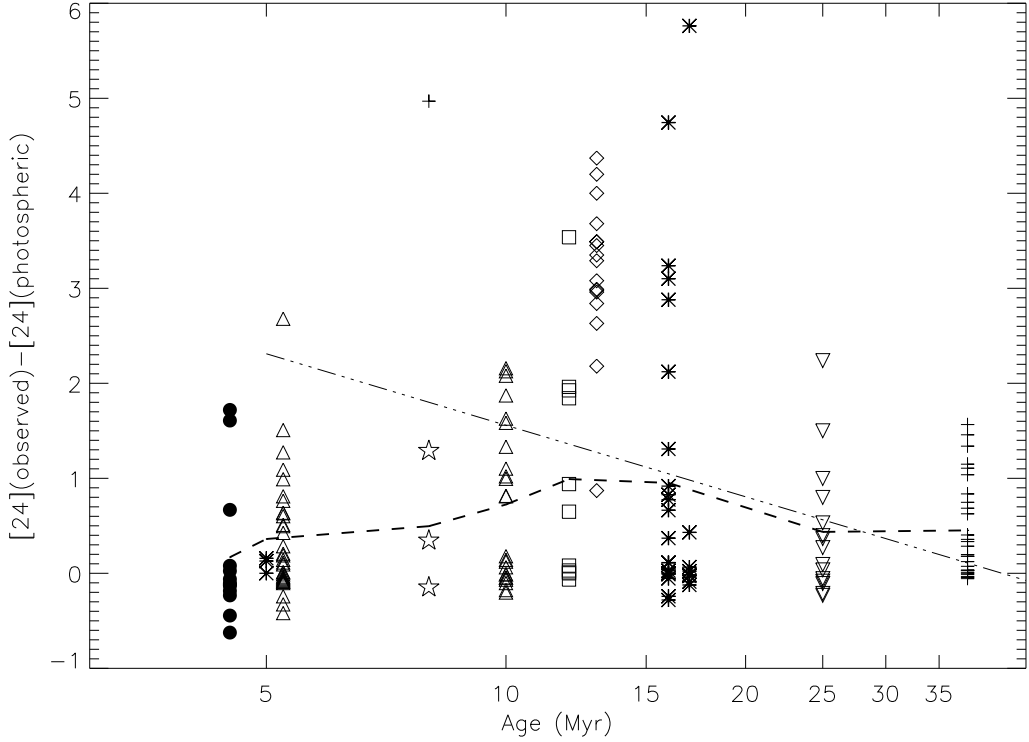}
\caption{Amplitude of 24 $\mu m$ excess around BAF stars
vs. time for 4--40 Myr old clusters.  Besides NGC 2362 (filled dots), we include
Orion Ob1a and b (10 and 5 Myr; triangles), $\eta$ Cha (8 Myr; stars), the $\beta$ Pic Moving Group (12 Myr;
squares), h and $\chi$ Persei (13 Myr; diamonds), Sco-Cen (5, 16, and 17 Myr;
asterisks), NGC 2232 (25 Myr; inverted triangles), and NGC 2547 (38 Myr; small crosses).
HR 4796A, from \citet{Lo05} is also shown as a cross at t$\sim$ 8 Myr and
[24]-[24]$_{phot}$ $\sim$ 5.  The data
are consistent with a rise in debris emission from 5 Myr to 10 Myr, a peak from
$\sim$ 10 Myr to 20 Myr as shown previously by \citet{Cu08a}.  
For stars older than $\approx$ 40 Myr, the distribution
of 24 $\mu m$ luminosity defines an envelope which is consistent with a t$^{-1}$ decline
\citep{Ri05}.  The dashed line shows the mean excess. Predictions for the decline in emission from
a steady-state collisional evolution model (dash-three dots) follow a t$^{-1}$ decline.}
\label{24evotime}
\end{figure}

 Figure \ref{24evotime} shows the individual amplitudes of 24 $\mu m$ excesses 
as a function of time.  In support of \citet{Cu08a}, there is a rise in the level of debris emission
from $\sim$ 5 Myr to 10 Myr and a peak at $\sim$ 10--20 Myr \citep[see also ][]{Cpk08}.
\citet{Ri05} show  that the emission at later ages ($\gtrsim$ 30--50 Myr) is 
consistent with a 1/t decline (dot-dashed line).
The mean level of excess (dashed line) for $\lesssim$ 20 Myr-old clusters
clearly deviates from a t$^{-1}$ decline.

MIPS data for NGC 2362 stars supports the 
trend of debris disk evolution from 5 Myr to 10 Myr 
first identified by \citet{Cu08a}.  Three 
clusters/associations in Figure \ref{24evotime} have ages of 
$\sim$ 5 Myr: Orion OB1b, Upper Scorpius, and NGC 2362.  
All three clusters/associations have levels of excess 
that are weaker than those for $\sim$ 10-15 Myr-old clusters, 
Orion Ob1a, the $\beta$ Pic Moving Group, h and $\chi$ Persei, 
and Upper Centaurus Lupus.  Debris disks in 5 Myr-old clusters 
also have lower 24 $\mu m$ luminosities than those in 
25--40 Myr-old clusters NGC 2232 and NGC 2547 \citep{Cpk08, Ri05}.
The conclusion that debris disks surrounding early-type, high-mass 
stars exhibit a 'rise' and 'fall' trend \citep{Cu08a}
instead of a monotonic decline in emission \citep[e.g.][]{Ri05} 
is now very robust.

The observed behavior of debris disk evolution agrees well
 with predictions from sophisticated debris disk models.
\citet{Kb08} investigate the evolution of debris emission from 
icy planet formation in disks surrounding $\gtrsim$ 1 M$_{\odot}$ 
stars.  For $\sim$ 2 M$_{\odot}$ stars with a disk mass of
$\sim$ 2--3$\times$ a scaled Minimum Mass Solar Nebula, 24 $\mu m$ debris 
emission rises rapidly from $\sim$ 5 Myr to $\sim$ 8 Myr after 
the start of the simulation which is sustained for $\approx$ 20 Myr.  
Spitzer observations indicate a $\approx$ 3 Myr lifetime for 
primordial disks around these stars \citep{He07, Ck08}.  The 
\citeauthor{Kb08} models would then predict an observed peak 
in 24 $\mu m$ debris emission by $\sim$ 8--11 Myr that lasts through 
$\approx$ 30 Myr.  These expectations agree with
the trend shown in Figure \ref{24evotime}. 

Within the context of the \citet{Kb08} models, the trend in 24 $\mu m$ 
debris emission is directly linked to the growth of icy planets.
The \citeauthor{Kb08} simulations start with a swarm of $\approx$ 0.1--10 km-sized 
planetesimals.  These bodies rapidly grow to $\approx$ 100--1000 km sizes 
within several Myr in the \textit{runaway growth} phase \citep{Ws93}.  
Once icy bodies grow to $\gtrsim$ 1000 km sizes, they stir the 
leftover planetesimals to much higher velocities.  These higher velocities 
reduce the gravitational focusing factor of the growing planets, and the 
planets enter the much slower \textit{oligarchic} growth phase.  The higher 
velocities also cause more energetic collisions between planetesimals, causing 
more fragmentation and increasing the dust production rate.  
Because small dust is needed to make debris disks visible in the mid-IR, \citet{Kb08} identify 
the onset of oligarchic growth with the peak in debris disk luminosity.
Therefore, the \citeauthor{Kb08} models associate the systematically 
weaker excesses for $\approx$ 5 Myr-old stars with an early stage in planet 
formation where icy bodies are $\lesssim$ 1000 km in size and actively growing.  
Similarly, the models associate the stronger excesses surrounding 10--30 Myr-old 
stars with later stages in planet formation where some icy bodies are $\gtrsim$ 1000 
km in size and are growing at slower rates.

\subsection{Constraints on the Primordial-to-Debris Disk Transition for Solar and Subsolar-Mass Stars} 
The large number of stars with disks in an intermediate stage between 
primordial disks and debris disks allows us to investigate the 
primordial-to-debris disk transition in detail.  Our results challenge  
conventional ideas about two aspects of the primordial-to-debris disk transition: the 
morphology of disks in such an intermediate stage and the timescale for 
transition.

In agreement with previous Spitzer work, particularly \citet{La06}, our results indicate 
the existence of at least two evolutionary paths from 
primordial disks to debris disks.  Many MIPS-detected cluster members have 
transition disk morphologies with
 SEDs showing weak, optically-thin emission through $\approx$ 5--8 $\mu m$ 
but strong, optically-thick emission by $\approx$ 24 $\mu m$.  However, 
many more stars are surrounded by homologously depleted disks whose mass 
is likely lost at all disk radii simultaneously.  While 24 $\mu m$ 
emission for transition disks is stronger than for homologously depleted disks, 
homologously depleted disks typically have stronger $\lesssim$ 8 $\mu m$ emission.
It is unlikely that either of these morphologies is 
 the evolutionary precursor of the other.  \textit{Thus, strictly speaking, there is 
more than one type of 'transition disk'}.  Both morphologies can be thought of 
as 'evolved primordial disks' since the disk characteristics suggest they are 
more likely to be composed of first generation dust, not second generation dust 
(i.e. debris disks).   

Our results also place new constraints on the timescale for the 
transition from primordial disks to debris disks.
According to studies of disks in young, $\lesssim$ 3 Myr-old clusters, 
most notably Taurus-Auriga, the paucity of transition disks relative to 
primordial disks implies that the transition 
timescale is $\lesssim$ 10$^{5}$ years \citep{Sk90, Har90, Sp95, Ww96}.  
Our work and other recent work \citep[e.g.][]{Haisch01, He07} indicates
that the lifetime of primordial disks surrounding 
many young stars is $\sim$ 3--5 Myr.  In this picture, the lifetime of 
the primordial disk phase is $\gtrsim$ 30--50$\times$ the lifetime of the transition disk phase. 
However, if disks spend a factor of 30--50$\times$ longer in the primordial disk phase, then 
primordial disks should always be far more numerous than transition disks.
If, on the other hand, the transition disk and primordial disk lifetimes are comparable, 
then transition disks should be equal to or greater in number than primordial disks by $\approx$ 5 Myr.

Based on these data, \textit{the traditional arguments favoring a $\lesssim$ 10$^{5}$ yr transition 
timescale are incorrect}.  For the MIPS-detected \citet{Dh07} sample, 
 transition disks are equal in number to primordial disks.  When the \citet{Ir08} sample of 
candidate members is included, transition disks are about twice as numerous as 
primordial disks.  Many disks are also homologously depleted and represent a separate 
evolutionary path from primordial disks to debris disks.  The combined population of MIPS-detected 
homologously depleted disks and transition disks in the \citeauthor{Dh07} sample 
are more numerous than primordial disks by a factor of $\sim$ 4.  When the \citet{Ir08} 
data are included, they dominate over the primordial disk population by 6 to 1.
Even if the homologously depleted disks with the strongest mid-IR emission (IDs 36, 85, 204, 214, and 
251) were all reclassified as primordial disks, primordial disks would still comprise 
less than half of the total disk population for late-type stars.
Because many homologously depleted disks may have 24 $\mu m$ fluxes lying just below the 
MIPS detection limit, the population of disks in between primordial and debris disks 
may be even more numerous.  

Our results strongly favor a longer, $\sim$ 1 Myr timescale for the 
primordial-to-debris disk transition for disks with ages $\sim$ 5 Myr.  
Besides this work, there is evidence in favor of such a trend from other 
recent Spitzer surveys.  Spitzer observations of some $\lesssim$ 5 Myr-old clusters reveal 
a substantial population of transition disks and homologously depleted disks consistent 
with $\sim$ 1 Myr timescale for the primordial-to-debris disk transition.
MIPS-detected disks showing evidence for substantial evolution in 2--3 Myr-old IC 348, labeled \textit{anemic} 
disks in \citet{La06}, comprise $\approx$ 40-45 \% of all MIPS-detected disks \citep[][]{Ck08}, a 
fraction much higher than that for Taurus-Auriga.  SED analysis of these sources shows that 
about two thirds are homologously depleted disks and one third are transition disks \citep{La06, Ck08}. 
Spitzer photometry and IRS spectroscopy of low-mass stars in the $\approx$ 1 Myr-old 
Coronet Cluster also reveal a population of transition disks and homologously depleted 
disks whose high frequency is in conflict with a $\lesssim$ 10$^{5}$ year transition 
timescale \citep{SiA08}.

These evolved primordial disks also dominate over primordial disks in at least two samples of $\gtrsim$ 5 Myr-old 
stars.  Most sources with optically-thick mid-IR emission from the FEPS sample 
(ages $\sim$ 3--30 Myr) have SEDs indicating that their disks' inner regions are cleared of 
dust to distances probed by $\sim$ 3.6--8 $\mu m$ \citep{Hi08}.  In four of the six stars 
shown in \citet{Hi08}, the SEDs rise from 8--13 $\mu m$ to $\sim$ 24 $\mu m$ as in 
transition disks.  The remaining two stars have SEDs showing excess emission weaker than that 
of typical primordial disks and indicative of homologously depleted 
disks studied here.  The 2MASS/IRAC survey of h and $\chi$ Persei (13--14 Myr) fail to detect any stars 
with clear K-band excess emission indicative of primordial disks but find that $\sim$ 4--8\% 
have 8 $\mu m$ excess emission \citep{Cu07a}.  While most of these stars likely harbor warm debris disks, 
several sources analyzed further in \citet{Cu07c} and one analyzed in \citet{Cu08a}, 
have accretion signatures and/or mid-IR colors suggestive of evolved primordial disks.

\section{Summary and Discussion}
In this paper, we analyze Spitzer data for the $\sim$ 5 Myr-old cluster NGC 2362 to 
explore the primordial-to-debris disk transition and the formation timescale for 
gas giant planets.  From the list of confirmed/candidate cluster members, we 
analyze mid-IR colors, compare source SEDs to disk models representating a range 
of evolutionary states, and compare SEDs to signatures of gas accretion to 
constrain the evolutionary state of disks.  Then we compare the properties of 
disks surrounding high-to-low mass stars to constrain the timescale for gas 
giant planet formation.  From the sample of early-type, high/intermediate-mass 
stars with debris disks, we investigate the evolution of debris emission from 
planet formation.  Finally, we study the morphology of disks evolving from 
primordial disks to debris disks and place limits on the timescale for this transition.  

\subsection{Overview of Results}
Our study yields the following major results:
\begin{itemize}
\item For all stellar masses, NGC 2362 contains very few stars surrounded by primordial disks 
with strong near-to mid-IR emission characteristic of 1 Myr-old stars in Taurus.
The disk population is dominated by evolved primordial disks, whose SEDs are indicative 
either of so-called transition disks (inner holes/gaps) or homologously depleted disks.

\item NGC 2362 lacks any stars with masses $\gtrsim$ 1.4 M$_{\odot}$ that have primordial disks 
which comprise the building blocks of giant planets.  These stars also lack evidence for 
having transition or homologously depleted disks.  Because most disks surrounding 1 Myr-old stars show 
evidence for protoplanetary disks, gas giant planets surrounding $\gtrsim$ 1.4 M$_{\odot}$ 
stars likely form in $\approx$ 1--5 Myr.

\item Most disks surrounding solar/subsolar-mass stars are either transition disks or 
homologously depleted disks and thus are 
likely in the process of dispersing.  Some disks may be debris disks, indicating that any 
gas giant planet formation has ceased.  Thus, gas giant planets may form by $\sim$ 4--5 Myr for 
stars with a wide range of masses.

\item Both canonical transition disks and 
homologously depleted disks represent an evolutionary pathway 
from primordial disks to debris disks \citep[see also][]{La06}.   Furthermore, there are 
a large number of these evolved primordial disks compared to primordial disks.  
This work and previous work suggests that primordial 
disks have a typical lifetime of $\sim$ 3--5 Myr.  Thus, the timescale for the 
primordial-to-debris disk transition must be $>>$ 0.1 Myr
 and may be roughly comparable to the primordial disk lifetime (e.g. $\approx$ 1 Myr).  A
 $\approx$ 1 Myr timescale for this evolution is inconsistent with published 
models predicting a rapid transition.
\end{itemize}

\subsection{On the Primordial-to-Debris Disk Transition}
Despite sample incompleteness in MIPS for the faintest cluster stars, 
the presence of debris disks around the high/intermediate-mass stars 
and presence of at least many primordial disks and evolved primordial 
disks around lower-mass stars in NGC 2362 is consistent with the stellar-mass dependent 
evolution of primordial disks found in other recent Spitzer studies.  
\citet{La06} showed that disks in 2--3 Myr-old IC 348 exhibit a spectral type/stellar-mass 
dependent frequency of warm dust from disks with the frequency peaking for
early M stars.  \citet{He07} also find a stellar-mass dependent frequency of 
warm dust in 3 Myr-old $\sigma$ Orionis.  

Morever, our work suggesting that debris disks may emerge faster around 
high/intermediate-mass stars supports the study of 5 Myr-old Upper Scorpius 
from \citet{Ca06}.  \citet{Ca06} find that high/intermediate-mass stars in 
Upper Scorpius not only have a lower frequency of disk emission compared 
to low-mass stars but also have a lower 16 $\mu m$ disk luminosity.
While \citet{Ca06} do not model the disk emission from high/intermediate-mass stars 
to demonstrate that these stars are surrounded by debris disks, they 
do note that the weak 16 $\mu m$ emission and lack of accretion signatures for these stars
are consistent with debris disk characteristics.  This work and that of \citeauthor{Ca06} 
combined then provide strong evidence that debris disks dominate the disk population 
of high/intermediate-mass stars by $\sim$ 5 Myr.  Analysis of the disk population 
around high-mass stars in IC 348 indicate that debris disks may emerge 
even sooner \citep{Ck08}.

A $\approx$ 1 Myr timescale for the primordial-to-debris disk transition 
undermines the viability of UV photoevaporation as the primary physical process by 
which most primordial disks disappear as presented in published models from 
\citet{Al06} and \citet{Cl01}.  In standard photoevaporation models \citep[e.g.][]{Al06}, 
once stellar accretion rates drop below the photoevaporative mass loss rate active 
disk dispersal can commence.  Once this condition is met, 
 primordial disks are completely dispersed within $\sim$ 0.01--0.1 Myr.
\citet{Al06} showed that disks dispersing by UV photoevaporation 
follow clear evolutionary tracks in their mid-IR colors:
stars with colors consistent with bare photospheres and 
 colors consistent with primordial disks should \textit{always} be more frequent than 
those with intermediate colors.  In contrast, our results find that disks in an intermediate 
stage dominate over primordial disks.  The inside-out dispersal of disks in photoevaporation 
models is also not consistent with the morphology of homologously depleted disks, which 
comprise the plurality of MIPS-detected disks in NGC 2362.  

The only way in which the NGC 2362 data is consistent with a very rapid transition timescale expected from 
standard photoevaporation models is if the spread in primordial 
disk lifetimes is very narrow ($<<$ 1 Myr) and with a mean value of $\approx$ 4 Myr (R. Alexander, pvt. comm.).
However, if this were the case for all clusters, plots of the disk frequency 
vs. time for clusters \citep[e.g.][]{Haisch01, He07} would look more like a step function than a 
steady decline.  Analysis of Spitzer data for other clusters both younger (Cha I, IC 348, $\sigma$ Ori) 
and older (TW Hya Association, 25 Ori) suggest a larger spread in lifetimes 
 \citep{Luh08, La06, He07, Lo05, He07b}.   For each cluster, especially those older than $\sim$ 2--3 Myr, 
there exists a wide diversity of disk properties.  Some stars around 2--3 Myr old stars have very weakly emitting, 
highly evolved disks or lack any evidence for a disk \citep{La06}.  Conversely, $\sim$ 5--10 Myr old clusters have some stars
with primordial disks, homologously depleted disks, or canonical transition disks \citep[e.g.][]{Lo05, He07b}.
For photoevaporation models to be consistent with observed disk properties, initial parameters (e.g. EUV flux, 
$\alpha$ viscosity parameter) must be fine-tuned to yield a far longer disk transition timescale.  It is 
unclear whether these models can modified in such a way.

Because our results do not find evidence for a rapid transition timescale, the supposedly rapid timescale 
for all disks as inferred from Taurus data must have alternate explanations.  Stars in Taurus show
an apparently rapid transition timescale because of two main factors: 1) their youth and
 2) binarity.  First, previous Spitzer studies indicate that the typical primordial disk lifetime is at least $\sim$ 3 Myr 
\citep{La06, He07}, which is older than the median age of Taurus ($\sim$ 1--2 Myr).  
 Therefore, most disks in Taurus (1--2 Myr) are too young to have begun leaving the 
primordial disk phase.  The frequency of transition or homologously depleted disks compared to primordial disks 
for Taurus should then be low as observed.  While it is possible that some stars (e.g. those in Taurus) 
make the primordial-to-debris transition in $\sim$ 10$^{5}$ years, our results indicate that a 
more typical transition timescale is far longer.

Second, binarity  affects primordial disk evolution.  \citet{Ik08} show 
that CoKu Tau/4, long regarded a prototypical transition disk, 
is actually a short-period binary system.  Its lack of warm dust emission is due to tidal interactions with the faint
binary companion and not from physical processes typically associated with transition disks (e.g. planet formation, photoevaporation).  
Recent high-contrast imaging of Taurus stars without disks indicate that a large number of them also have
short-period binaries (A. Kraus et al. in prep.).  Because binary companions are responsible for clearing these disks, 
the true number of stars lacking a disk because of standard disk dispersal processes (grain growth, accretion, planet 
formation, photoevaporation) is likely \textit{far} lower than previously thought.  Therefore,
 \textit{the small fraction of transitional disks in Taurus and other young star-forming regions 
does \textbf{not} necessarily mean that the typical transition timescale due to disk processes is rapid}.

\subsection{Implications for the Formation of Gas Giant Planets and Future Work}
Combined with results from previous work, the most straightforward 
interpretation of our results is that gas giant planets 
 have to form sometime between 1 Myr and 4--5 Myr around most stars 
or else they will not form at all.  We acknowledge that the strength 
of this conclusion is limited by using indirect tracers for the presence 
of gas-rich primordial disks for many stars.  It is also possible that the absolute age calibration 
of NGC 2362 is uncertain by several Myr.  Nevertheless, a short, 1--5 Myr 
formation timescale for gas giant planets is in strong agreement with timescales inferred 
from properties of our solar system's planets and satellites.  For instance, 
a 1--5 Myr timescale is consistent with the results 
of \citet{jcast07} who argue that Saturn formed as early as 
 2.5 Myr after the formation of calcium-aluminum-rich inclusions (CAIs), 
whose formation may have been contemporaneous with the birth of the Sun and the solar 
nebula.

A short formation gas giant formation timescale presents a challenge 
for core accretion models of gas giant planet formation.
Many detailed models of planet formation \citep[e.g.][]{Alibert05a} 
assume that circumstellar disks with initial masses comparable to the Minimum Mass 
Solar Nebula are depleted by only factors of $\sim$ 2--5 
by $\sim$ 5 Myr.  
Our analysis suggests that primordial disks may not last this long.
Other models \citep[e.g.][]{Dodson08} claiming to 
form gas giants within $\sim$ 3--5 Myr rely on simple
treatments of core growth that neglect important effects such as 
the evolution of the planetesimal velocity distribution and collisional fragmentation 
of planetesimals.  To date, only the model of \citet{Kb08b} has 
successfully modeled the formation of gas giant planet cores in $\lesssim$ 5 Myr 
 while properly incorporating these first-order effects.
The relative timescales for core growth and the dispersal of primordial disks 
may explain trends in exoplanet properties \citep{Cu09}.

In spite of the failure of most core accretion models to form gas giant planets 
in less than 5 Myr while accounting for important effects, radial velocity surveys indicate that gas giants 
are common \citep{Cumming08}.  Moreover, gas giants may be most frequent around 
high/intermediate-mass stars \citep{johnjohn07}.  Our results imply that the 
formation timescales for gas giants around these stars are probably much shorter compared 
to those for lower-mass stars.  If gas giants form by core accretion then their formation, especially 
around high/intermediate-mass stars, must be very rapid and efficient.

Recent and upcoming Spitzer surveys provide an opportunity to explore the
primordial-to-debris disk transition and planet formation timescales further.  Unpublished Spitzer IRS data 
as well as Cycle 5 IRAC and MIPS observations of massive clusters 
(e.g. Upper Scorpius and h and $\chi$ Persei) strengthen constraints 
on the primordial-to-debris disk transition at $\sim$ 5 Myr and  
 probe this transition at later, $\sim$ 10--15 Myr ages.  
Combined with this study, these surveys should result in strong constraints 
on the evolution of disks from the primordial disk-dominated epoch ($\lesssim$ 5 Myr) 
to the debris disk-dominated epoch ($\gtrsim$ 5--15 Myr) and yield the most 
definitive statements on the timescale for planet formation from the 
\textit{Spitzer Space Telescope} mission.

\acknowledgements{
We thank Richard Alexander for useful comments about UV photoevaporation and 
Adam Kraus for very informative discussions on the role of binarity in disk evolution.  
We also thank Carol Grady for pointing out constraints on formation of Saturn from 
the planetary science community.  Support for this work was provided by NASA 
through the Spitzer Space Telescope Fellowship Program (TPR).}

\begin{deluxetable}{lllllllllllllllllll}
 \tiny
\setlength{\tabcolsep}{0.02in}
\tabletypesize{\tiny}
\tablecolumns{16}
\tablecaption{Photometry Catalog for All Stars on the NGC 2362 Field}
\tiny
\tablehead{{Number}&{RA}&{DEC}&{J}&{H}&{K$_{s}$}&{[3.6]}&{$\sigma$([3.6])}&{[4.5]}&{$\sigma$([4.5])}&
{[5.8]}&{$\sigma$([5.8])}&{[8]}&{$\sigma$([8])}&{[24]}&{$\sigma$([24])}}
\startdata
1& 110.0939& -25.1644&  13.729&  13.486&  13.379&  13.267&   0.011&   0.000&   0.000&  13.185&   0.070&   0.000&   0.000&  99.000&   0.000\\
   2& 110.1280& -25.3220&  12.178&  11.345&  11.118&   0.000&   0.000&   0.000&   0.000&   0.000&   0.000&   0.000&   0.000&  10.021&   0.184\\
   3& 110.0985& -25.1656&  12.275&  12.206&  12.014&  11.964&   0.005&   0.000&   0.000&   0.000&   0.000&   0.000&   0.000&  99.000&   0.000\\
   4& 110.0921& -25.1472&  16.153&  15.419&  15.431&  15.186&   0.039&   0.000&   0.000&   0.000&   0.000&   0.000&   0.000&  99.000&   0.000\\
   5& 110.0960& -25.1629&  14.618&  14.142&  14.054&  13.977&   0.017&   0.000&   0.000&  13.434&   0.085&   0.000&   0.000&  99.000&   0.000
\enddata
\tablecomments{The table includes photometry for all stars on the NGC 2362 field with 
a detection in IRAC and/or MIPS.  
Values of 99.000 or 0.00 in the photometry column and 0.00 in the photometric uncertainty columns denote 
sources that were not observed in a given filter.} 
\label{allphot}
\end{deluxetable}

\begin{deluxetable}{lllllllllllllllllll}
 \tiny
\setlength{\tabcolsep}{0.02in}
\tabletypesize{\tiny}
\tablecolumns{19}
\tablecaption{Photometry Catalog for NGC 2362 Cluster Members}
\tiny
\tablehead{{ID}&{RA}&{DEC}&{V}&{R}&{I}&{J}&{H}&{K$_{s}$}&{[3.6]}&{$\sigma$([3.6])}&{[4.5]}&{$\sigma$([4.5])}&
{[5.8]}&{$\sigma$([5.8])}&{[8]}&{$\sigma$([8])}&{[24]}&{$\sigma$([24])}}
\startdata
1&109.5817&-24.9550&19.430&18.470&16.800&15.220&14.620&14.370&14.130&0.050&14.110&0.070&14.125&0.148&14.054&0.138&99.000&0.000\\
2&109.5905&-24.9578&17.420&16.450&15.650&14.670&13.950&13.770&13.580&0.040&13.590&0.050&13.491&0.086&13.705&0.097&99.000&0.000\\
3&109.5909&-24.8615&18.170&17.170&15.950&14.600&13.860&13.580&13.240&0.030&13.060&0.040&12.745&0.047&12.089&0.022&7.649&0.009\\
4&109.5942&-25.0283&19.680&18.530&16.920&15.540&14.910&14.520&14.280&0.050&14.250&0.070&13.979&0.129&13.885&0.117&99.000&0.000\\
5&109.6002&-24.8997&18.890&17.840&16.420&14.930&14.410&14.010&14.000&0.060&13.850&0.060&13.819&0.113&13.992&0.130&99.000&0.000\\
\enddata
\tablecomments{The table includes photometry for NGC 2362 cluster members identified by \citet{Dh07}.  
Values of 99.000 or 0.00 in the photometry column and 0.00 in the photometric uncertainty columns denote 
sources that were not observed in a given filter.} 
\label{dhmemphot}
\end{deluxetable}

\begin{deluxetable}{lllllllllllllllllll}
 \tiny
\setlength{\tabcolsep}{0.02in}
\tabletypesize{\tiny}
\tablecolumns{10}
\tablecaption{Photometry Catalog for NGC 2362 Candidate Members from \citet{Ir08}}
\tiny
\tablehead{{ID}&{RA}&{DEC}&{V}&{I}&{J}&{H}&{K$_{s}$}&{[3.6]}&{$\sigma$([3.6])}&{[4.5]}&{$\sigma$([4.5])}&
{[5.8]}&{$\sigma$([5.8])}&{[8]}&{$\sigma$([8])}&{[24]}&{$\sigma$([24])}}
\startdata
1& 109.5003 &-25.1379&  20.515&  17.525 & 15.985&  15.346&  14.938&  14.835 &  0.030 & 14.831 &  0.041&  14.773 &  0.265 & 14.637&   0.273 & 99.000  & 0.000\\
   2& 109.5145& -25.1769 & 16.786 & 15.526&  14.564 & 13.971&  13.854 &  0.000&   0.000&  13.793 &  0.020&   0.000&   0.000 & 14.005 &  0.135 & 99.000 &  0.000\\
 3& 109.5160 & -25.1295 & 21.548 & 18.795 & 99.000 & 99.000 & 99.000 & 16.067 &  0.079 & 16.044  & 0.111  & 0.000 &  0.000 &  0.000 &  0.000 & 99.000 &  0.000\\
   4& 109.5212& -25.1615 & 20.936&  18.074&  16.399&  15.728&  99.000&  15.392&   0.045&  15.228&   0.056&  14.925&   0.308 & 14.879 &  0.369&  99.000 &  0.000\\
   5& 109.5234& -25.2406 & 22.386&  18.925&  16.617&  15.570&  14.916&   0.000&   0.000&   0.000&   0.000&   0.000&   0.000 &  0.000 &  0.000&   9.800 &  0.064\\
\enddata
\tablecomments{The table includes photometry for candidate NGC 2362 cluster members identified by \citet{Ir08}.  
Values of 99.000 or 0.00 in the photometry column and 0.00 in the photometric uncertainty columns denote 
sources that were not observed in a given filter.} 
\label{irmemphot}
\end{deluxetable}

\begin{deluxetable}{lllllllclcll}
 \tiny
\setlength{\tabcolsep}{0.0125in}
\tabletypesize{\tiny}
\tabletypesize{\scriptsize}
\tablecolumns{11}
\tablecaption{Dust Removal Timescales and Dust Masses for Early-Type Stars with MIPS 24 $\mu m$ Excesses}
\tablehead{{ID}&{J-K$_{s}$}&{Spectral Type} & {{L$_{\star}$}/{L$_{\odot}$}}&{M/M$_{\odot}$}& 
{r$_{dust}$ (AU)}&{t$_{pr}$ (yr)}&{s$_{max}$}&{Disk}&{ max(M$_{dust}$})\\
{}&{}&{(Estimated)}&{(Estimated)}&{(Estimated)}&{(200, 120 K)}&
{(10, 30 AU)} &{($\mu m$)}&{Model}&{($\times$10$^{-10}$M$_{\odot}$)}}
\startdata
31&0.11&A5&30.0&2.3&10.7,29.8&2.3$\times$10$^{4}$, 2.1$\times$10$^{5}$&14.8&3020154-5-9&14\\
144&0.45&G5&10.0&2.0&6.2,17.2&7.0$\times$10$^{4}$, 6.3$\times$10$^{5}$&5.8&3008807-5-9&1.1\\
303&0.01&A0&35.3&2.5&11.6,32.3&2.0$\times$10$^{4}$, 1.8$\times$10$^{5}$&16.0&3017274-5-9&2.2\\
314&-0.05&B9&56.5&2.7&14.7,40.9&1.2$\times$10$^{4}$, 1.1$\times$10$^{5}$&23.8&3017274-5-9&2.2\\
\enddata
\label{noremnant}
\tablecomments{Constraints on the dust removal timescales and dust masses for early-type  
sources.  The Poynting-Robertson drag timescale is calculated assuming 10 $\mu m$-sized grains.  Disk models 
listed represent the file number from the \citet{Ro06} grid; dust masses for those models are listed as 
M$_{dust}$.  
}
\end{deluxetable}

\begin{deluxetable}{lllllllllll}
 \tiny
\setlength{\tabcolsep}{0.02in}
\tabletypesize{\tiny}
\tablecolumns{8}
\tablecaption{Evolutionary State of Confirmed/Candidate NGC 2362 
Members with Evidence for Disk Emission}
\tiny
\tablehead{{ID}&{ST}&{A$_{v}$ Range}&{EW(H$_{\alpha}$)}&{log(M$_{disk}$/M$_{\odot}$)}&{R$_{in, min}$ (AU)}&{$\beta$}&{Disk Type}&{Notes}}
\startdata
\cutinhead{Dahm and Hillenbrand (2007) sample with MIPS detections}
   3& M2&0-0.65& NA&-4.51$\pm$ 0.48& 4.10 & 1.05&Transition\\
   36 &K5 &0-0.53&-6.7&-3.02 $\pm$ 1.06&0.07 & 1.02&Homologously Depleted\\
   41 &K1& 0-0.33&-0.1&-7.39 $\pm$ 0.42 & 0.12 & 1.05&Homologously Depleted\\
   63 &M0$^{1}$& 0.31?&-2.9&-6.62 $\pm$ 1.02 & 0.06 & 1.06 &Homologously Depleted\\
   85 &M2 & 0-0.93&-45.4& -3.57 $\pm$ 0.93 & 0.02 & 1.14 & Homologously Depleted\\
   111&K7& 0-0.45&-18.8& -2.07 $\pm$ 0.36 & 0.08 & 1.13 &Primordial\\
   139&M2& 0-0.8&NA& -2.09 $\pm$ 0.39 & 0.11 & 1.13 &Primordial\\ 
   168&K3& 0.55-0.9&-0.2& -5.63 $\pm$ 0.85 & 5.20 & 1.09 &Transition\\
   177&M2& 0-0.2&-3.0 & -3.12 $\pm$ 0.97 & 0.06 & 1.11 &Homologously Depleted\\
   187&M2& 0.03-1.3&-117.7& -3.58 $\pm$ 0.7 & 0.27 & 1.04 &Primordial$^{2}$\\
   194&K5&1.02-1.8&-27.5 & -3.91 $\pm$ 0.57 &7.57 & 1.12& Transition\\
   196&M0$^{1}$& 0.31?& NA& -5.95 $\pm$ 0.54 & 3.63 & 1.04 &Transition\\
   202&M3& 0-1.08&-2.6 & -2.41 $\pm$ 0.12 & 0.03 & 1.18 & Primordial\\
   204&M3& 0-1.05&-10.4 & -3.87 $\pm$ 0.85 & 0.03 & 1.16 &Homologously Depleted\\
   213&M3& 0-0.8&-10.2 & -4.35 $\pm$ 1.03 & 0.04 & 1.10 &Homologously Depleted\\
   214&K7& 0-0.43&-3.5 & -3.63 $\pm$ 1.00 & 0.06 & 1.10 &Homologously Depleted\\
   219&M0& 0.78-1.68&-5.9 & -4.24 $\pm$ 1.14 & 0.03 & 1.07 &Homologously Depleted\\
   229&M1.5& 0-0.95&-3.8 & -3.81 $\pm$ 1.05 & 0.03 & 1.07 &Homologously Depleted\\
   251&M2$^{1}$& 0.31?&NA& -3.98 $\pm$ 0.82 & 0.02 & 1.15 &Homologously Depleted\\
   267&M2$^{1}$& 0.31?&NA& -4.35 $\pm$ 0.79 & 0.02 & 1.12 &Homlogously Depleted\\
   295&M2$^{1}$& 0.31?&NA& -4.36 $\pm$ 1.12 & 1.29 & 1.08 &Transition\\
\cutinhead{Irwin et al. (2008) sample with MIPS detections}
   IR-185&M2$^{1}$& 0.31?&NA&-4.12 $\pm$ 1.05 & 0.75 & 1.07 &Transition\\
   IR-459&M2$^{1}$& 0.31?&NA& -4.59 $\pm$ 1.57 & 0.05 & 1.06 &Homologously Depleted\\
   IR-526&M2$^{1}$& 0.31?&NA& -3.62 $\pm$ 0.76 & 0.03 & 1.14 &Primordial\\
   IR-583&M2$^{1}$& 0.31?&NA& -3.77 $\pm$ 0.94 & 0.6 & 1.12 &Transition\\
   IR-663&M2$^{1}$& 0.31?&NA& -4.41 $\pm$ 0.93 & 0.02 & 1.10 &Homologously Depleted\\
   IR-732&M2$^{1}$& 0.31?&NA& -3.30 $\pm$ 1.27 & 0.04 & 1.05 &Homologously Depleted\\
   IR-750&M2$^{1}$& 0.31?&NA& -3.10 $\pm$ 0.72 & 4.32 & 1.09 &Transition\\
   IR-809&M2$^{1}$& 0.31?&NA& -4.93 $\pm$ 1.07 & 0.03 & 1.08 &Homologously Depleted\\
   IR-817&M2$^{1}$& 0.31?&NA& -3.22 $\pm$ 0.63 & 1.95 & 1.11 &Transition\\
   IR-851&M2$^{1}$& 0.31?&NA& -3.25 $\pm$ 0.51 & 0.21 & 1.14 &Transition\\
   IR-931&M2$^{1}$& 0.31?&NA& -4.24 $\pm$ 1.27 & 0.39 & 1.03 & Transition\\
   IR-958&M2$^{1}$& 0.31?&NA& -5.09 $\pm$ 1.19 & 0.04 & 1.07 &Homologously Depleted\\
   IR-1046&M2$^{1}$& 0.31?& NA& -2.9 $\pm$ 0.95 & 0.04 & 1.13 &Primordial\\
   IR-1091&M2$^{1}$& 0.31?&NA& -4.39 $\pm$ 1.64 & 1.45 & 1.06 &Transition\\
\cutinhead{Disk Candidates from Dahm and Hillenbrand (2007) without MIPS detections}
   12& M3& 0.9-1.72&-20.2& -&-&-&Homologously Depleted& 3\\
   27& M3& 0.2-1.03&-7.2& -&-&-&Homologously Depleted&3 \\
   39& M2$^{1}$& 0.31?&NA &-&-&-&Homologously Depleted& 3\\
   100&K5& 0.3-1.1&NA& -&-&-&Primordial & 4\\
   227&M2$^{1}$& 0.31?&$<$ 0& -&-&-&Homologously Depleted& 3\\
\enddata
\tablecomments{Footnote (1) identifies stars whose spectral types are estimated from optical/near-infrared colors.  
Footnote (2): IDs 187 and IR-526 have low disk masses but are classified as primordial 
based on their mid-IR fluxes (compared to the median Taurus SED).
Note (3) in the last column -- Stars with probable 2--3 $\sigma$ detections at MIPS 24 $\mu m$.
Note (4) in the last column -- MIPS detection is blended with another star. 
The range in extinction, A$_{v}$, is calculated by comparing the observed V-I colors 
to intrinsic colors from \citet{Kh95}, assuming a 1 subclass uncertainty in the 
spectral type.  
Negative values for EW(H$_{\alpha}$) identify stars with 
H$_{\alpha}$ emission.  Sources lacking a measured EW(H$_{\alpha}$) have NA listed in the EW(H$_{\alpha}$) column.  
The column log(M$_{disk}$/M$_{\odot}$) refers to the unweighted average and standard deviation of all models that 
provide a good fit; R$_{in, min}$ is the minimum inner radius of all models with good fits.}
\label{diskevostate}
\end{deluxetable}


\begin{figure}
\centering
\plottwo{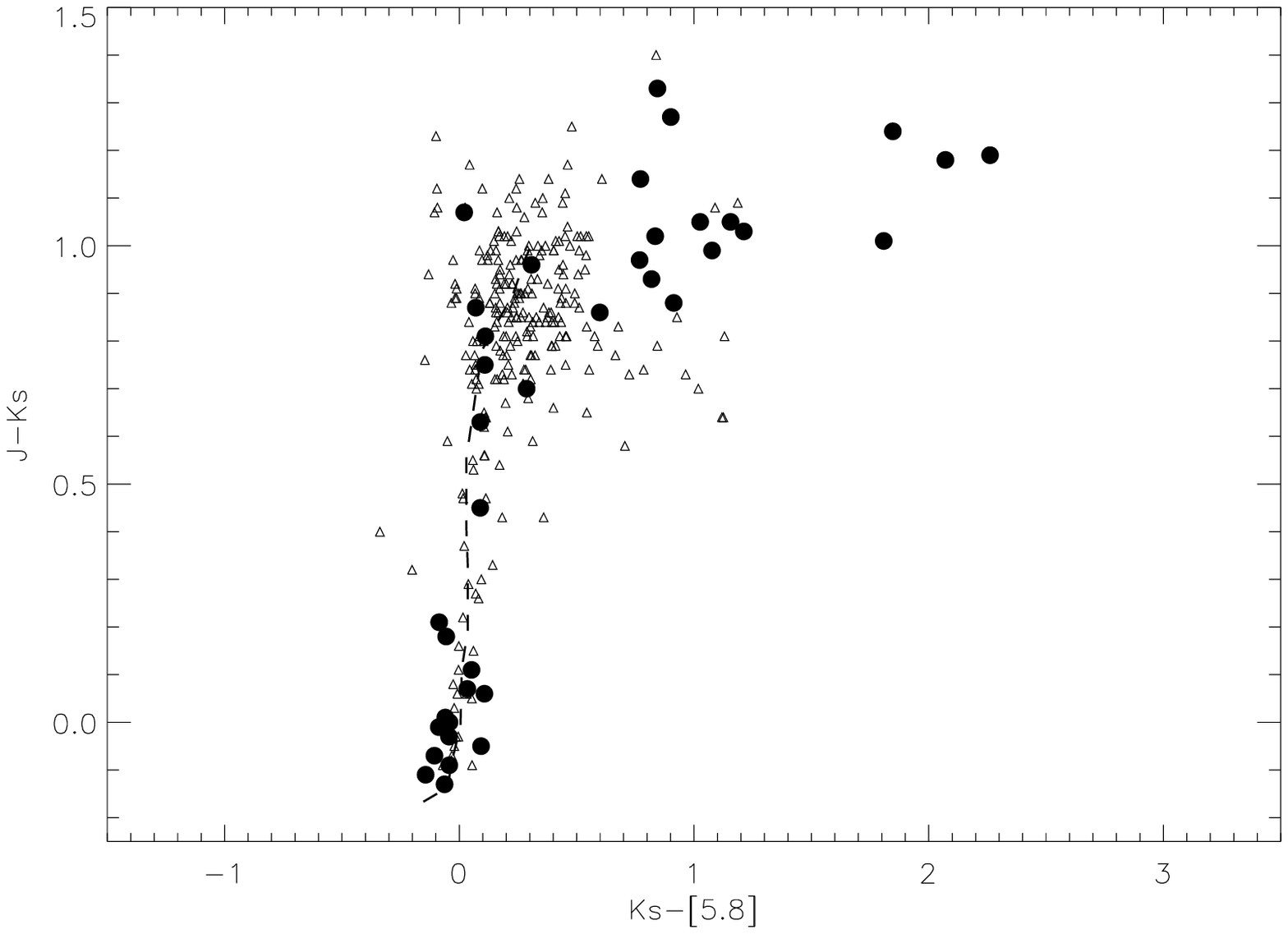}{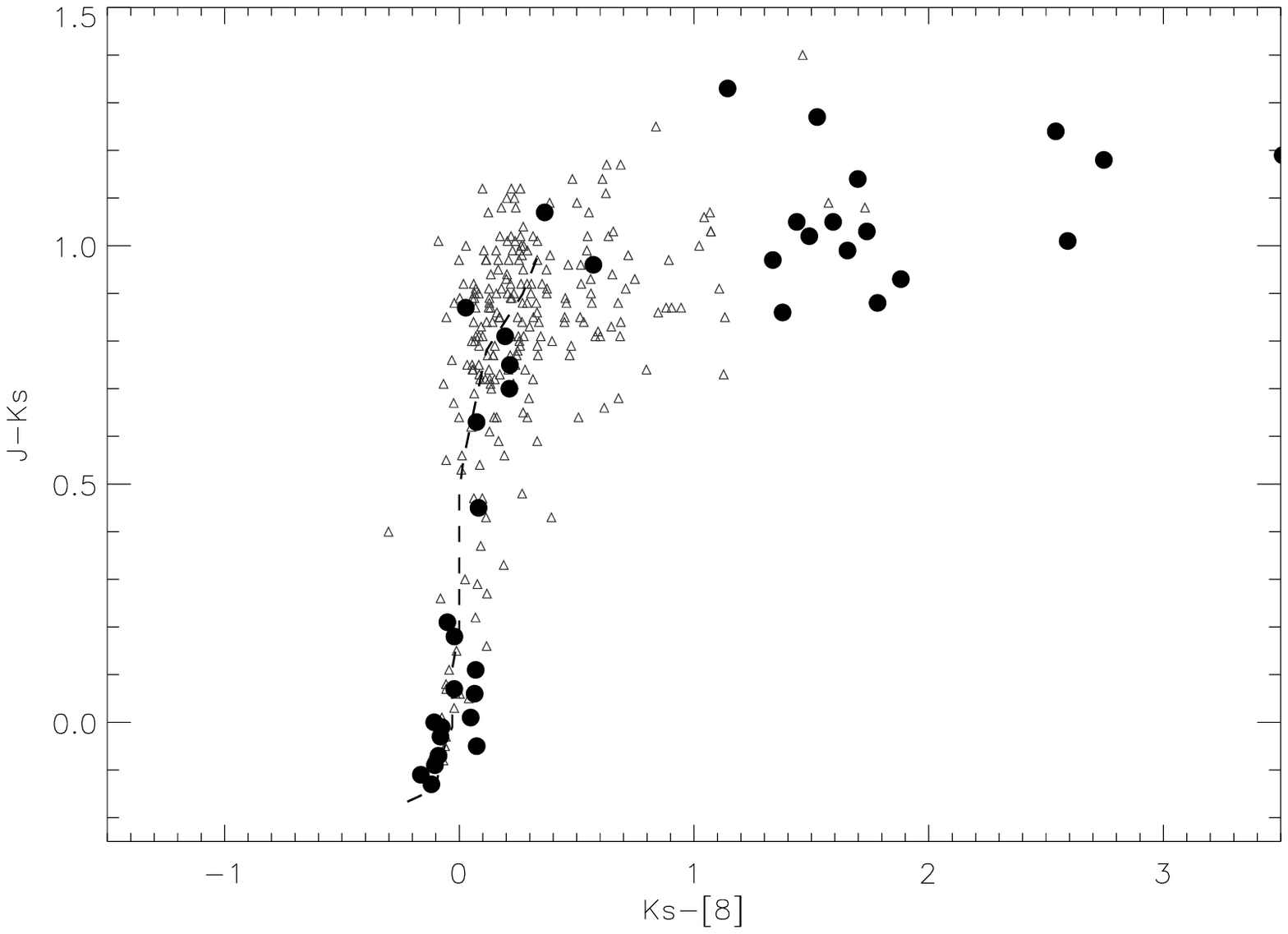}
\caption{J-K$_{s}$ vs. K$_{s}$-[5.8] (left) and K$_{s}$-[8] color-color 
diagrams for sources in the \citet{Dh07} membership list using 
photometry derived in this paper.  Sources lacking 
MIPS detections are shown as triangles.  Sources with MIPS detections are shown 
as dots.  Overplotted are the locii of photospheric colors (dotted lines) from the Kurucz-Lejeune 
stellar atmosphere models as calculated by the SENS-PET tool.  Most sources with 
IRAC excesses have MIPS detections.}
\label{jkk34dh}
\end{figure}
\begin{figure}
\centering
\plottwo{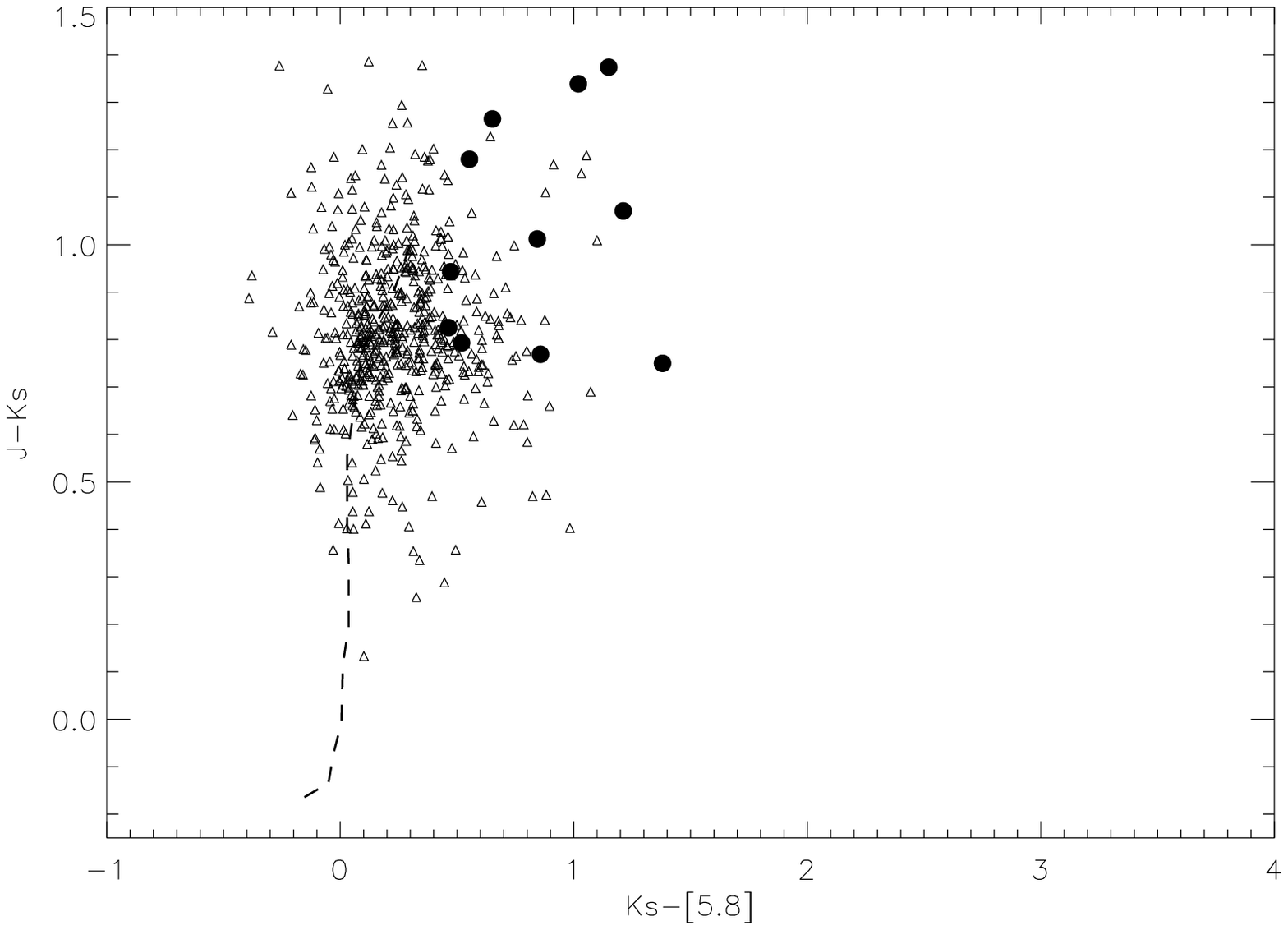}{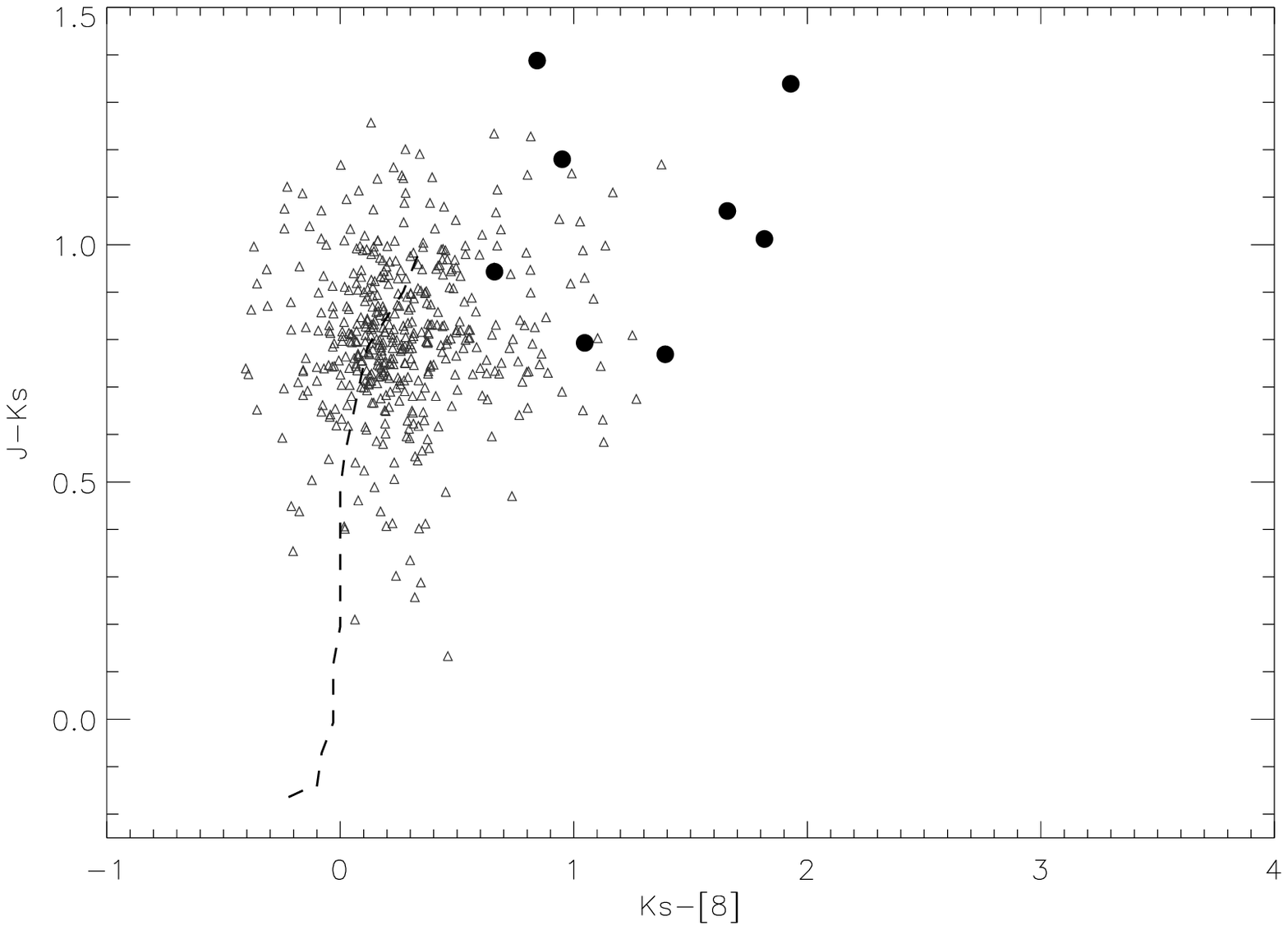}
\caption{Same as Figure \ref{jkk34dh} except for the list of candidate members 
from \citet{Ir08}.  As with the \citet{Dh07} sample, most sources with 
IRAC excesses have MIPS detections.}
\label{jkk34ir}
\end{figure}

\clearpage


\clearpage
\appendix
\section{A Comparison between our IRAC photometry and that from Dahm and Hillenbrand (2007)}
Our primary motivation for redoing IRAC photometry of NGC 2362 is to 
identify candidate cluster stars with mid-IR excesses at $\le$ 10 $\mu m$ 
from the \citet{Ir08} catalog and to construct full SEDs from V band through 
24 $\mu m$ to constrain the origin of the emission in all cluster stars.
To test the accuracy of our IRAC photometry, we compare it with
photometry from \citet{Dh07}.  Because 
\citet{Dh07} only publish photometry for cluster members, our analysis in 
this section is limited to the 337 \citet{Dh07} (candidate) members. 

\begin{figure}[t]
\plottwo{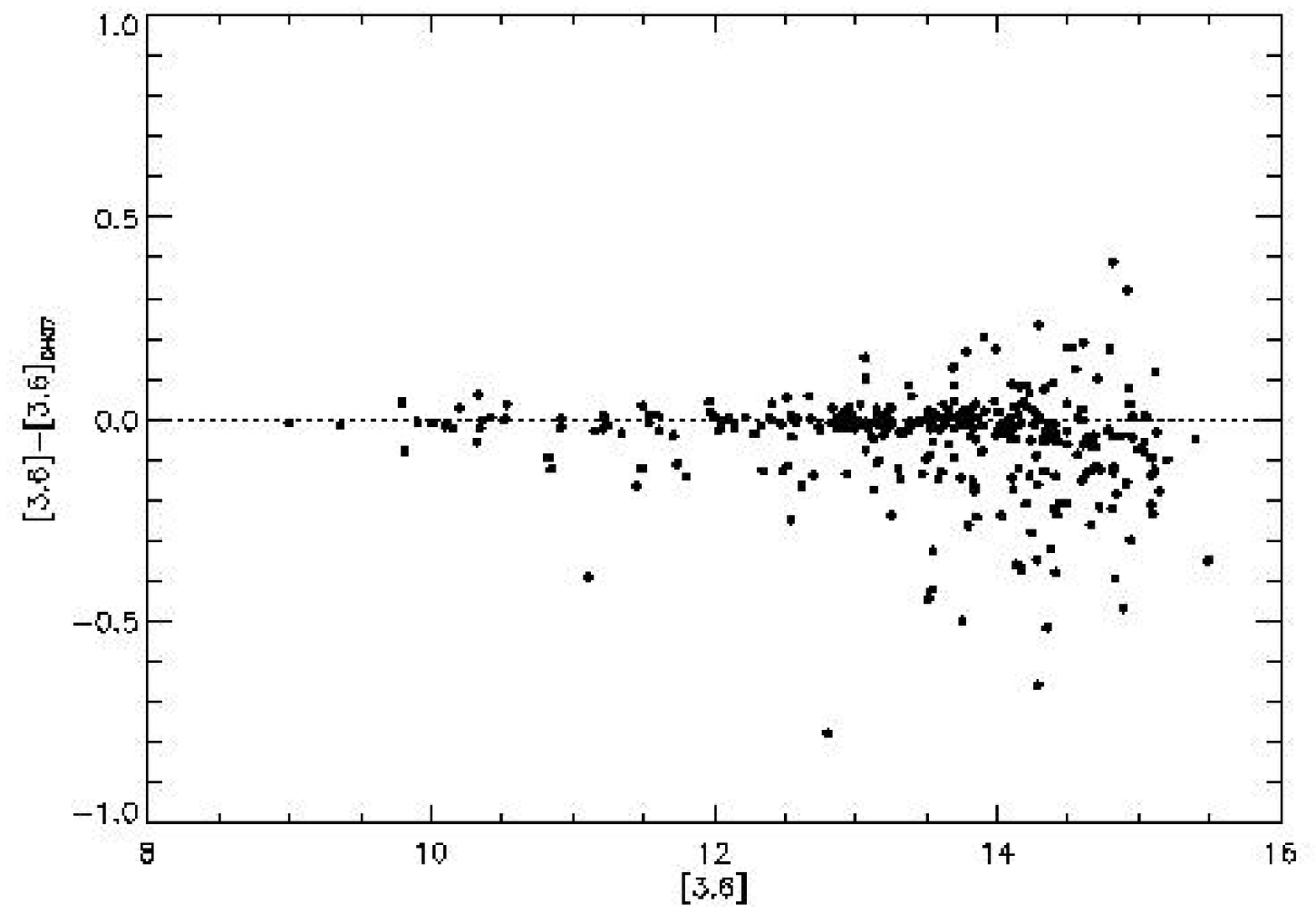}{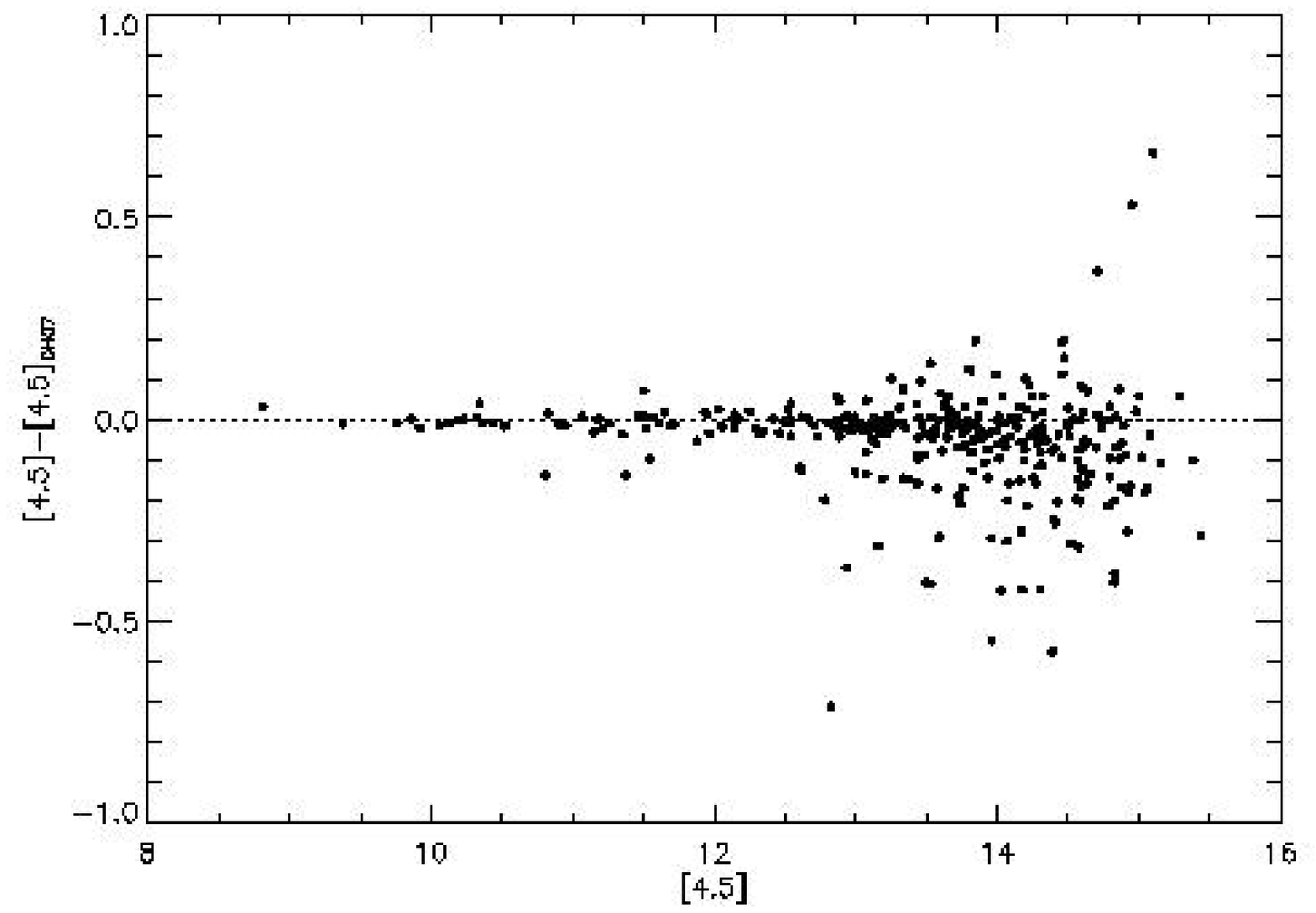}
\plottwo{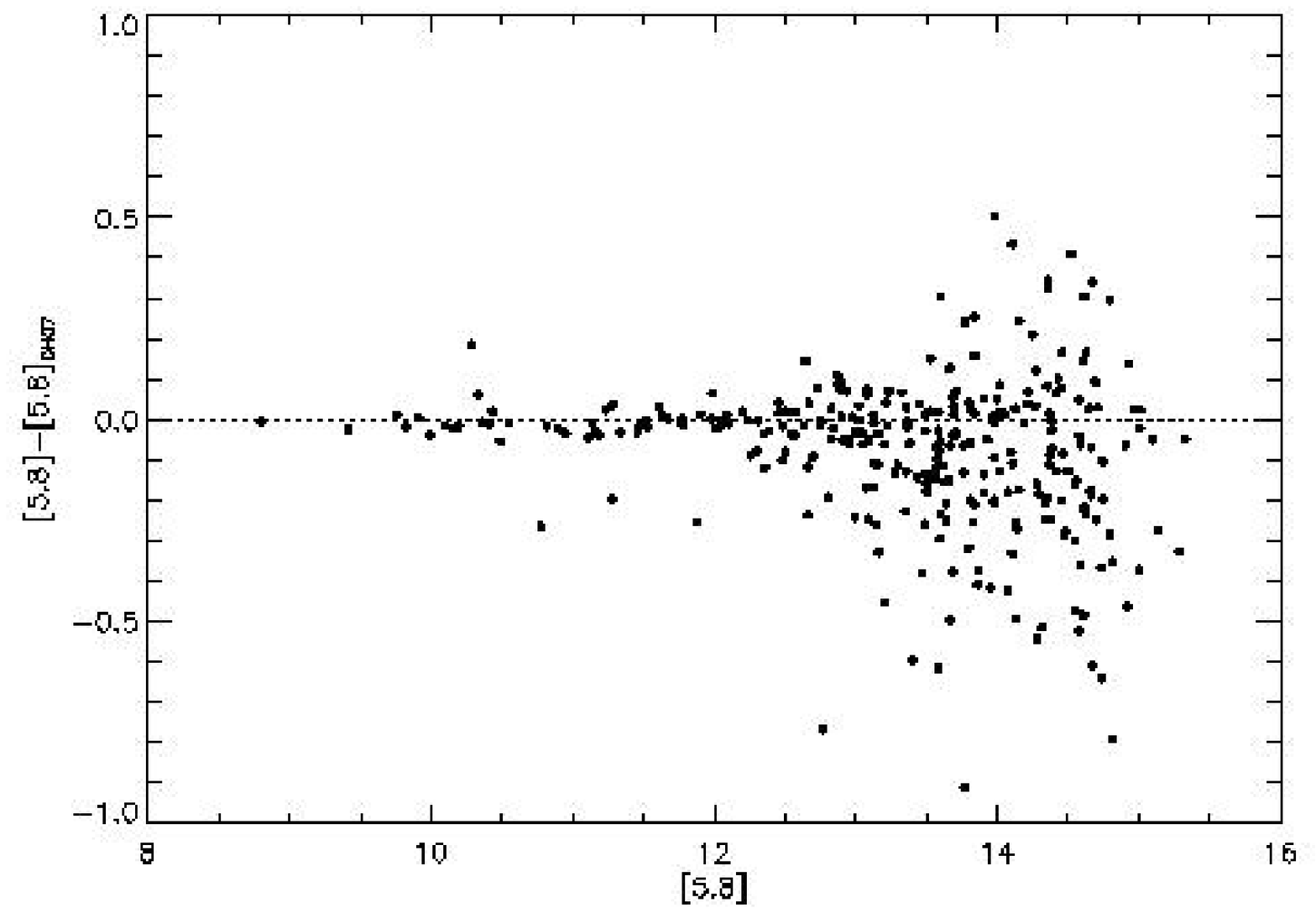}{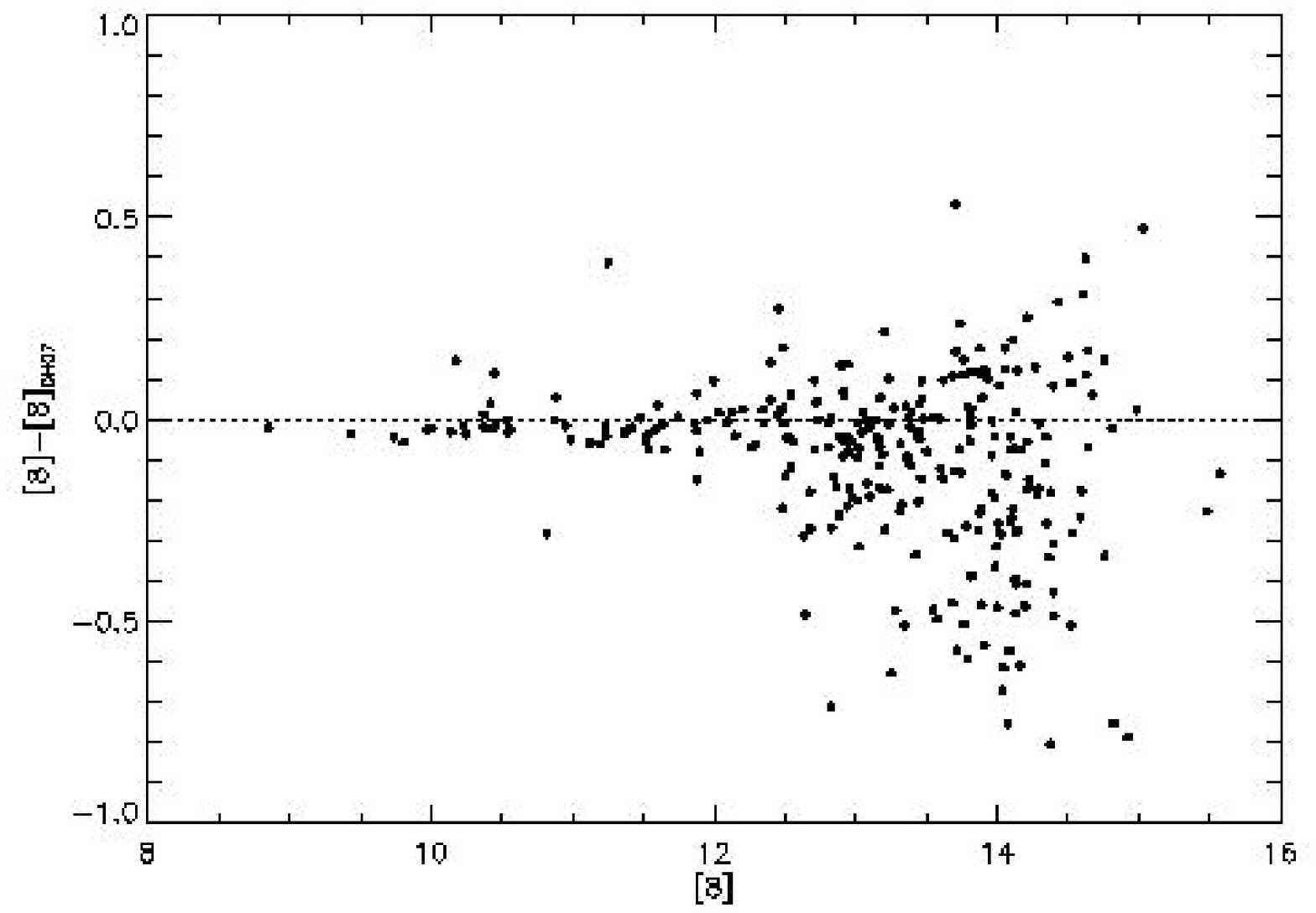}
\caption{Difference between our IRAC photometry and that from \citet{Dh07} as a function of our IRAC magnitude.  
The \citet{Dh07} photometry exhibits a deflection towards fainter magnitudes as a function of survey depth ($\delta$(IRAC) 
$<$ 0).}
\label{dhcompare1}
\end{figure}

\begin{figure}
\centering
\plottwo{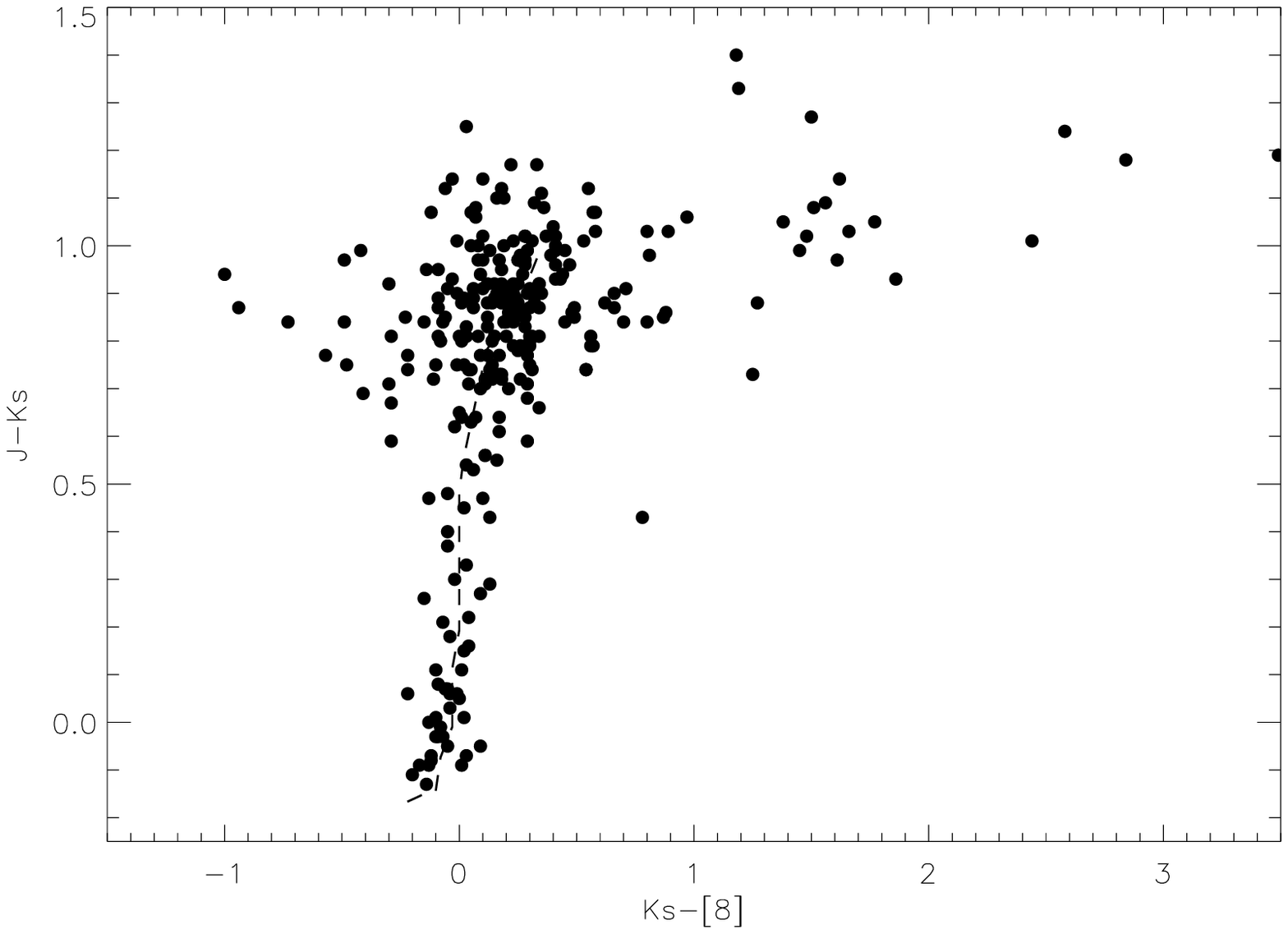}{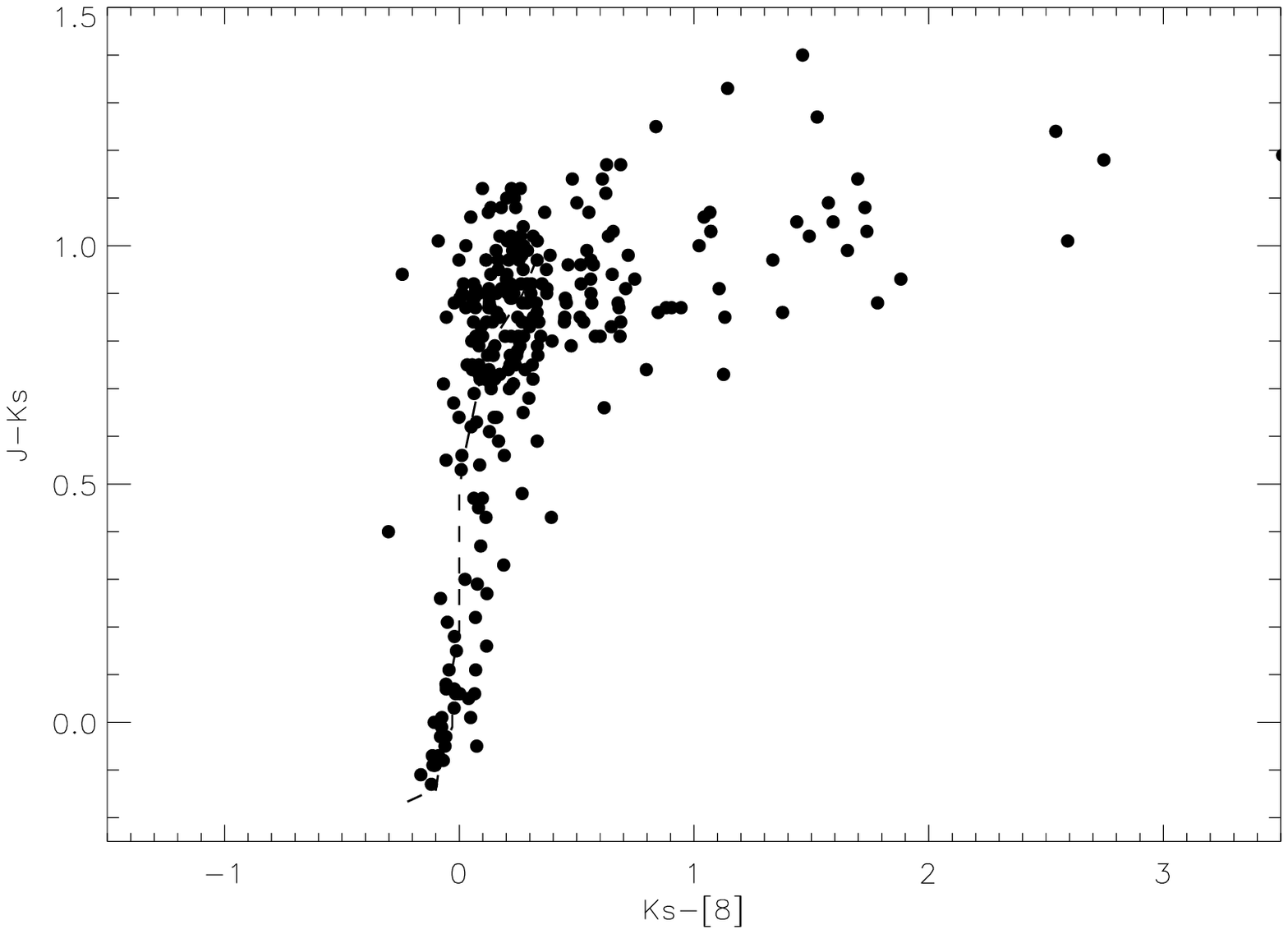}
\plottwo{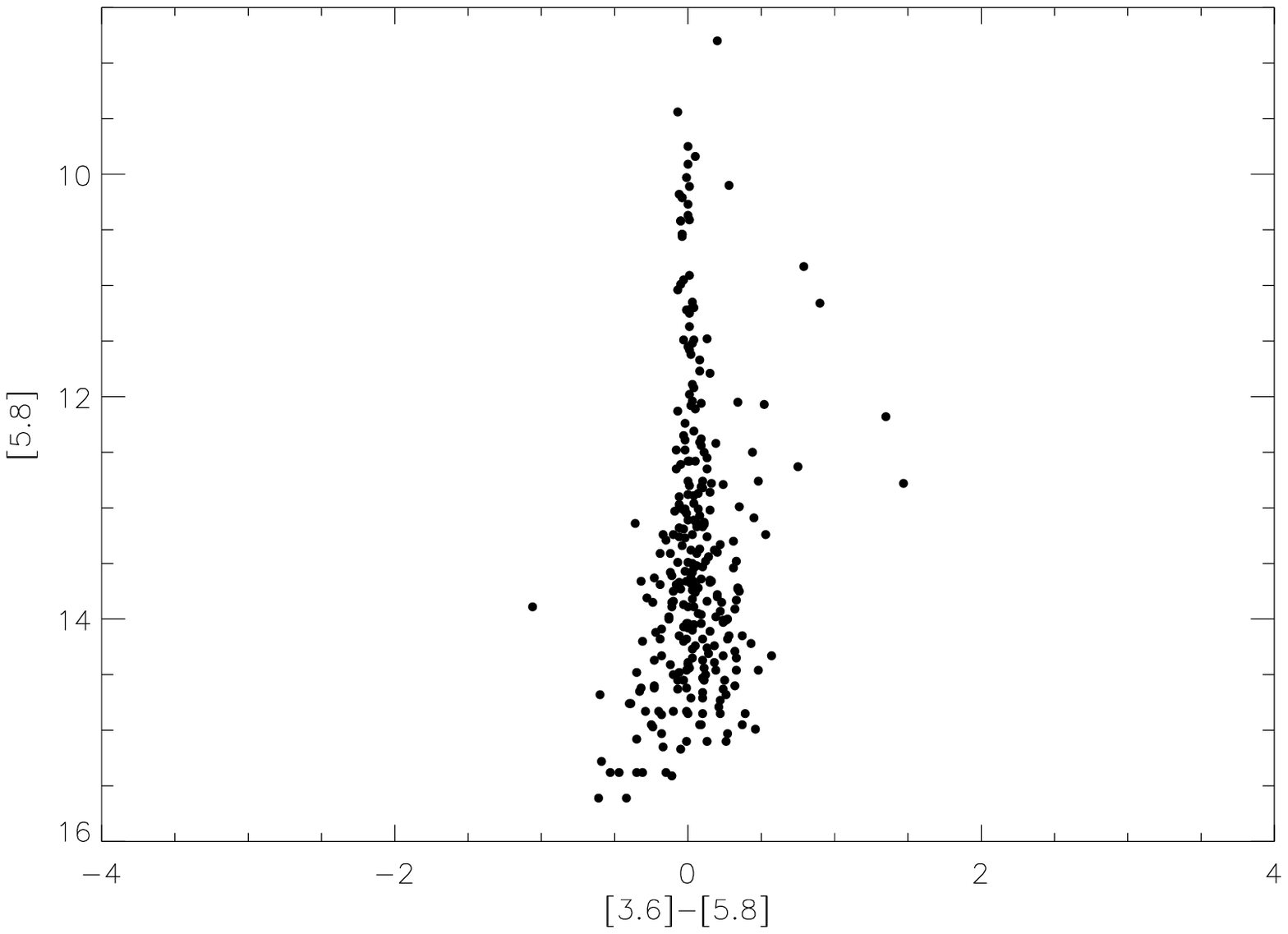}{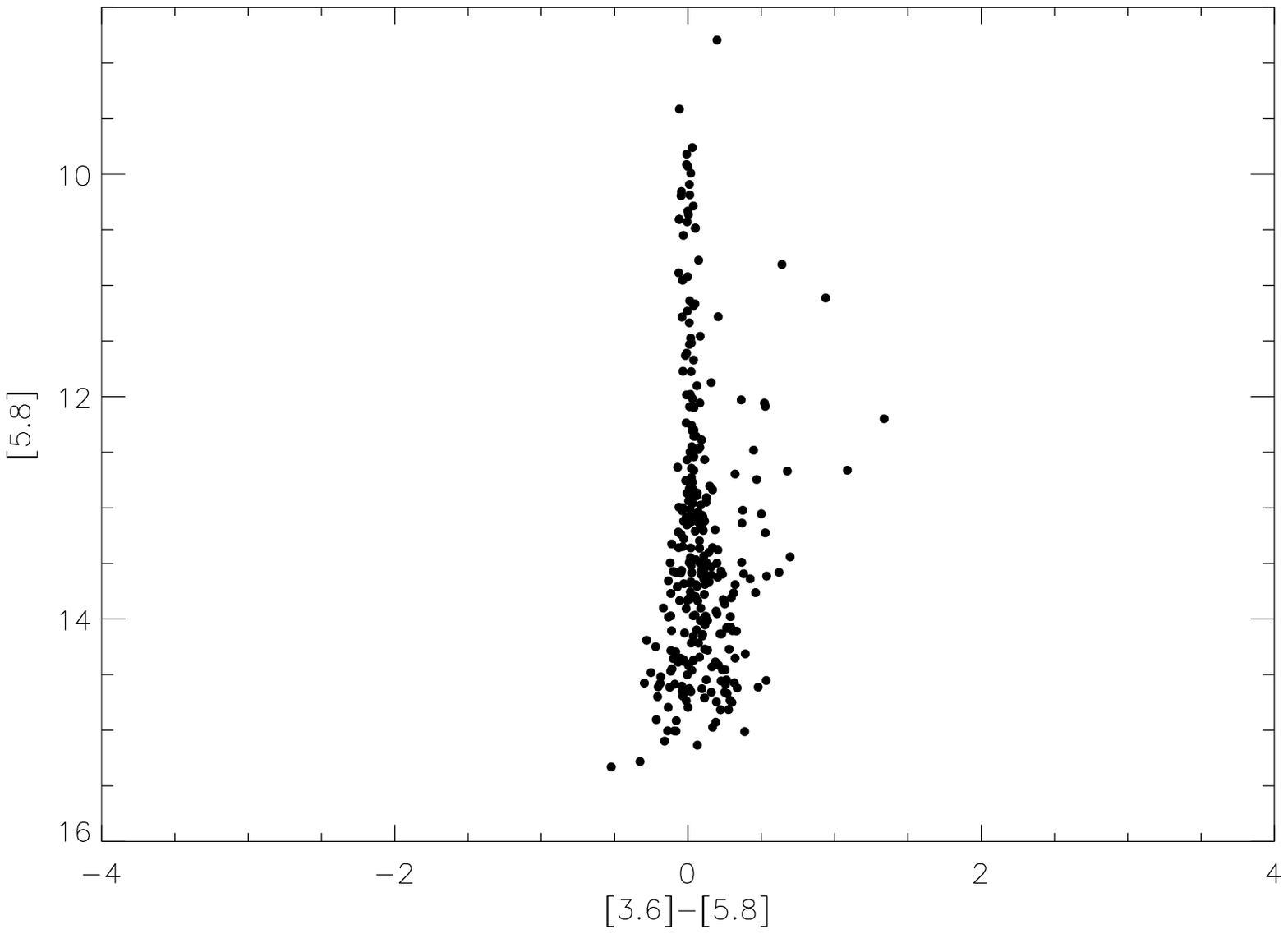}
\plottwo{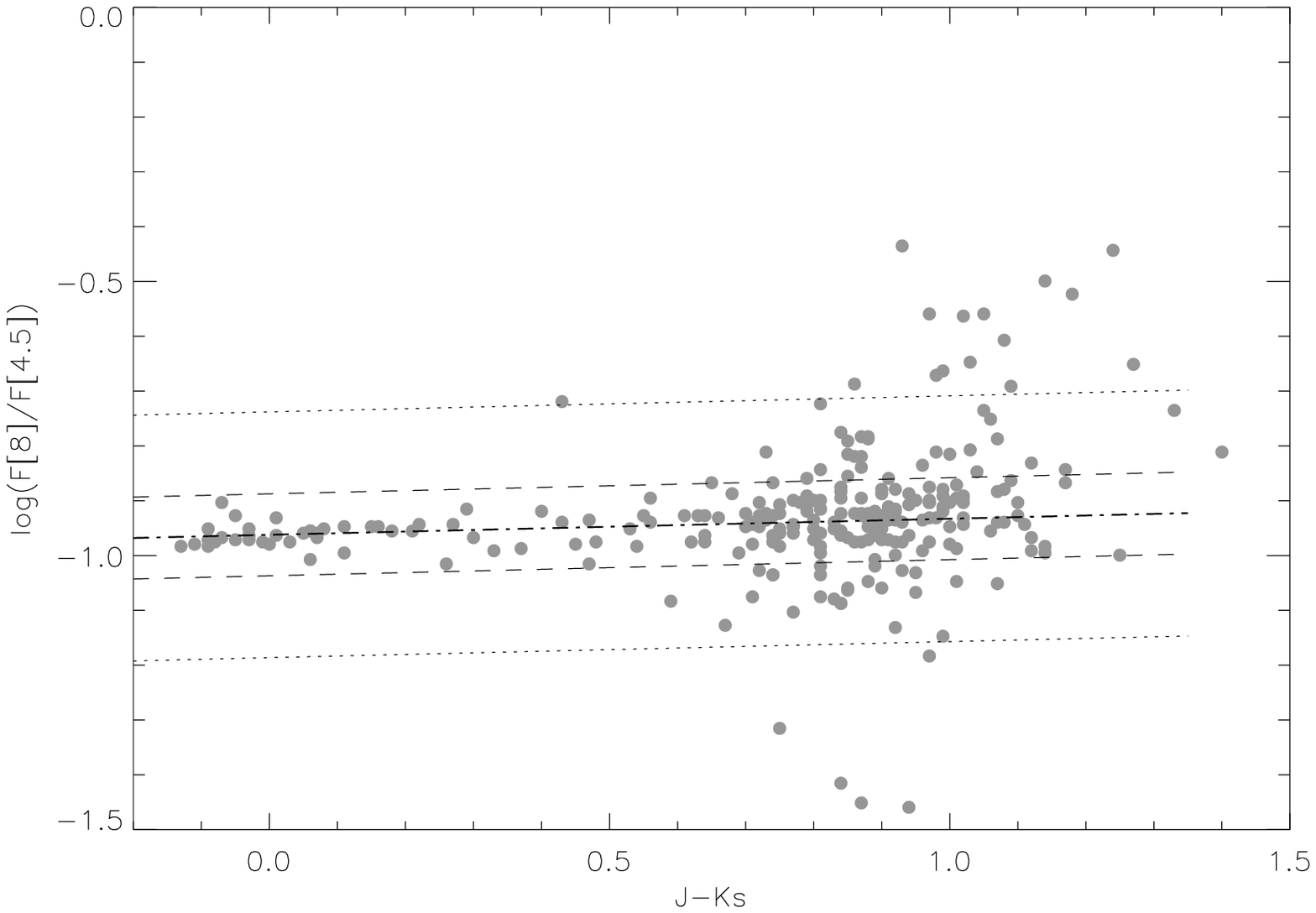}{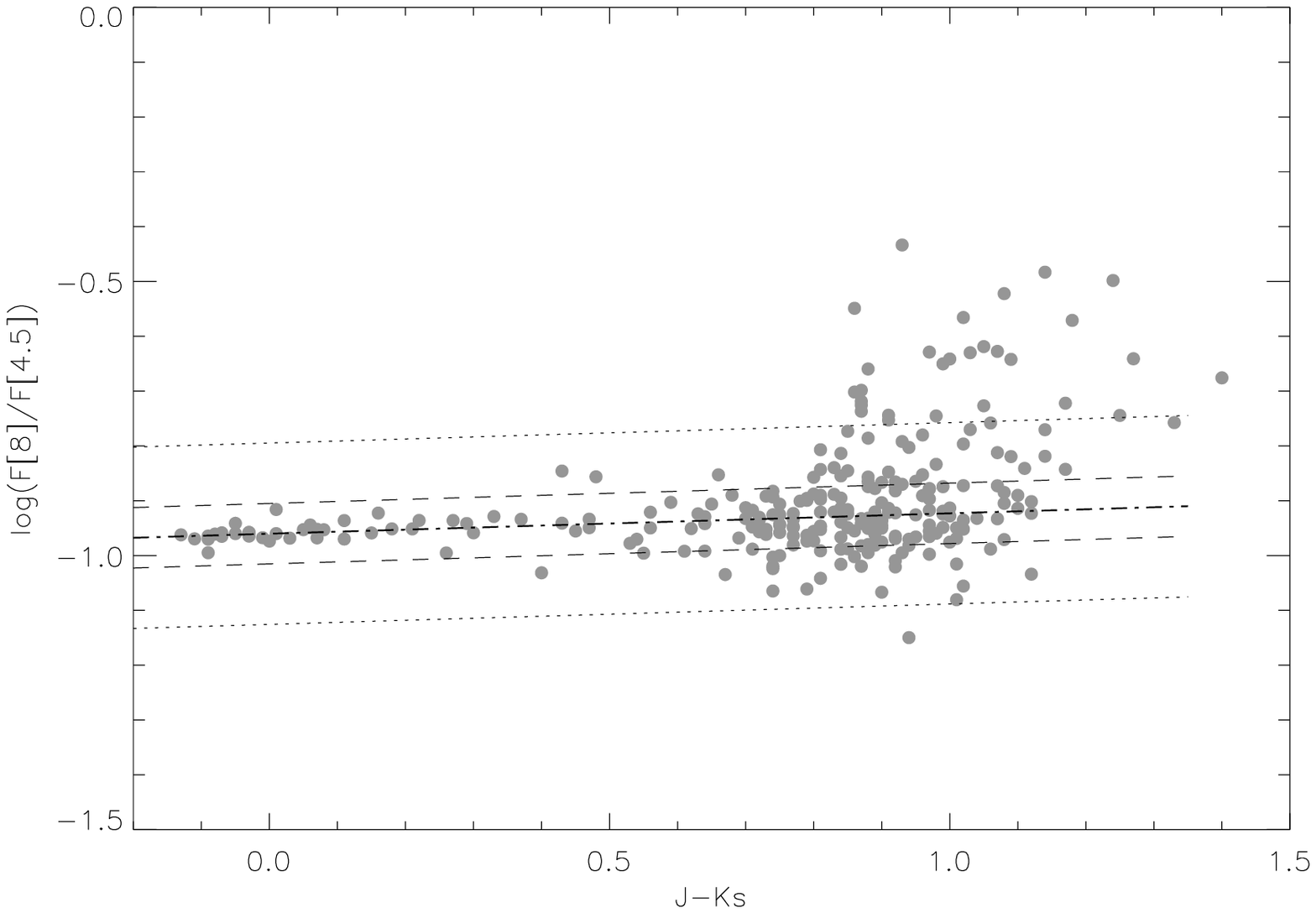}
\caption{IRAC colors from \citet[][left columns]{Dh07} and this work (right columns).  
(top) The J-K$_{s}$ vs. K$_{s}$-[8] color-color diagram for 
 sources in common between \citet[][]{Dh07} and this work.
(middle) [5.8] vs. [3.6]--[5.8] color-magnitude diagrams.  (bottom) 
The 8 $\mu m$ to 4.5 $\mu m$ flux-ratio diagram for the same samples.  The 
best-fit line (dash-dot line) and 3 $\sigma$ dispersion above and below the best-fit line (dashed lines)
are overplotted.  The \citet{Dh07} photometry yields many stars with unphysically blue 2MASS/IRAC colors (K$_{s}$-[8] 
$\le$ -0.2; [3.6]-[5.8] $\le$ -0.5) and unphysically 
small flux ratios ($\le$ -1.15).  The photometry from this work lacks many blue colors and exhibits 
less far scatter, especially at fainter magnitudes in the 8 $\mu m$ channel.}
\label{dhcompare2}
\end{figure}

Figures \ref{dhcompare1} and \ref{dhcompare2} contrast our photometry with  
\citet{Dh07}.  In Figure \ref{dhcompare1}, we show the difference in magnitude for 
each IRAC channel as a function of (our) magnitude.  Photometry from both groups 
appears to have little/no zero-point offset.  However, the \citet{Dh07} photometry 
exhibits a deflection towards fainter magnitudes for m(IRAC) $\ge$ 13 in all filters relative 
to our photometry.  This effect is especially pronounced in the [5.8] and [8], which clearly show a non-gaussian 
scatter.  

Figure \ref{dhcompare2} shows that the mid-IR colors also have systematic differences.  
The top panels compare the distributions of 
J--K$_{s}$ vs. K$_{s}$--[8] colors.  
We have far fewer sources with
 with unphysical K$_{s}$-[8] colors ($<$ -0.2): the blue limit of 
the K$_{s}$-[8] distribution tracks the photospheric locus (solid line) very well.
.  The K$_{s}$-[8] colors from \citet{Dh07} deflect blueward of the photospheric locus by up to 
0.5--1.  Our photometry includes only two stars with K$_{s}$-[8] $\sim$ -0.2 compared 
to seventeen in \citet{Dh07}.
The [5.8] vs. [3.6]--[5.8] distribution for our sources (middle 
panels) also contains fewer stars with very blue ([3.6]--[5.8] $<$ -0.5) colors and 
exhibits a smaller dispersion in color, especially for stars fainter than m[5.8]=13.

In the bottom panels of Figure \ref{dhcompare2}, we quantify the 
differences in photometry at 4.5 $\mu m$ and 8 $\mu m$ 
by fitting a line (dash-dot) to the 8 $\mu m$ to 4.5 $\mu m$ flux ratio (log(F[8]/F[4.5])) 
as a function of J--K$_{s}$ color \citep[see also ][]{Ca06, Cpk08} for 
sources with 5$\sigma$ detections in both filters.  As in \citet{Ca06} and \citet{Cpk08}, 
the best-fit line is determined by a linear least squares fit with 4$\sigma$ clipping to 
remove outlying sources, including stars with infrared excess.  The dispersion and 
rms residuals are then recalculated.  The \citet{Dh07} 
photometry has a 3$\sigma$ dispersion in log(F[8]/F[4.5]) of $\sim$ 0.23 and residuals 
in the linear fit of $\sim$ 5.5\%.  Five stars with larger photometric uncertainties 
have negative flux ratios that lie outside the dispersion (dashed lines).    
Our photometry has a dispersion of 0.16 and 4.3\% fit residuals with one 
outlying star.  Furthermore, four stars in the \citet{Dh07} photometry have 
extremely negative flux ratios, lying 5--6 $\sigma$ fainter than the photospheric 
locus.  If the distribution in flux ratios followed a normal distribution, 
we would expect, at most, one such source.  Our photometry contains just one 
3$\sigma$ outlier with a negative flux ratio.

The differences in photometry are primarily due to 
the smaller aperture radius employed here (2 pixels vs. 3 pixels in \citealt{Dh07}).
The sky background may be overestimated from the larger sky annulus needed for a 3 pixel aperture 
in densely populated regions, which results in artificially faint sources whose colors deflect 
blueward.  Differences in image processing methods may also explain differences in photometry.
Differences in the aperture radius and image processing methods have led to 
significant discrepancies in photometry for other Spitzer cluster data.  In these cases, 
more accurate photometry was also obtained using a smaller aperture radius \citep[e.g. Cha I;][]{Luh08}.

\clearpage
\section{Atlas of SEDs of MIPS-detected (candidate) cluster members}
\begin{figure}
\epsscale{0.95}
\centering
\plottwo{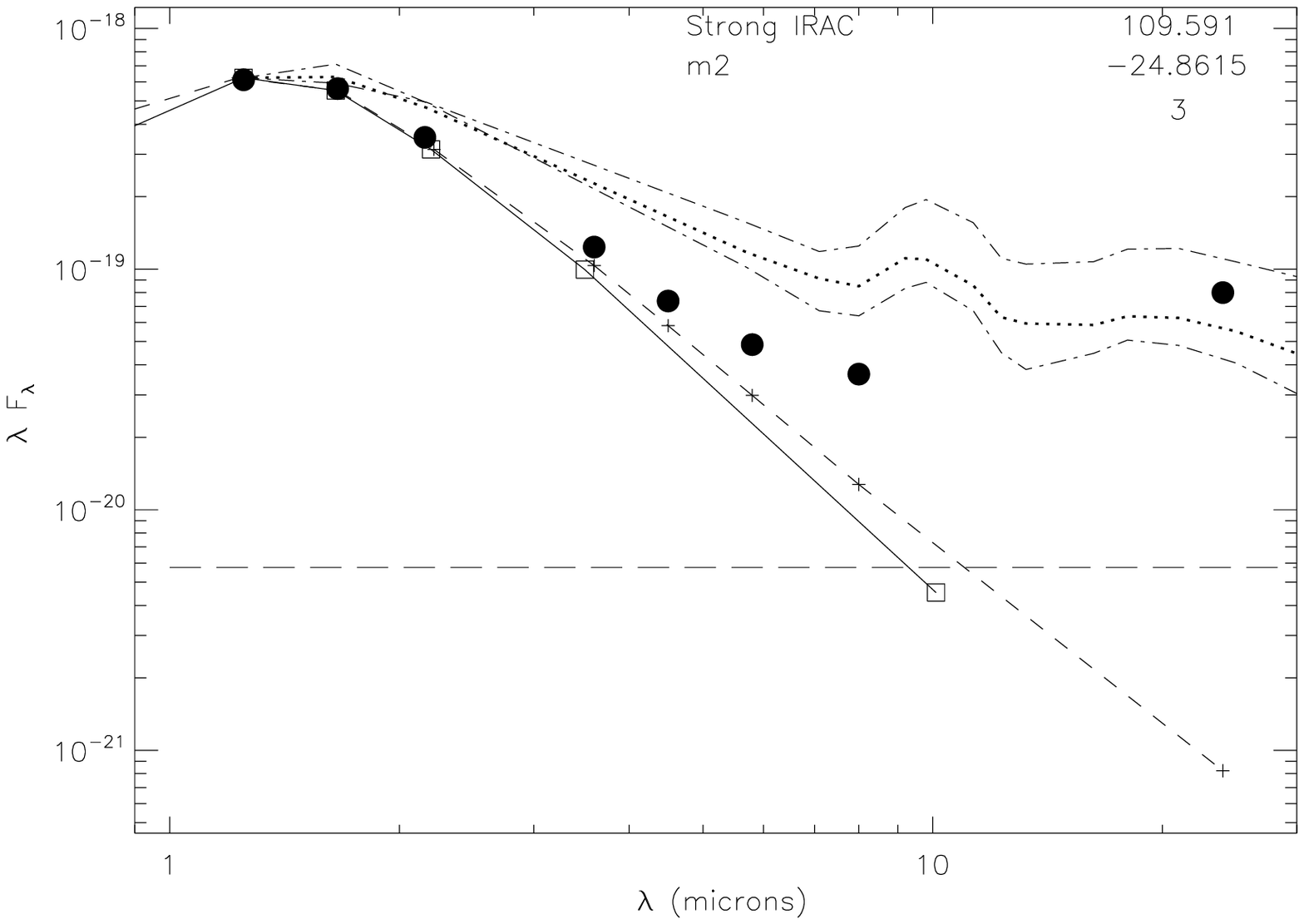}{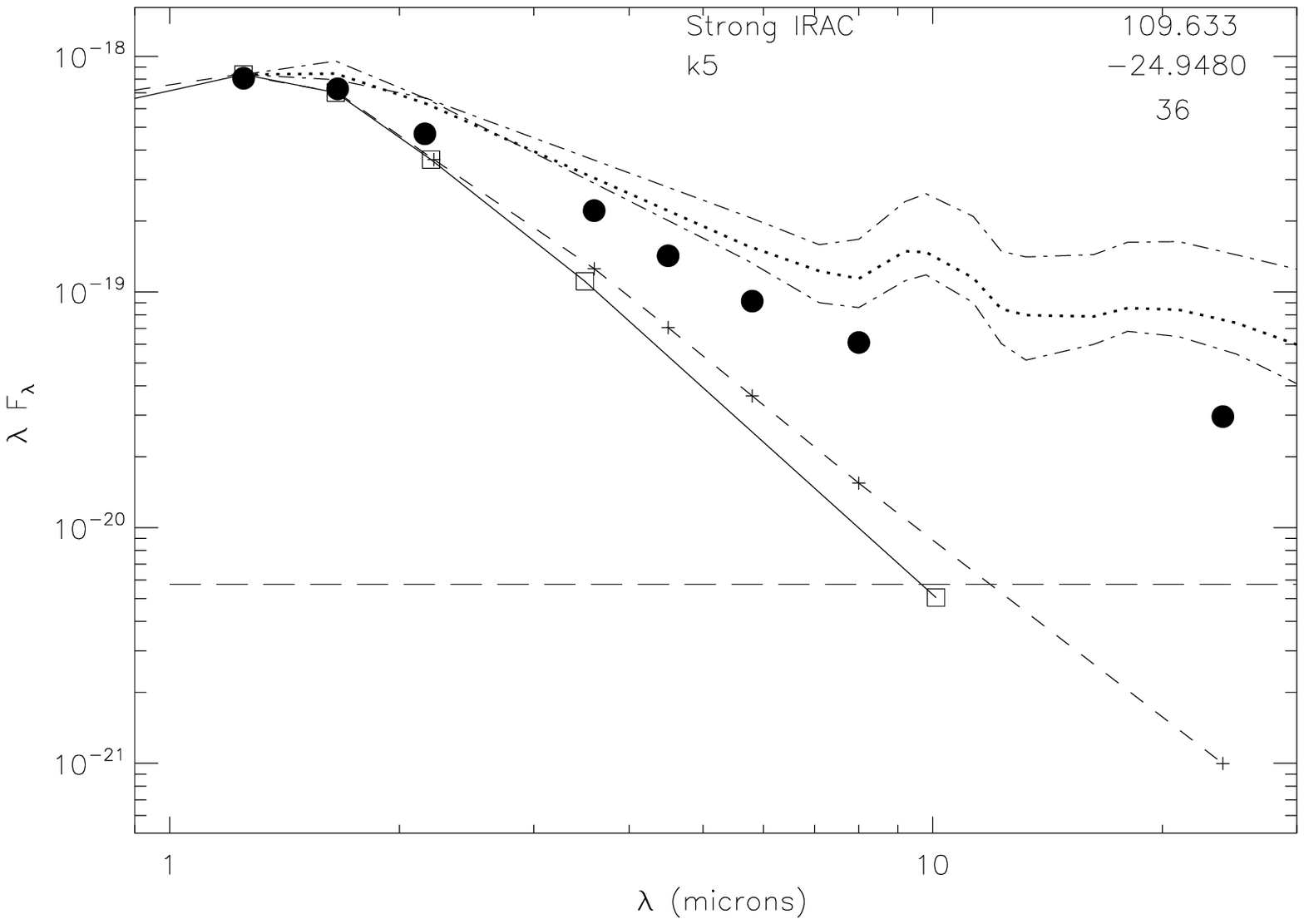}
\plottwo{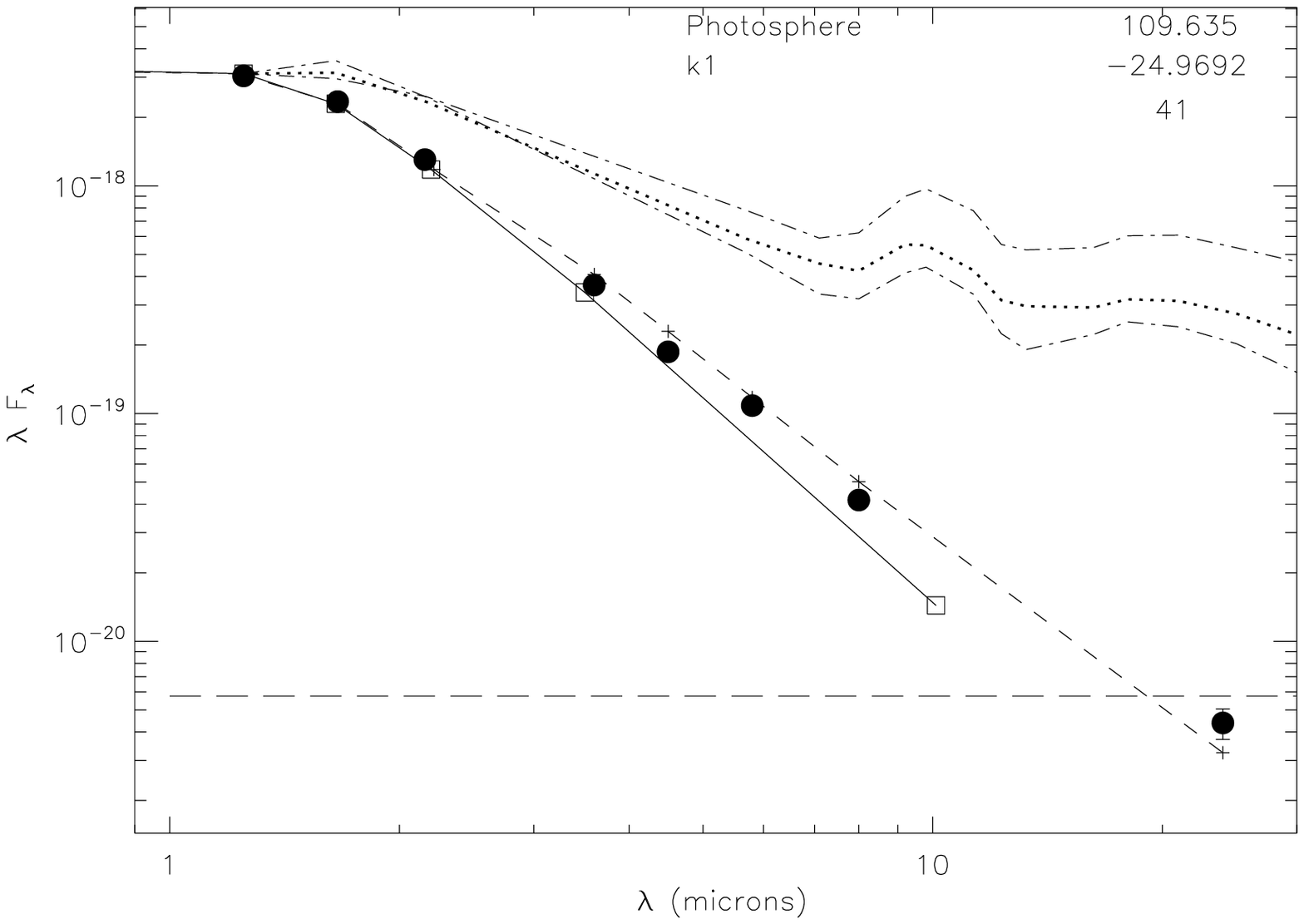}{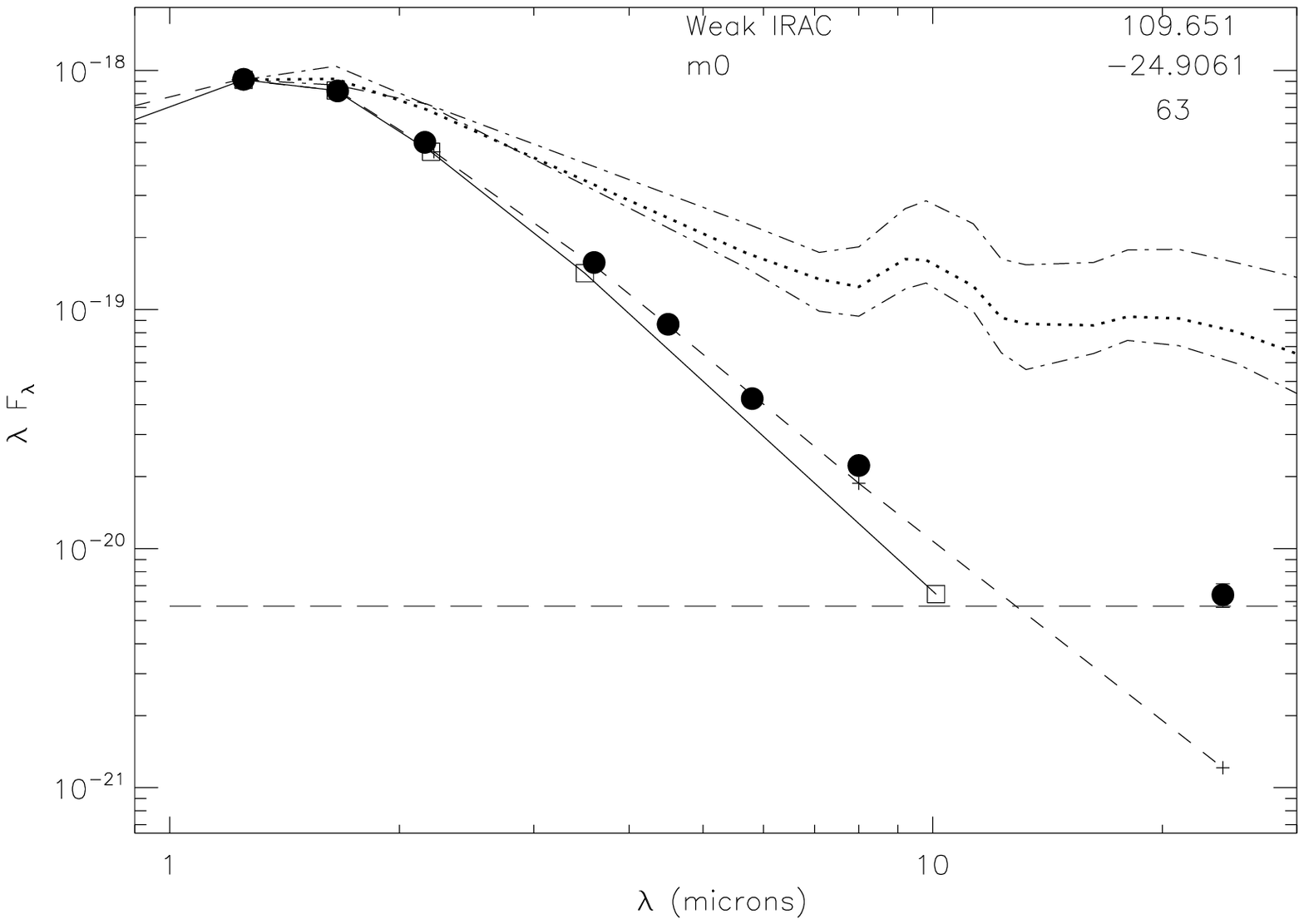}
\plottwo{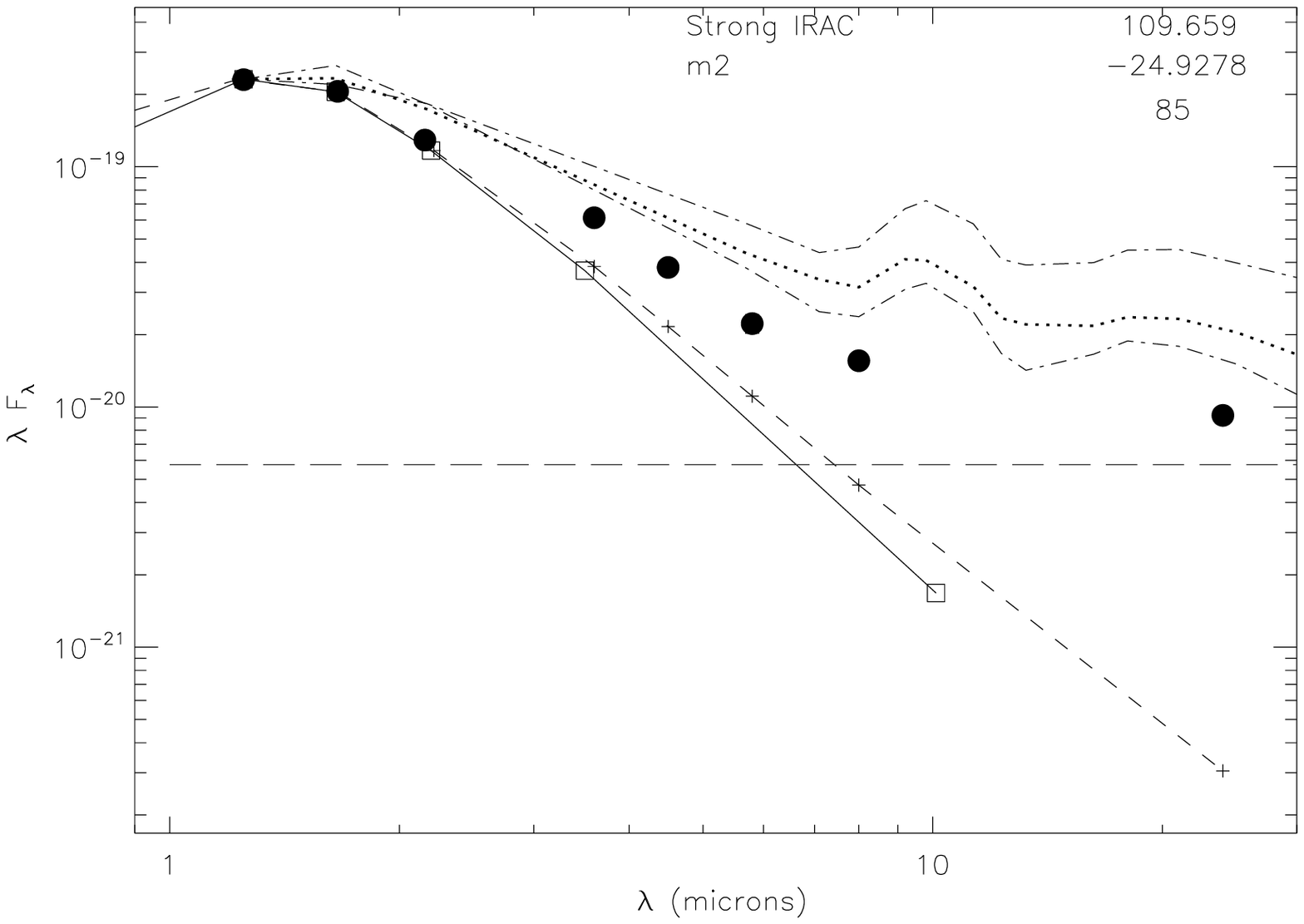}{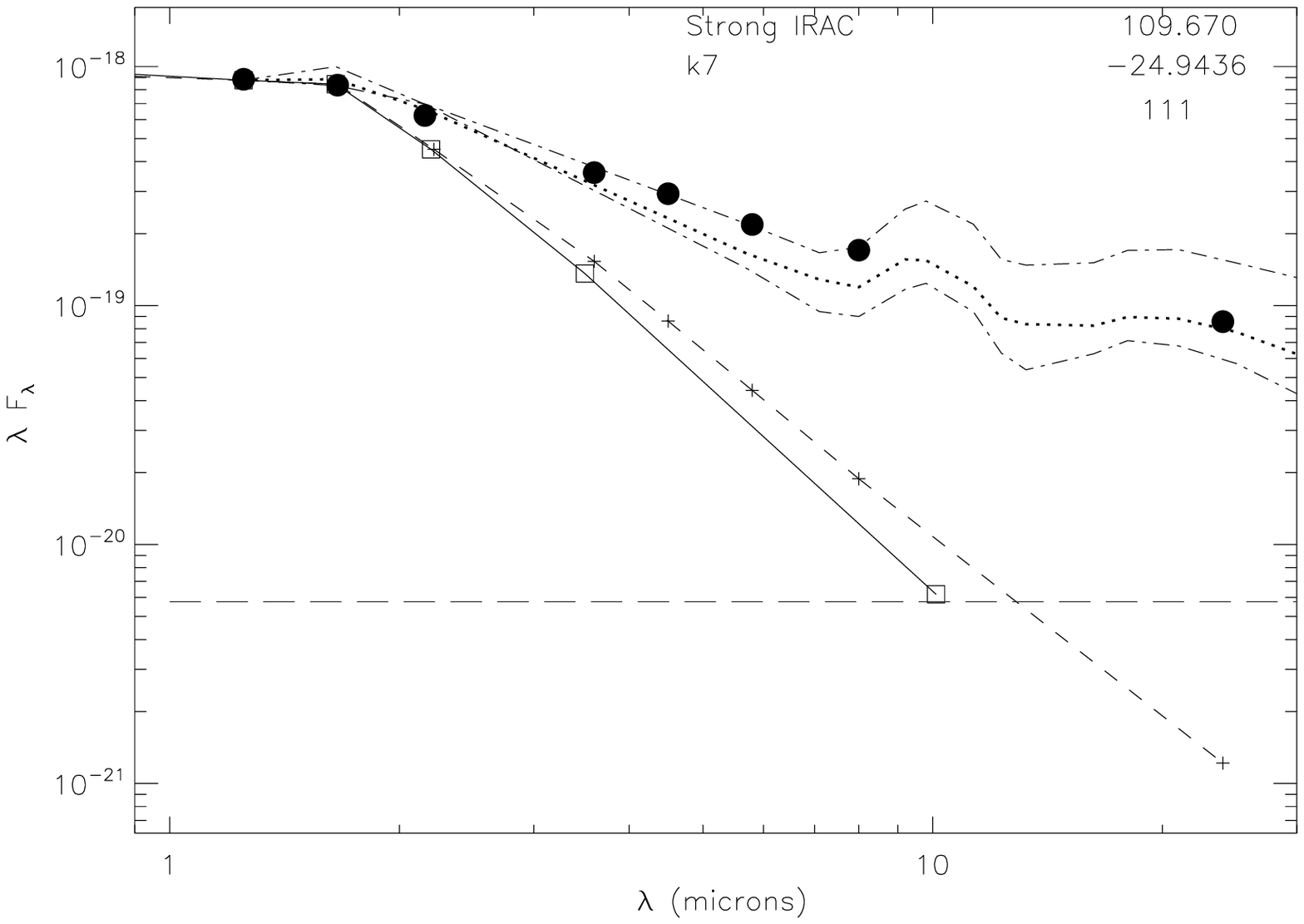}
\plottwo{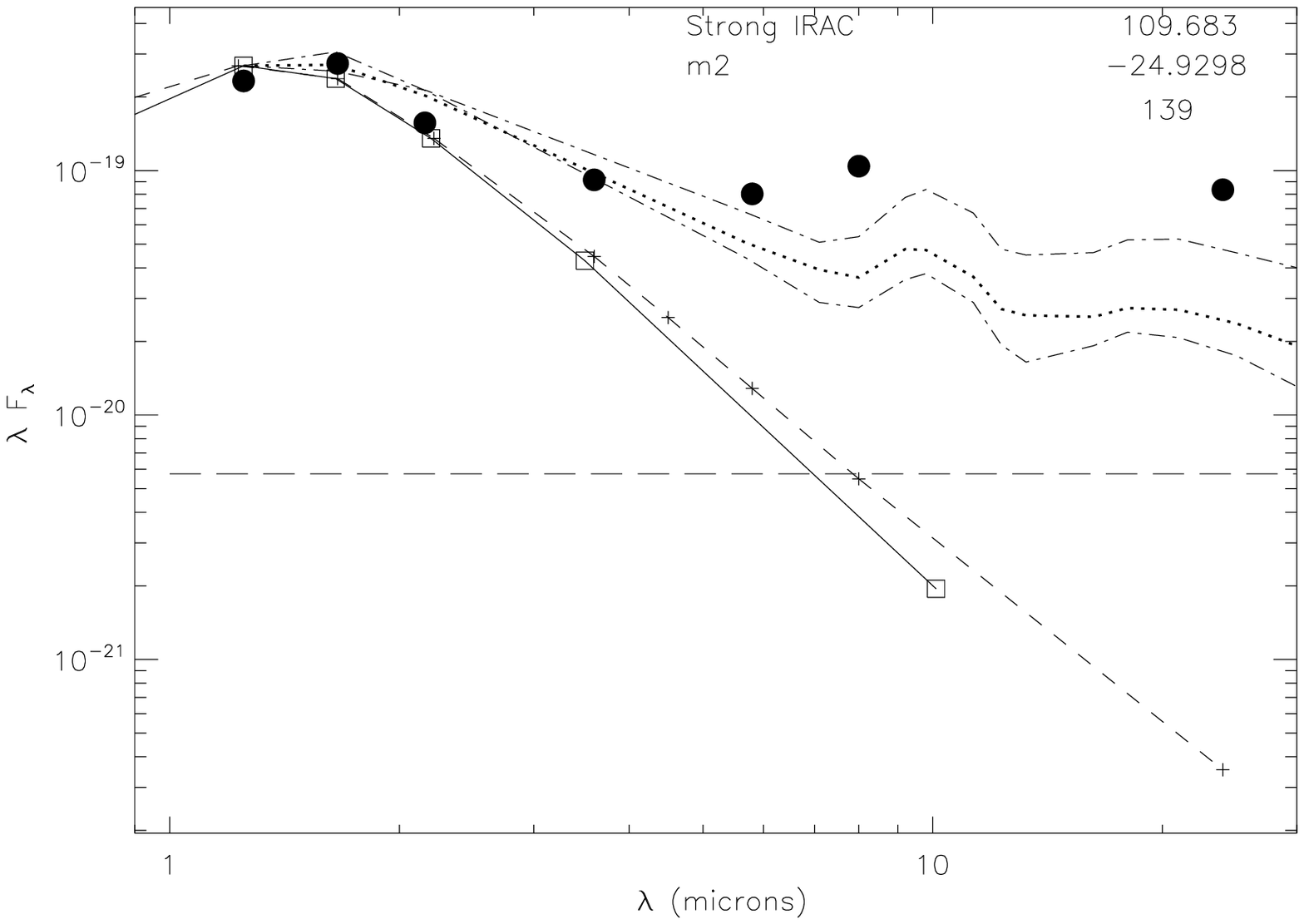}{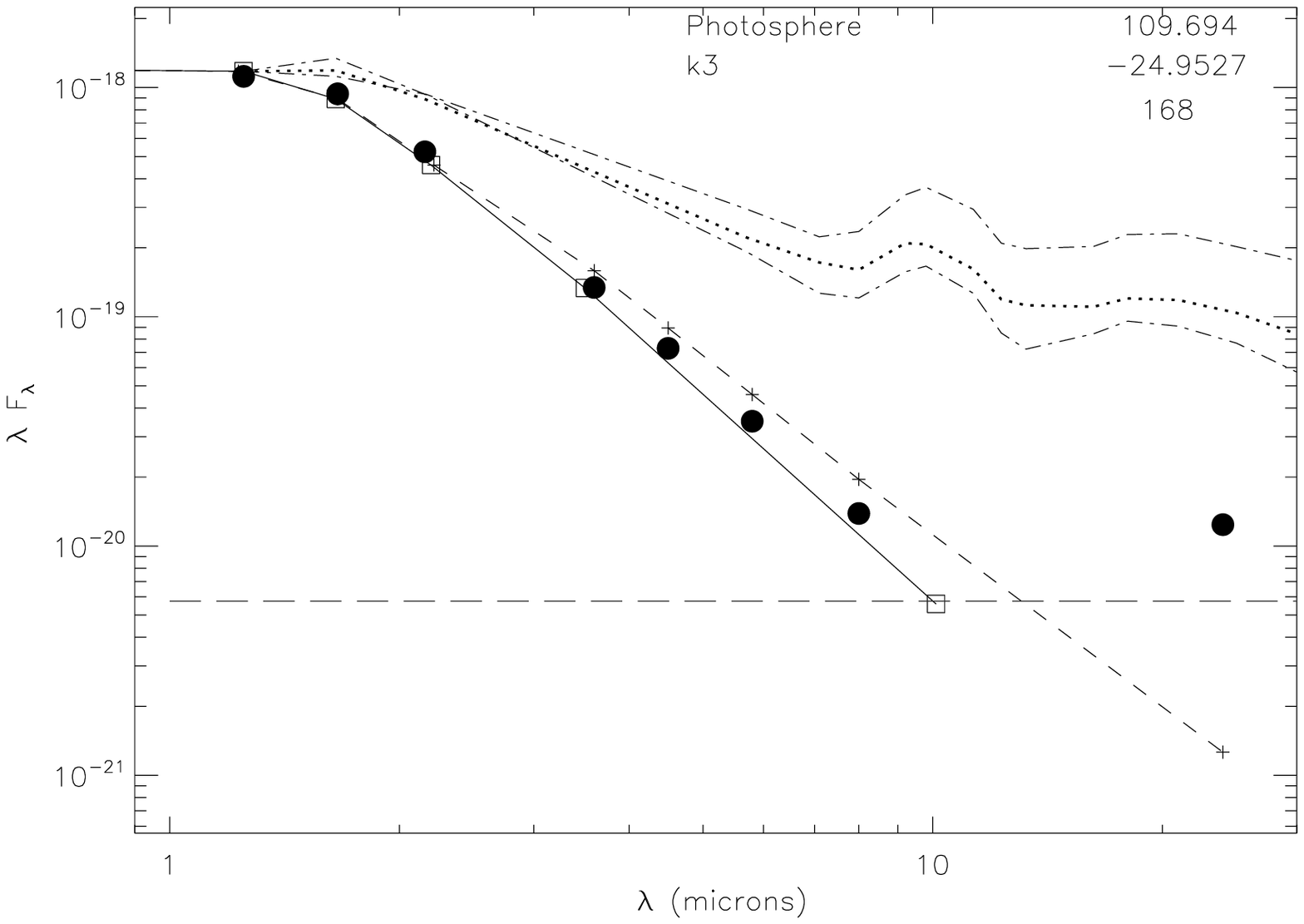}
\caption{Atlas of spectral energy distributions of late-type NGC 2362 members with MIPS 24 $\mu m$ excesses.  
Overplotted are the median Taurus SED (dotted line) with upper and lower quartiles (dot-dashed line).
A terrestrial zone debris disk model is shown as a dashed line with mid-IR flux slightly greater
than the photosphere (solid line, connected by open squares).
The MIPS 5$\sigma$ detection limit is shown as a horizontal grey dashed line.
}
\label{seds}
\end{figure}
\begin{figure}
\epsscale{0.99}
\centering
\plottwo{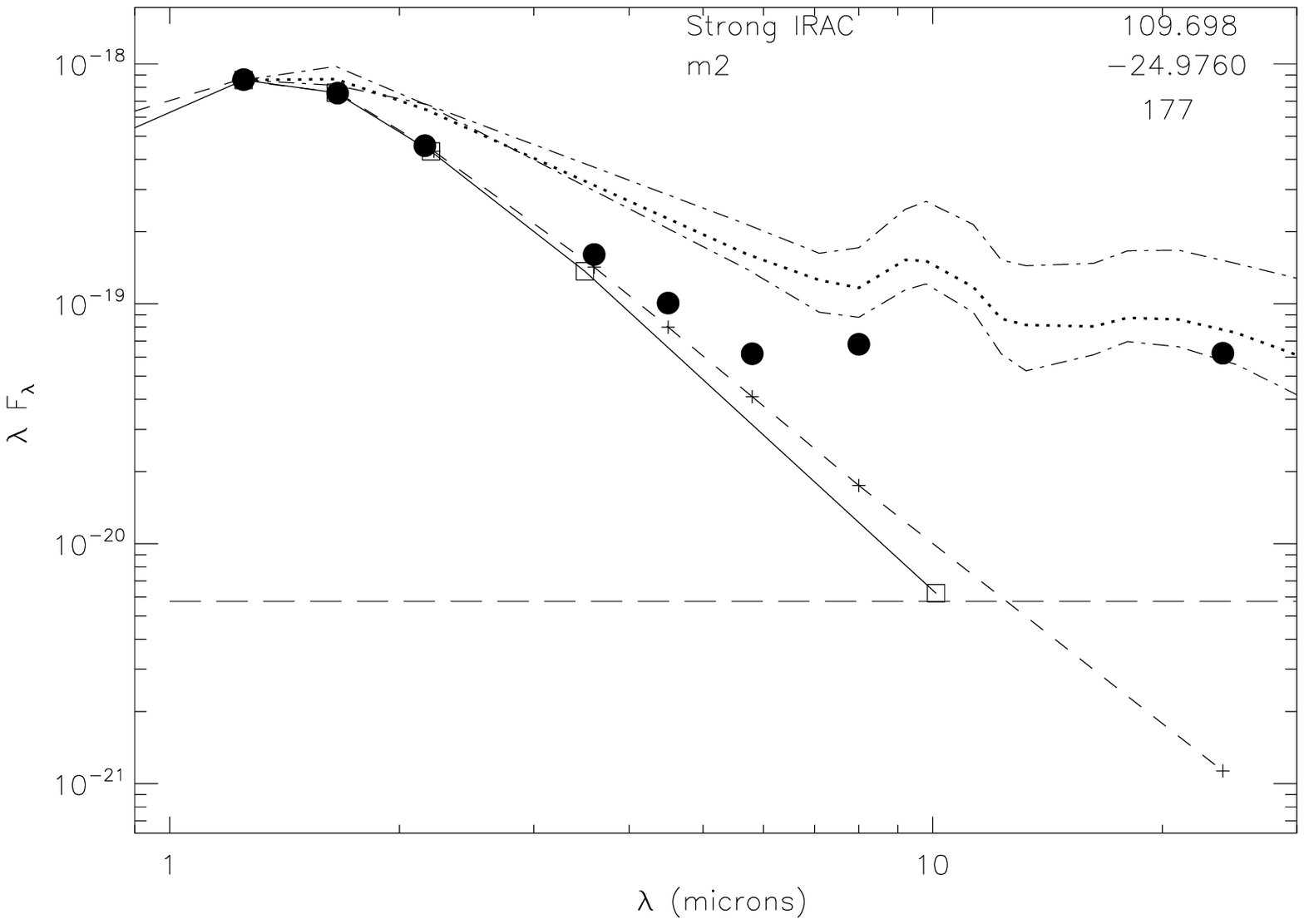}{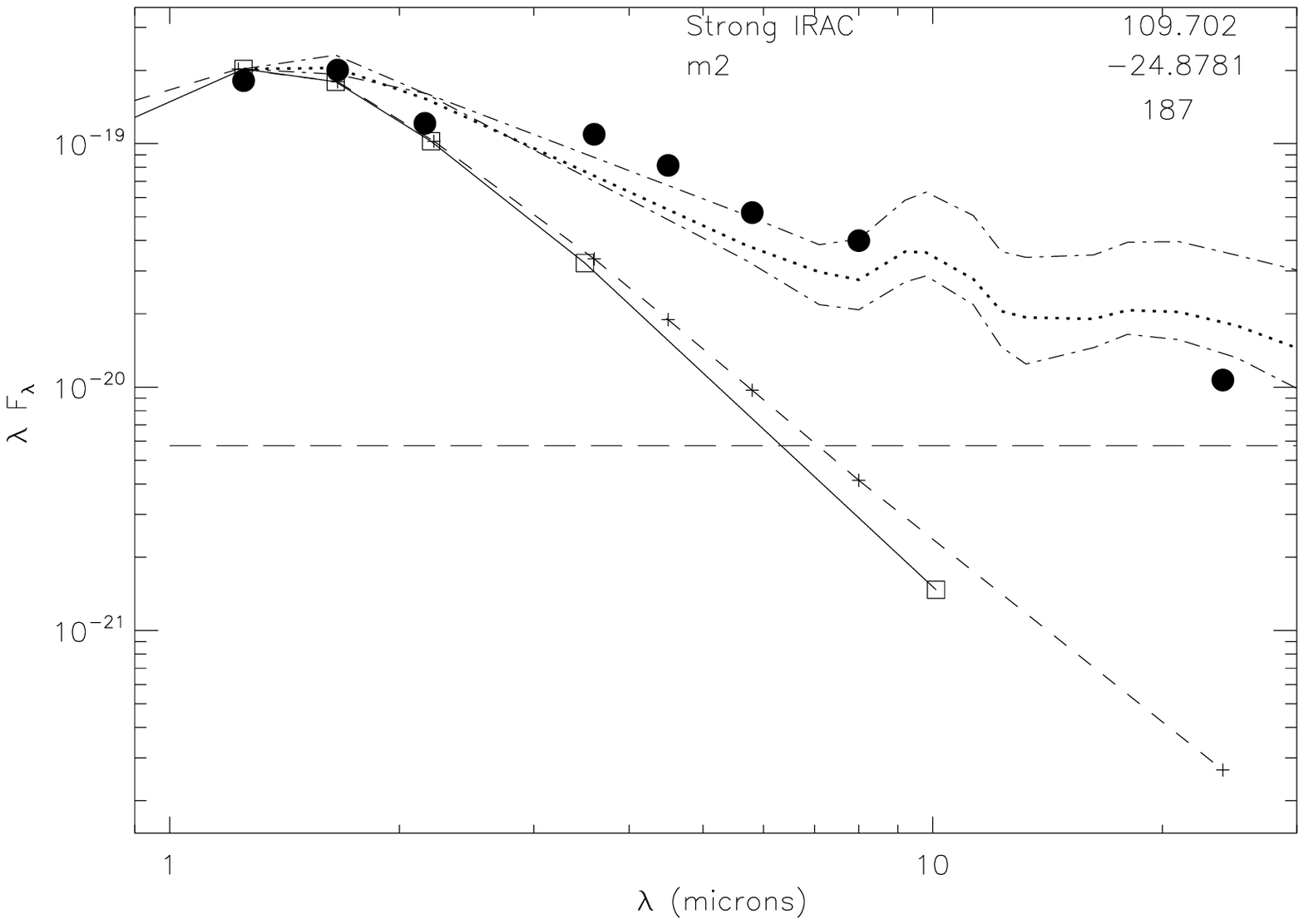}
\plottwo{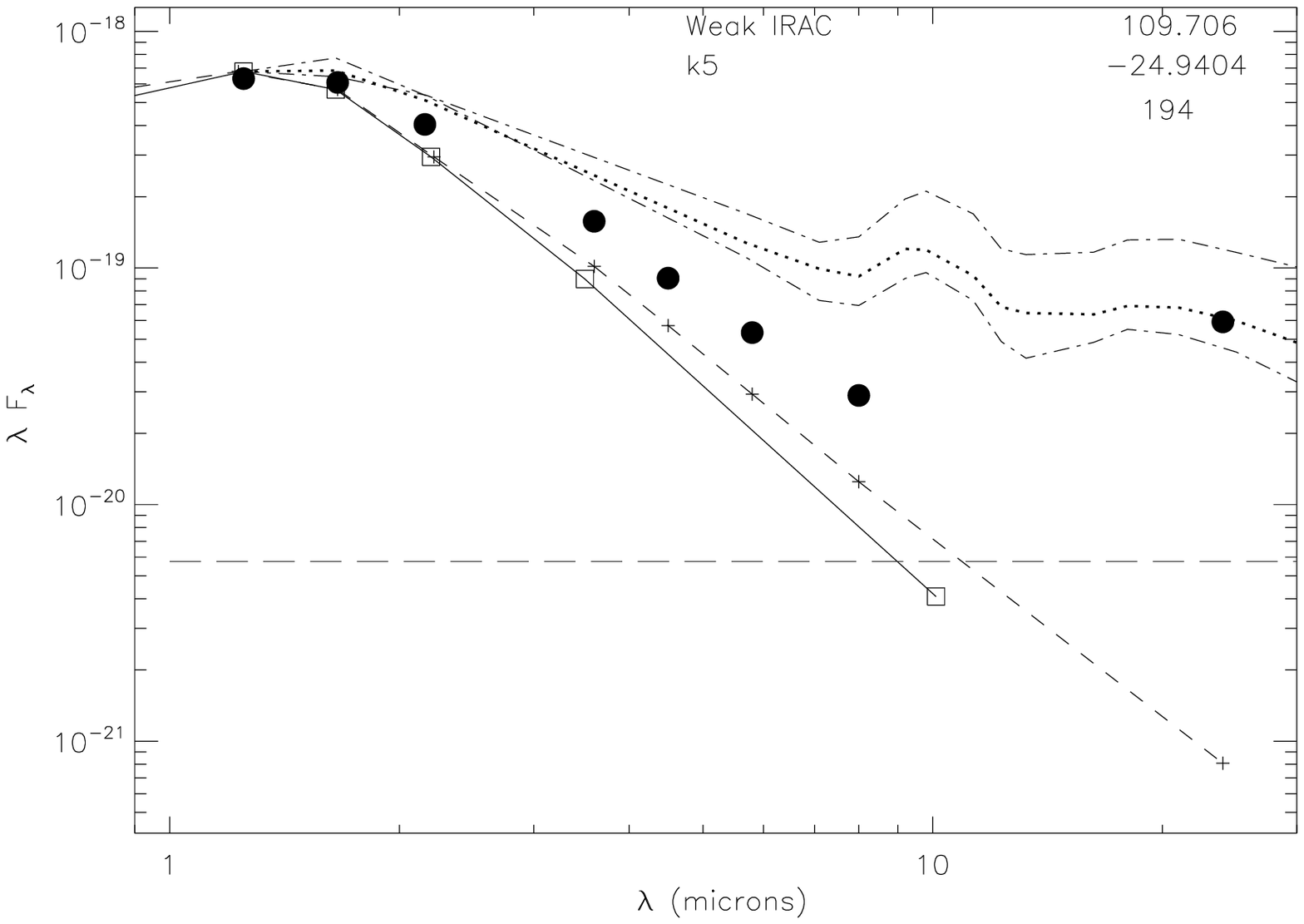}{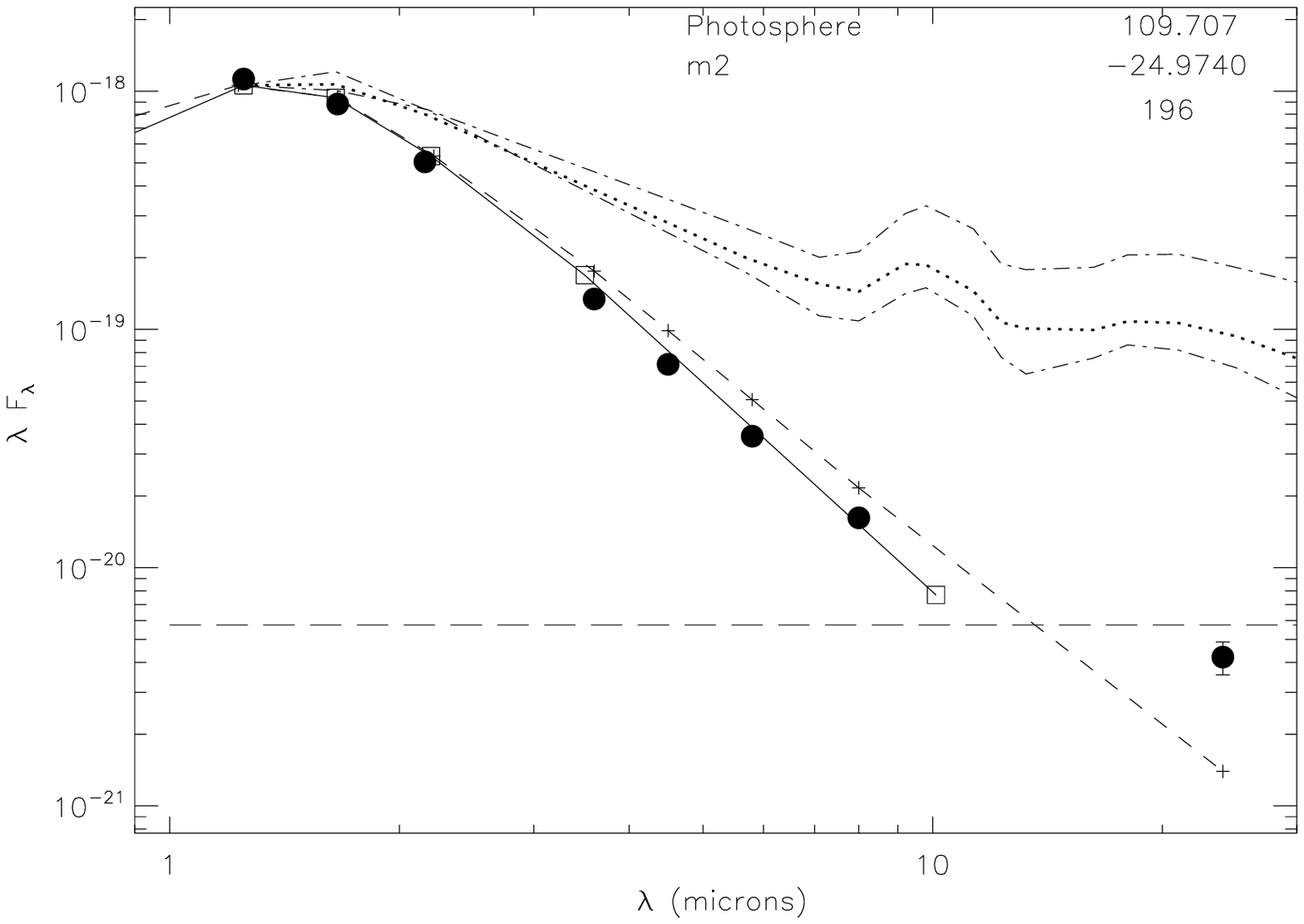}
\plottwo{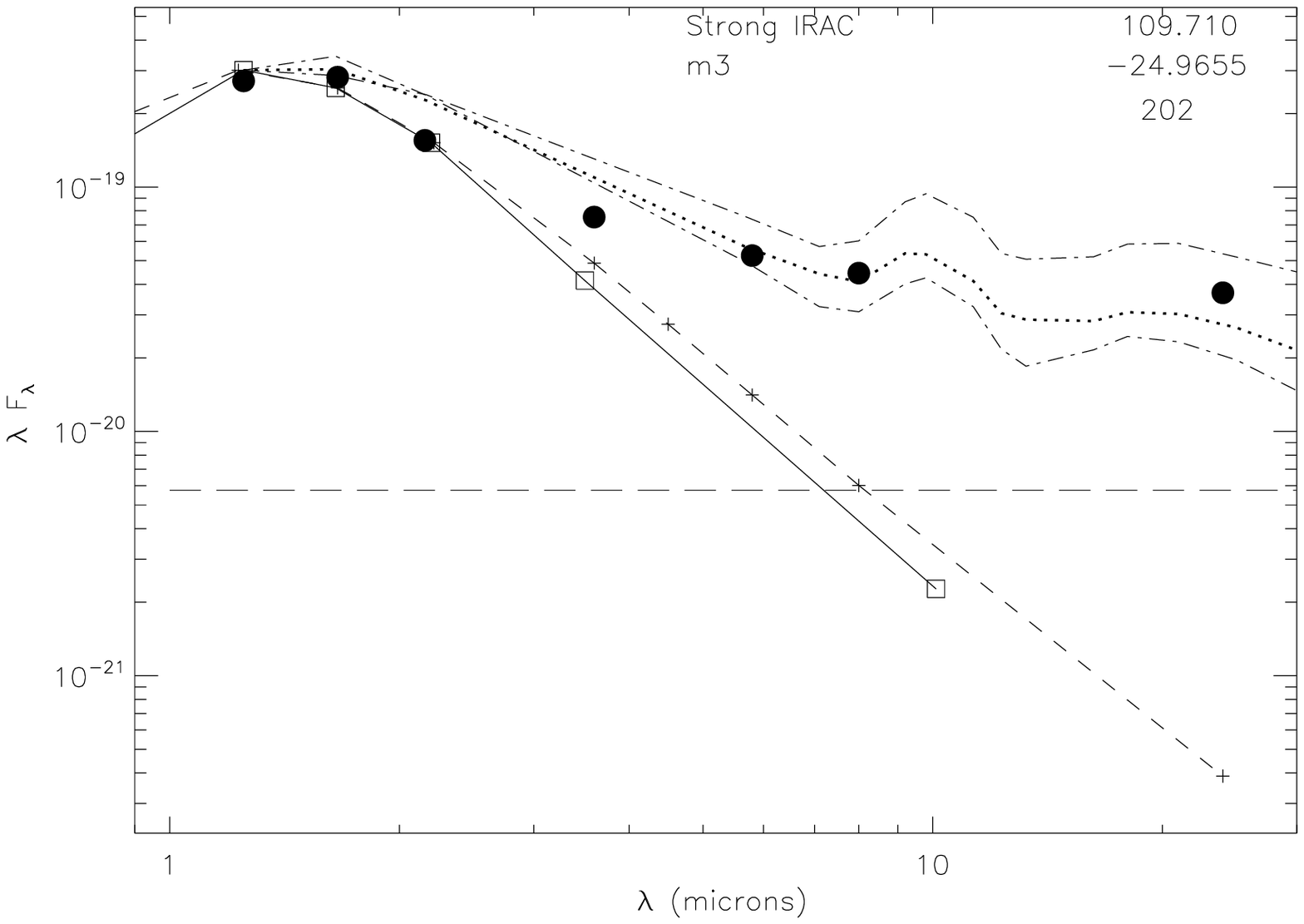}{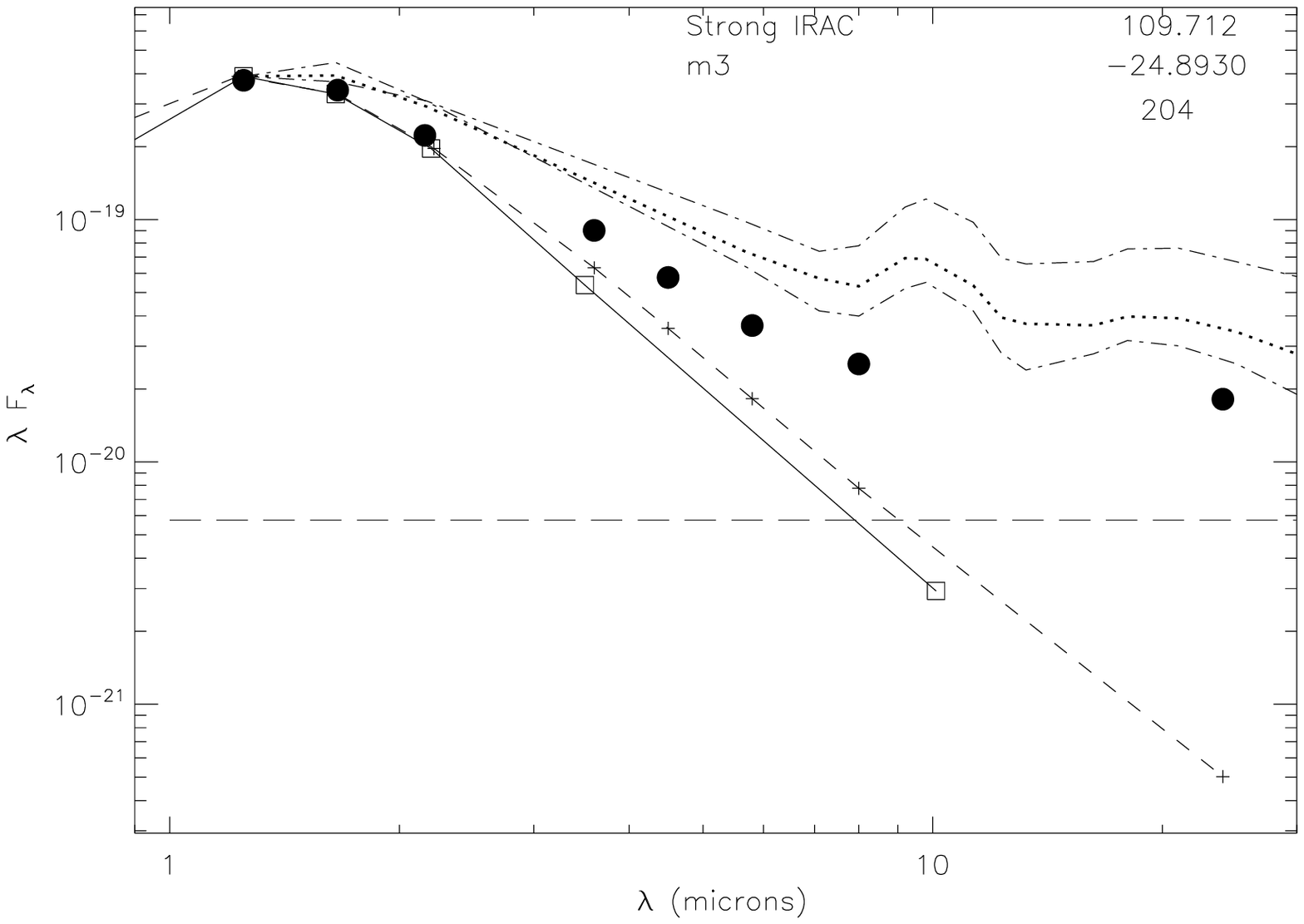}
\plottwo{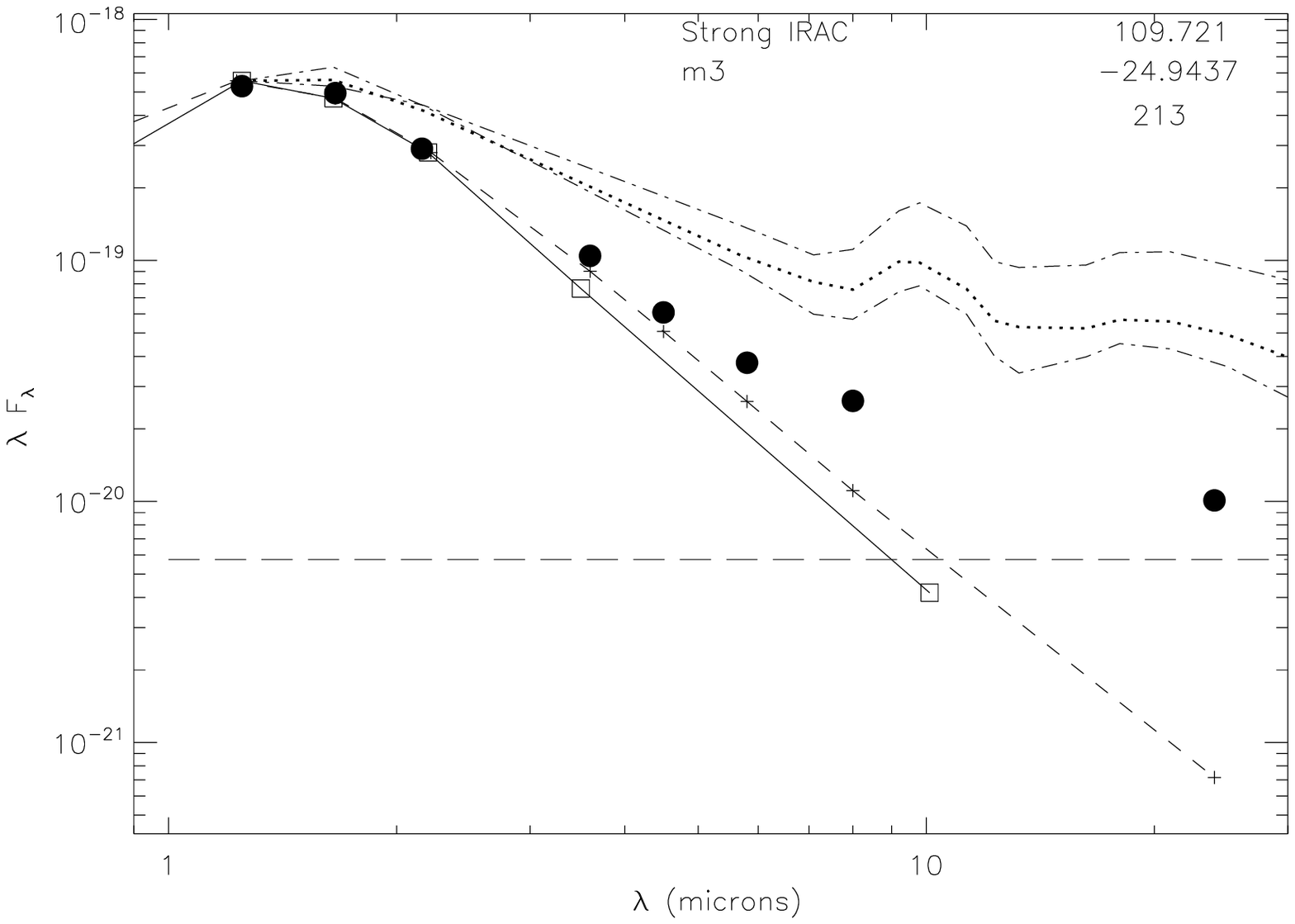}{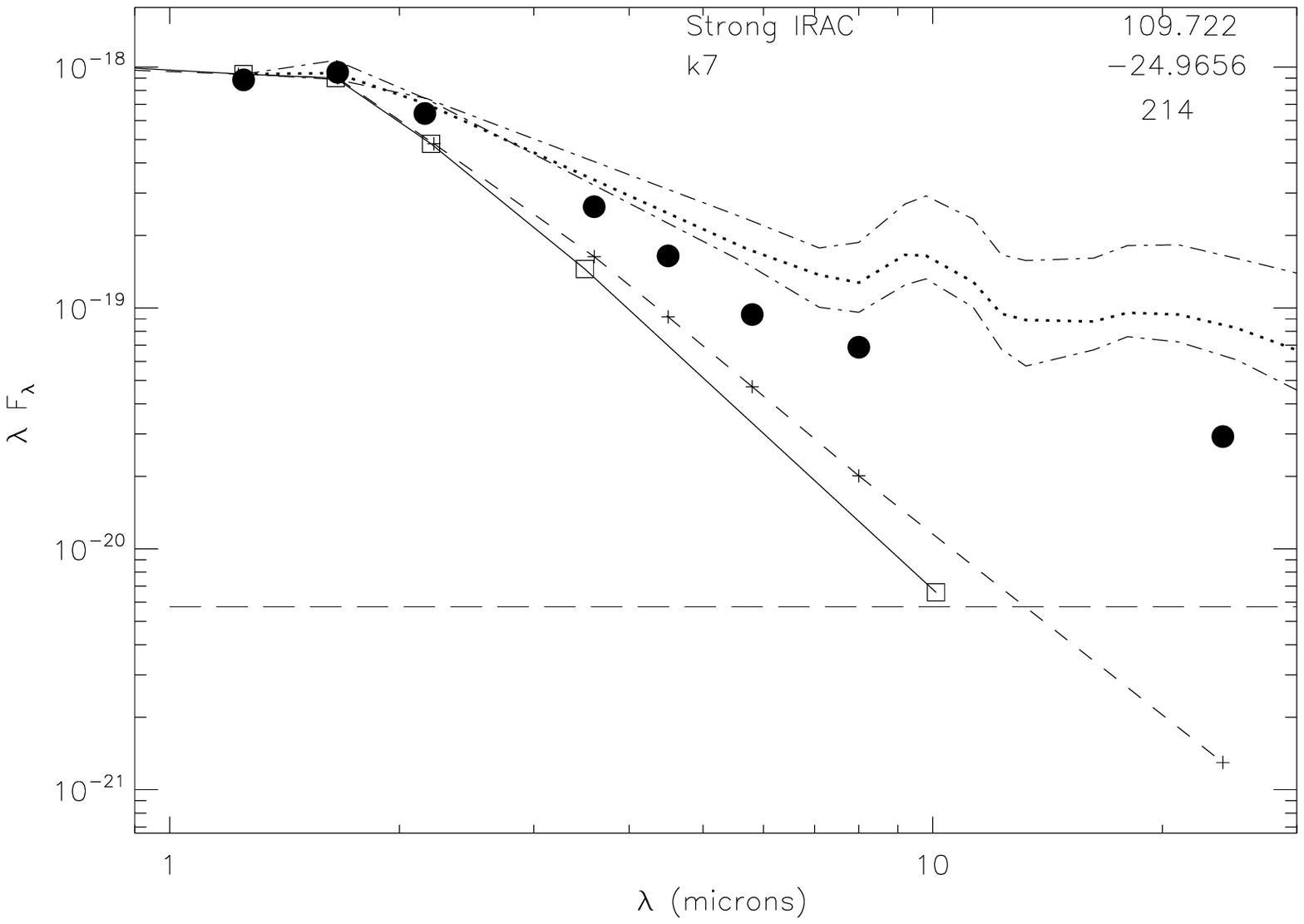}

\caption{Same as Figure \ref{seds}}
\label{seds2}
\end{figure}
\clearpage
\begin{figure}
\epsscale{0.99}
\centering
\plottwo{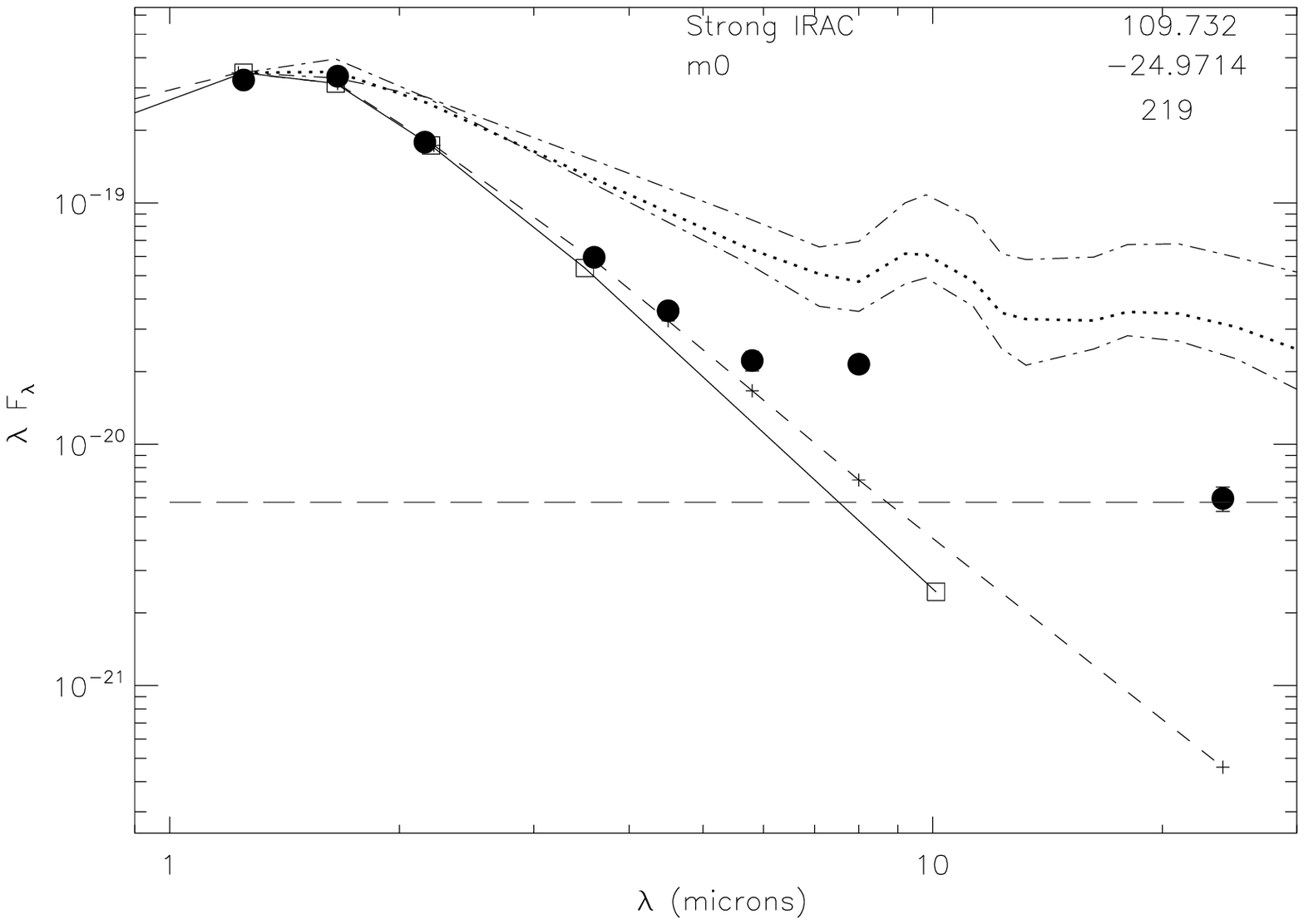}{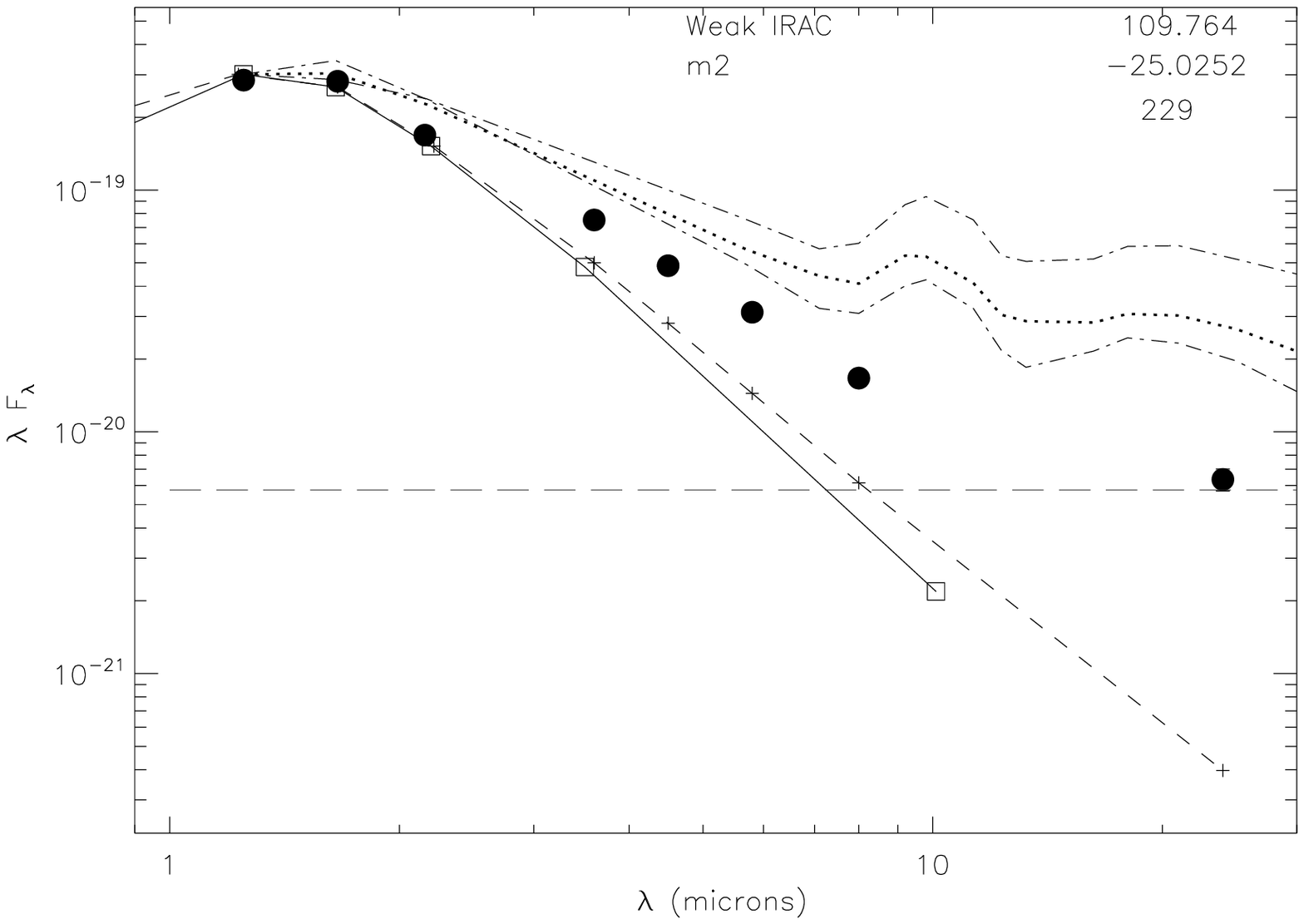}
\plottwo{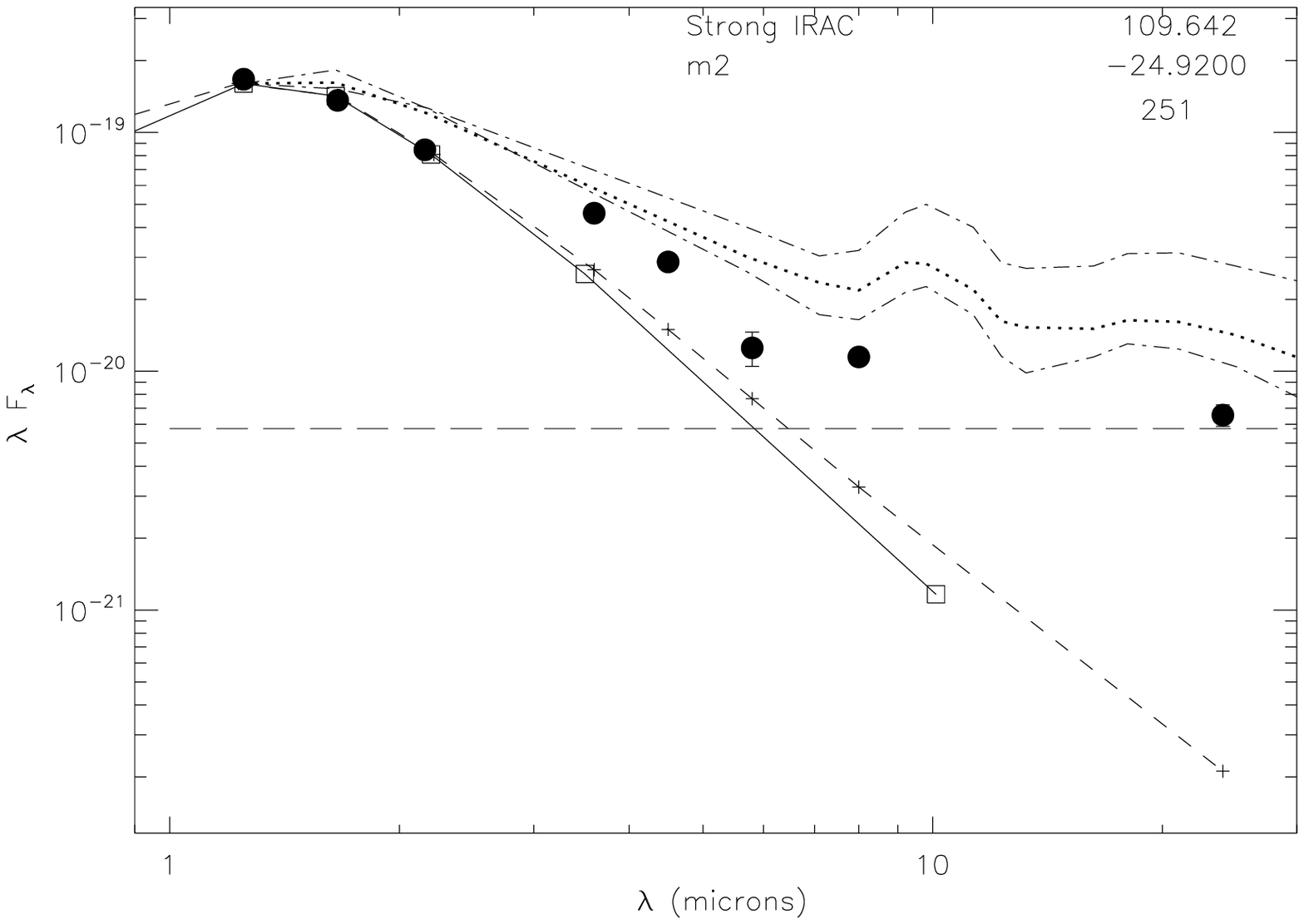}{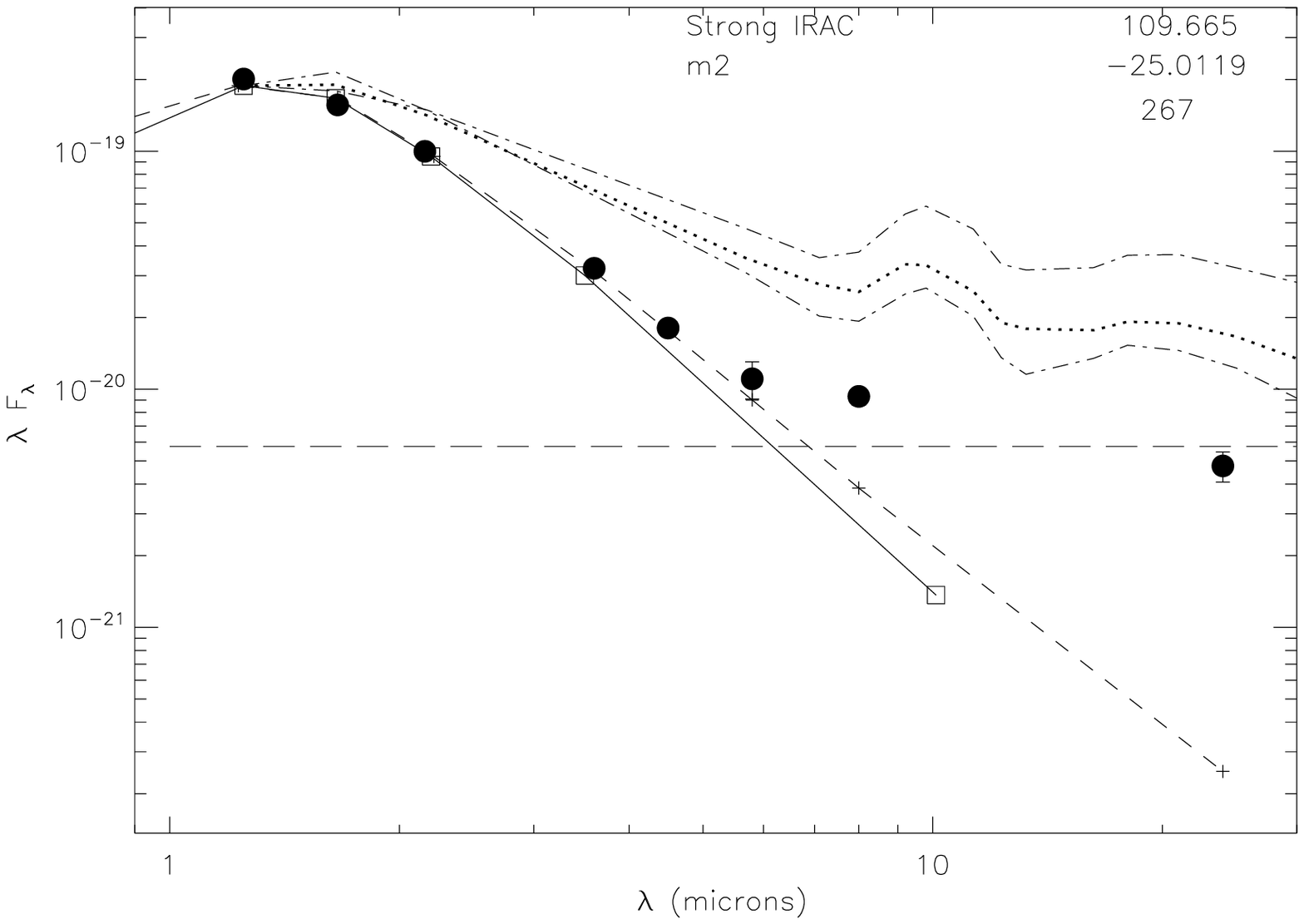}
\epsscale{0.5}
\plotone{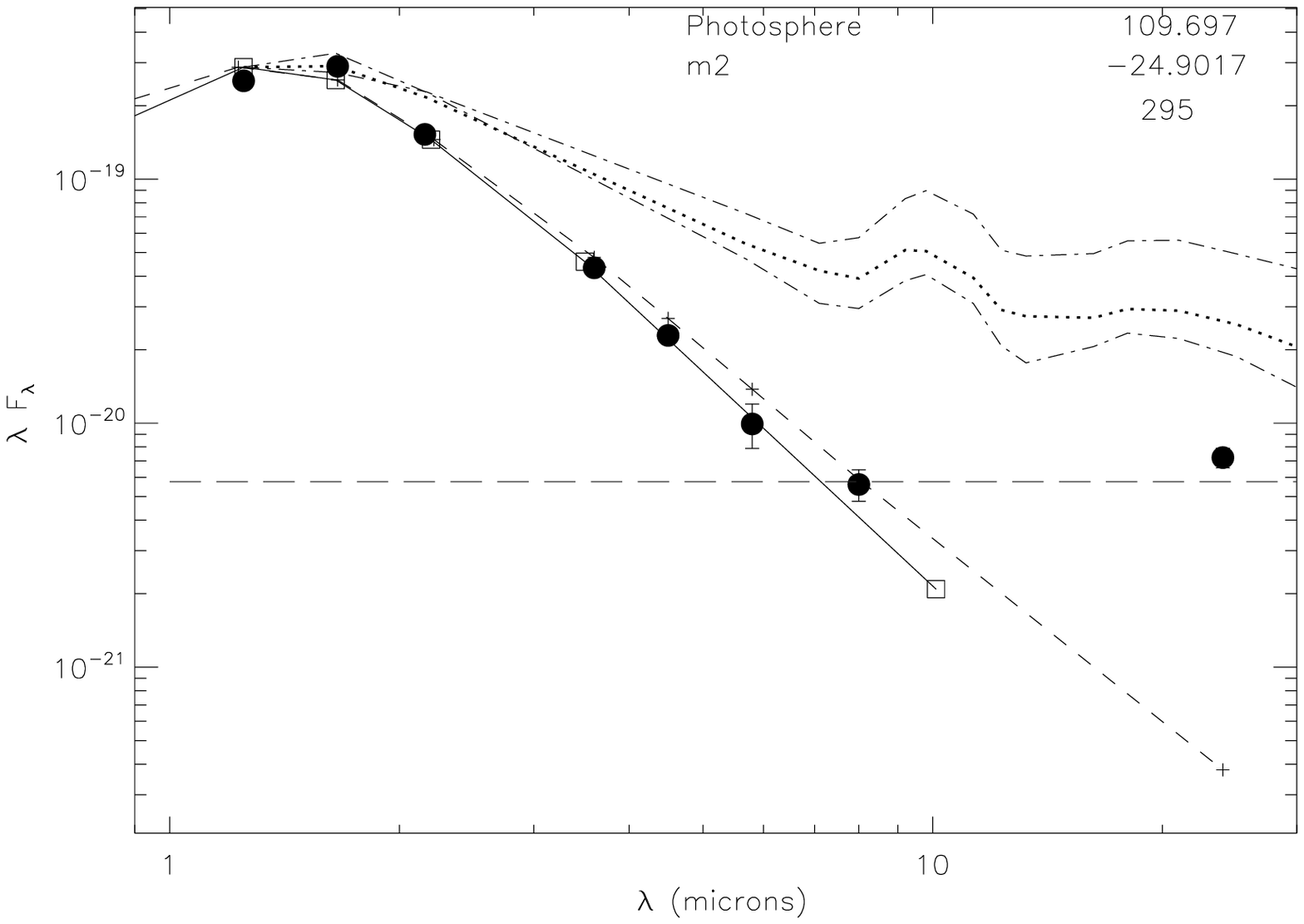}
\caption{Same as Figure \ref{seds} and \ref{seds2}}
\label{seds3}
\end{figure}
\clearpage
\epsscale{0.99}
\begin{figure}
\centering
\plottwo{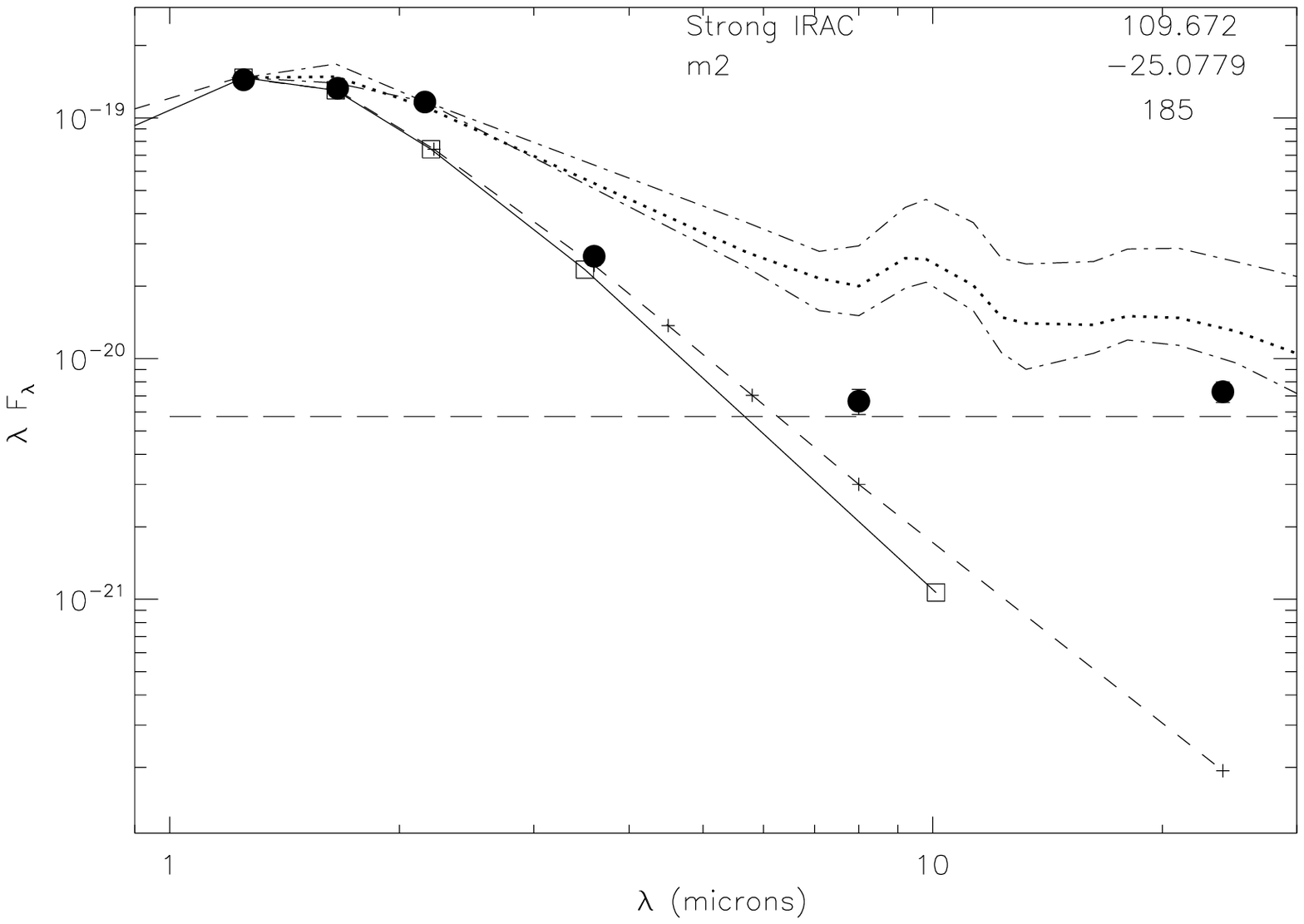}{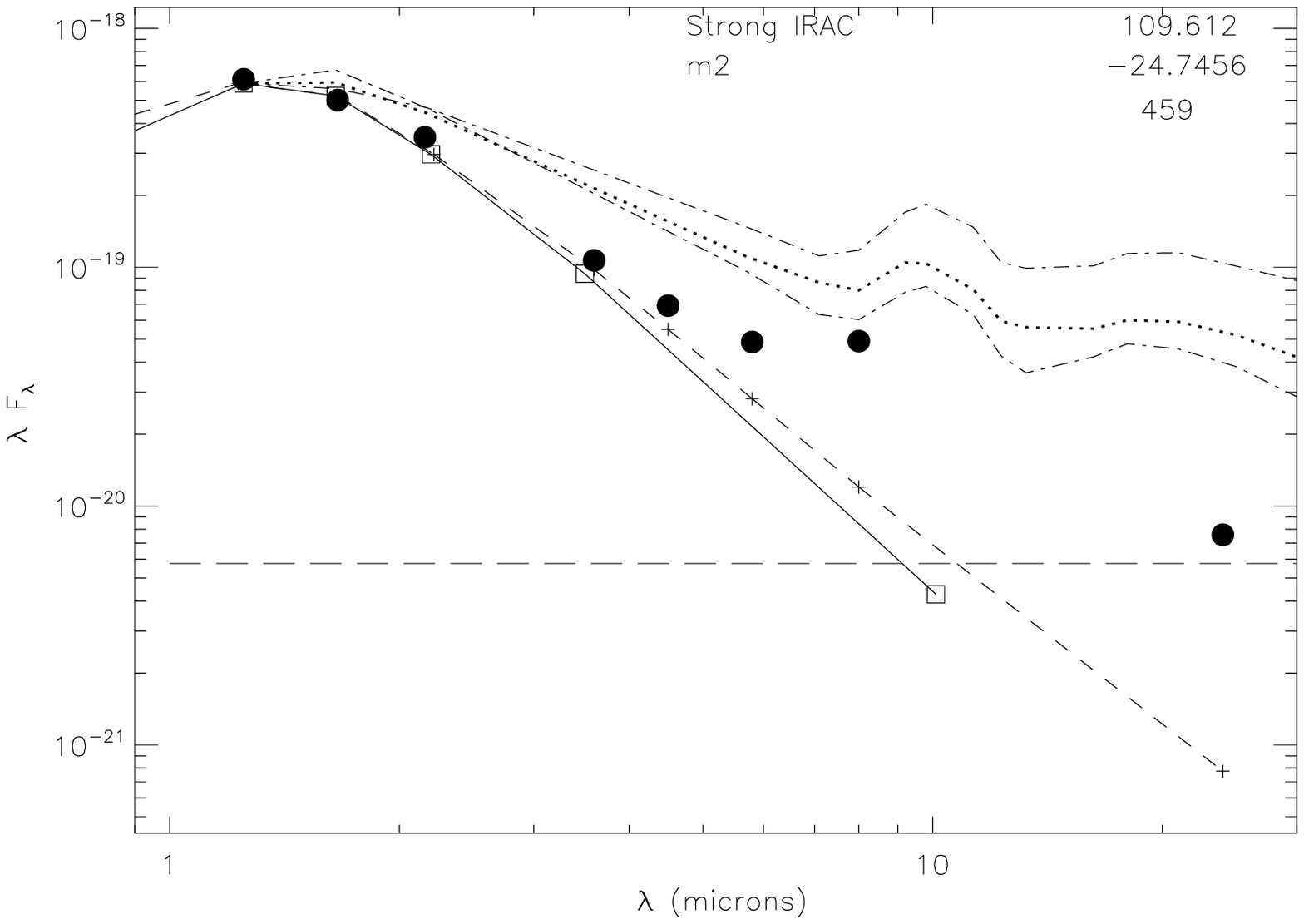}
\plottwo{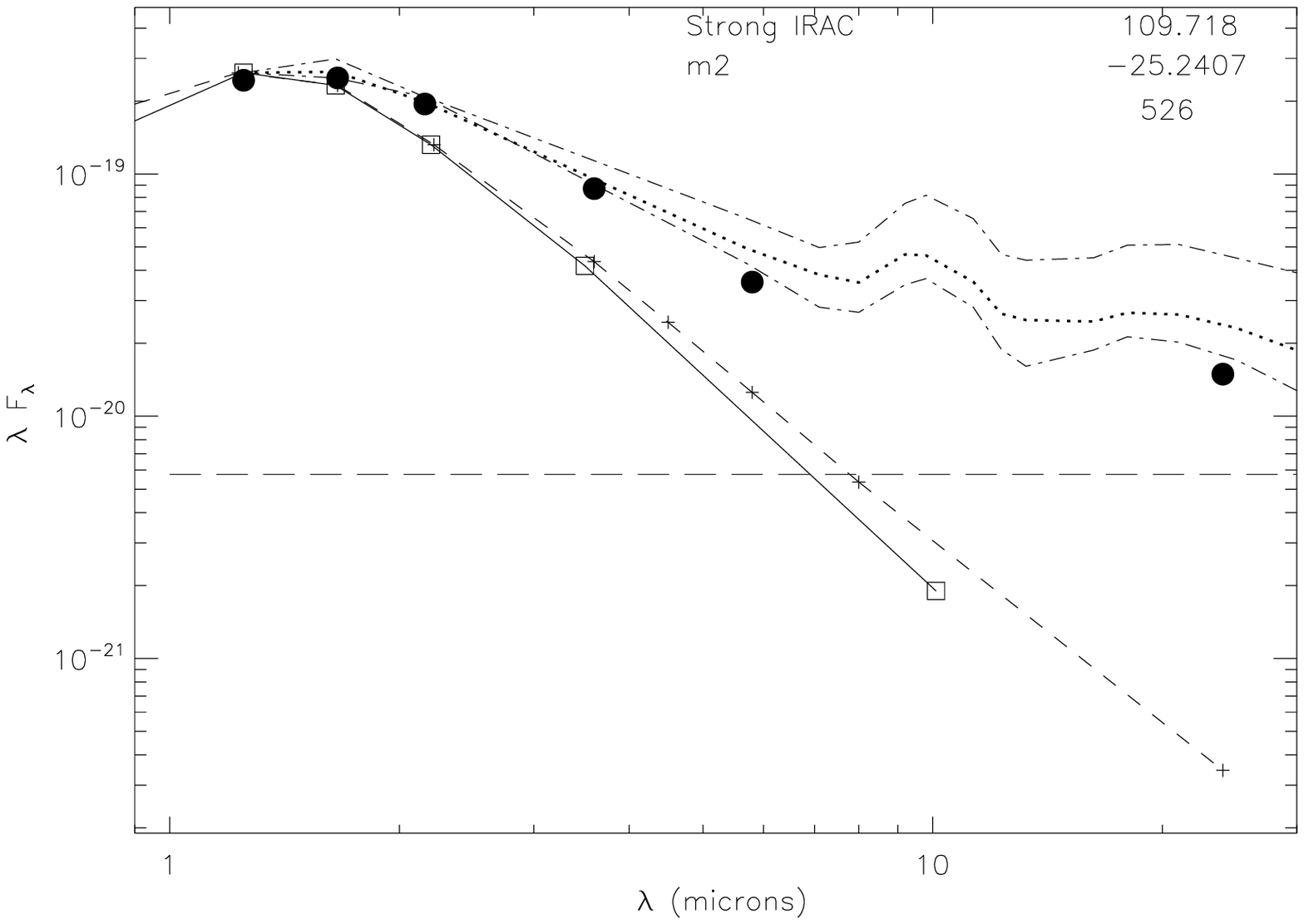}{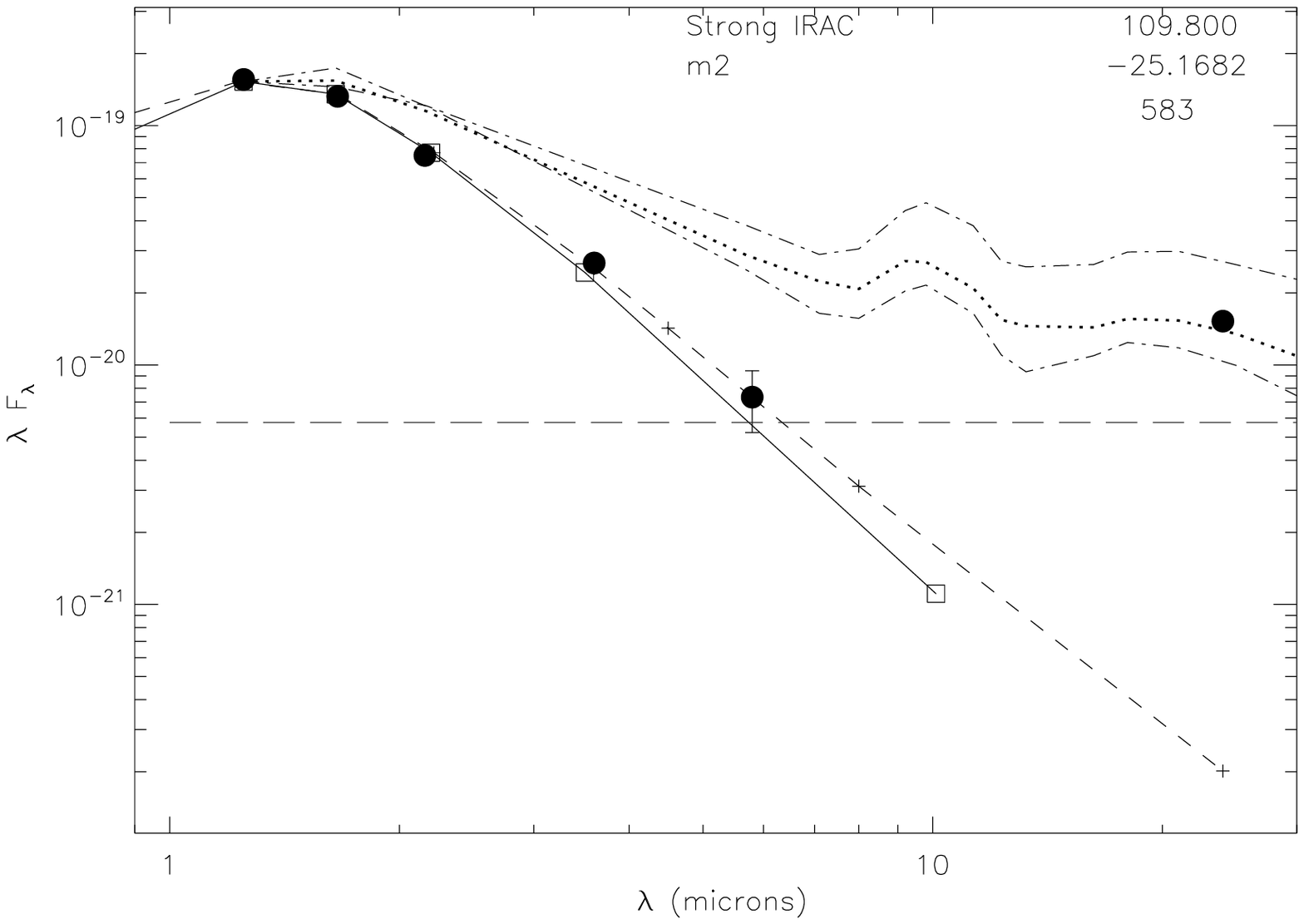}
\plottwo{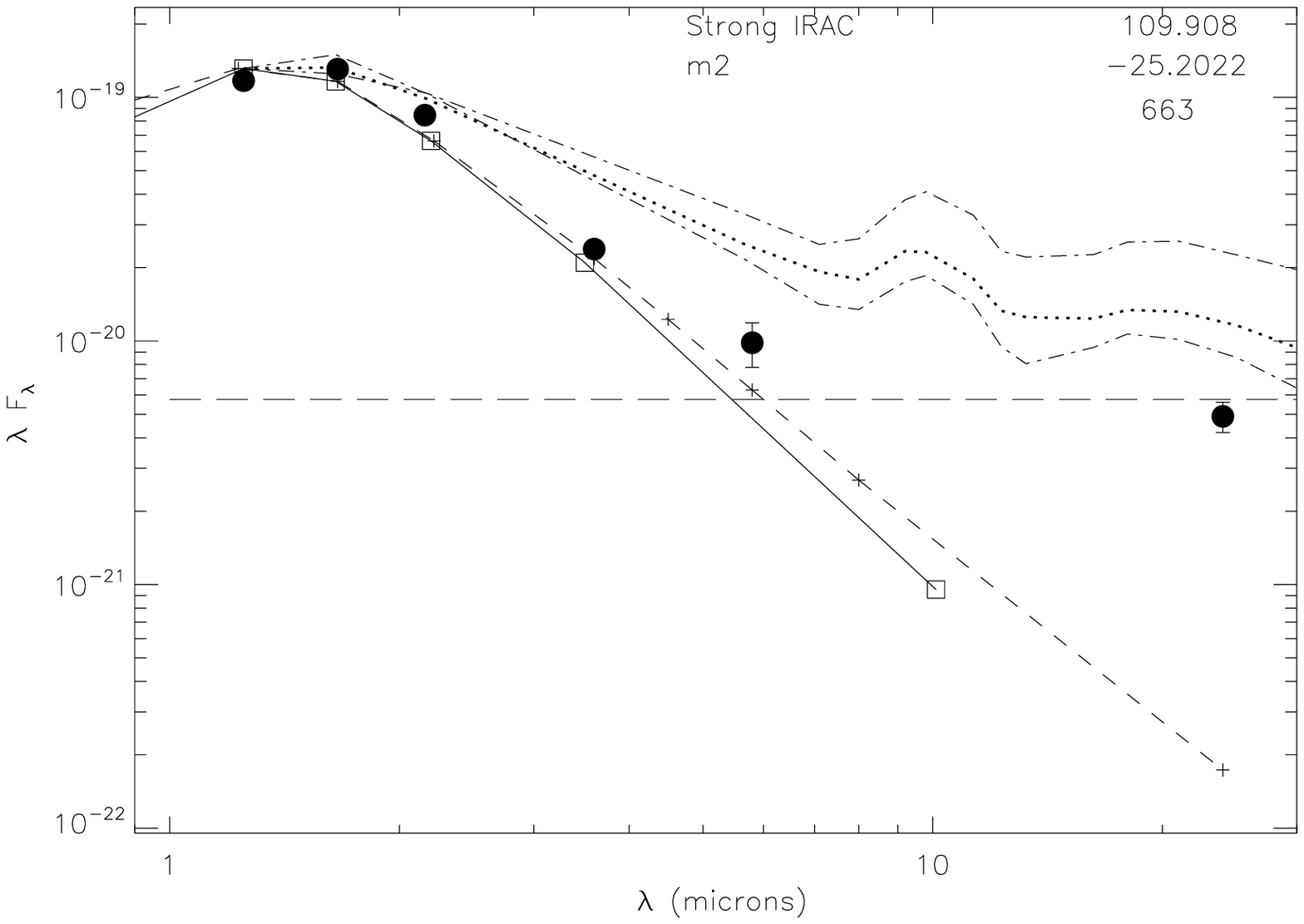}{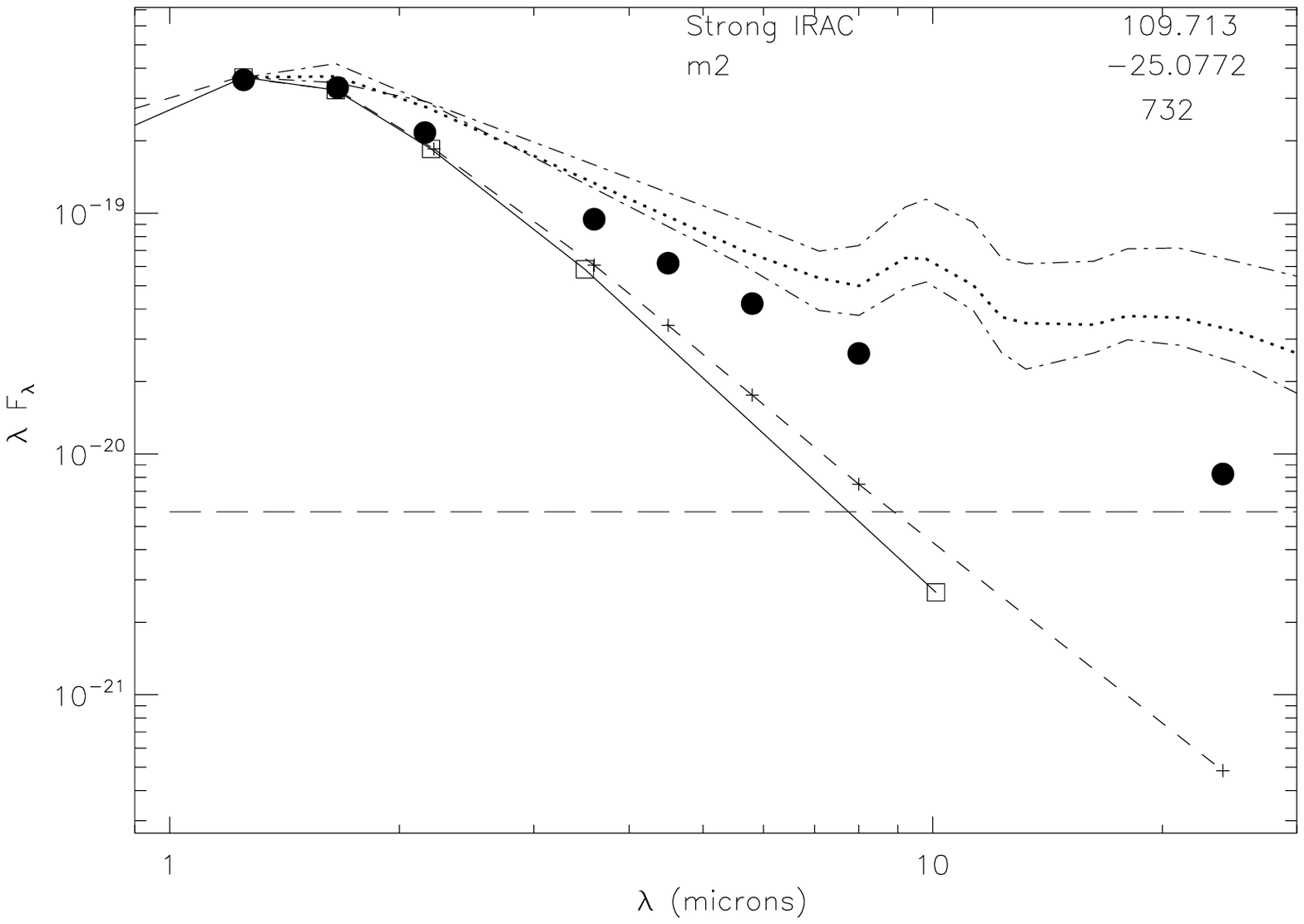}
\plottwo{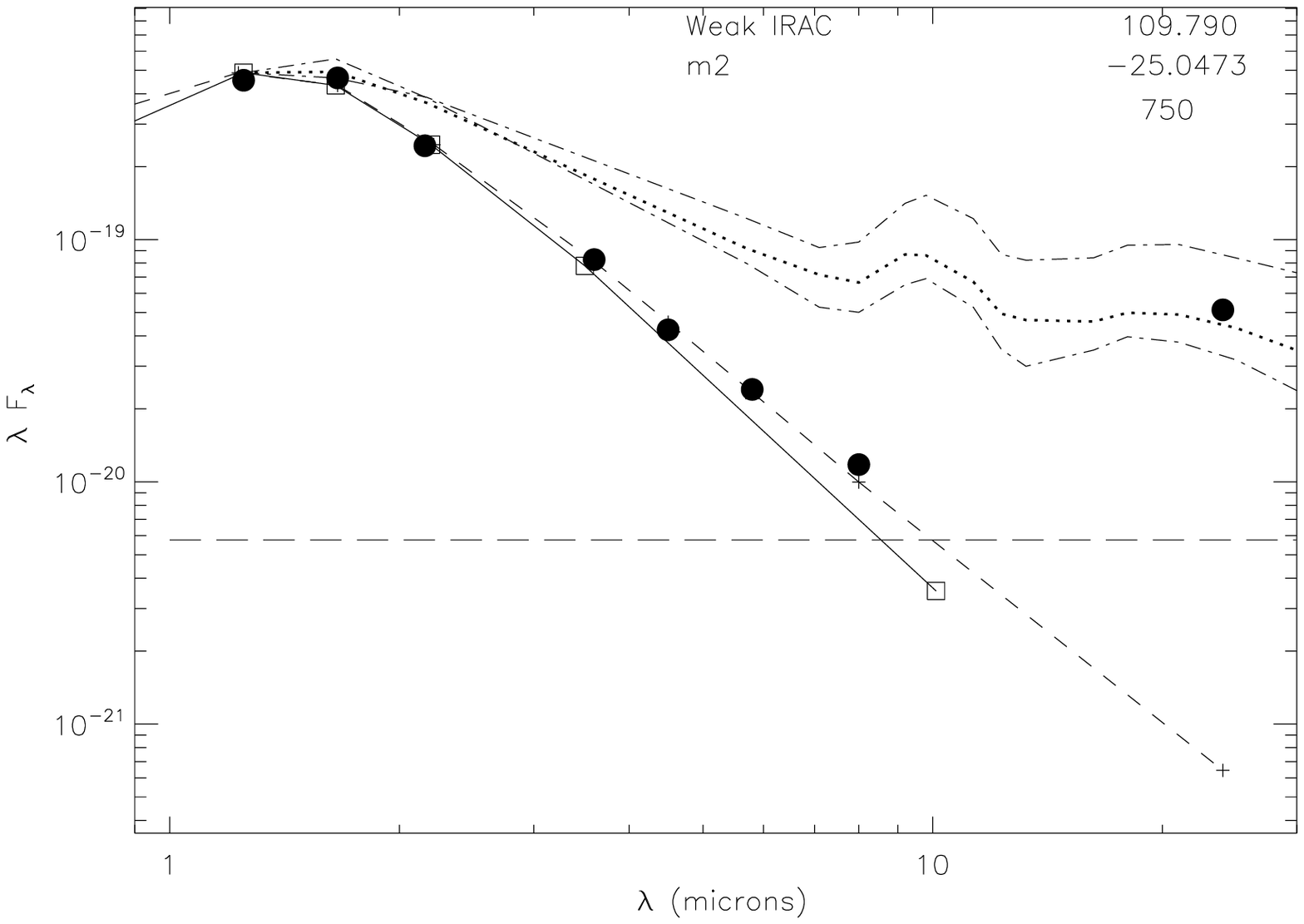}{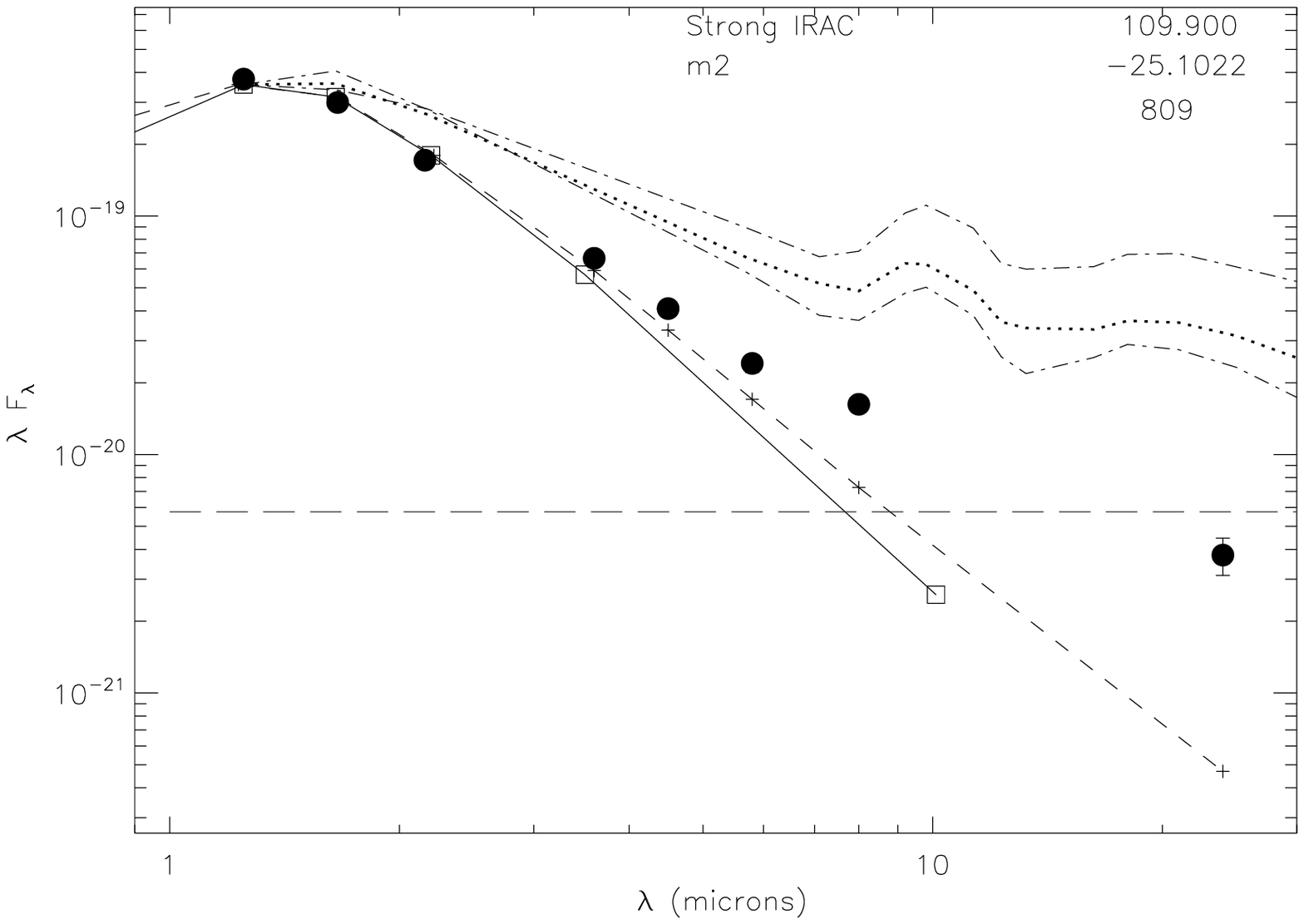}
\caption{Spectral energy distributions for MIPS-detected candidates from 
 the \citet{Ir08}.  Symbols are the same as in the preceeding figures.}
\label{sedsir1}
\end{figure}
\clearpage
\begin{figure}
\centering
\plottwo{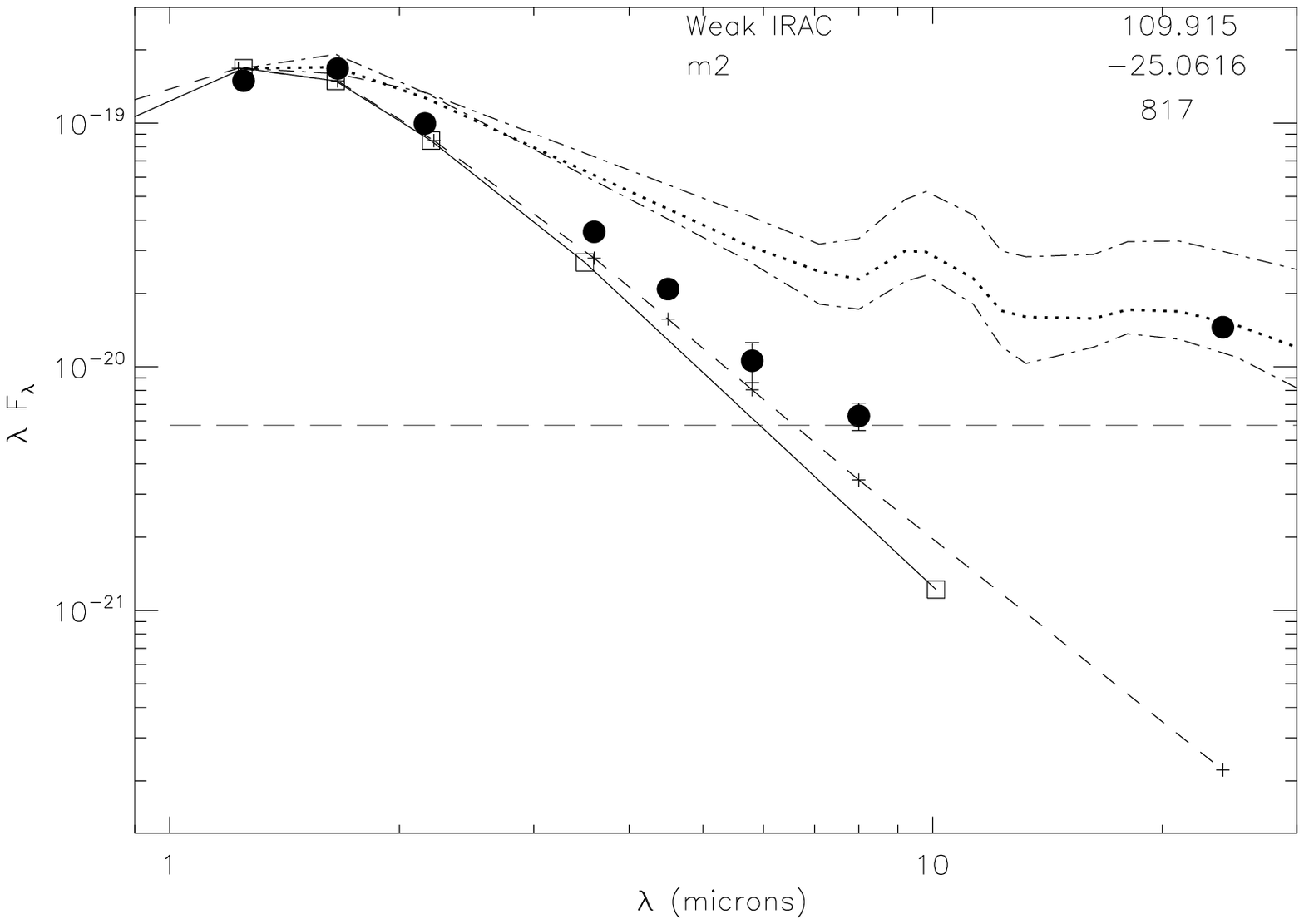}{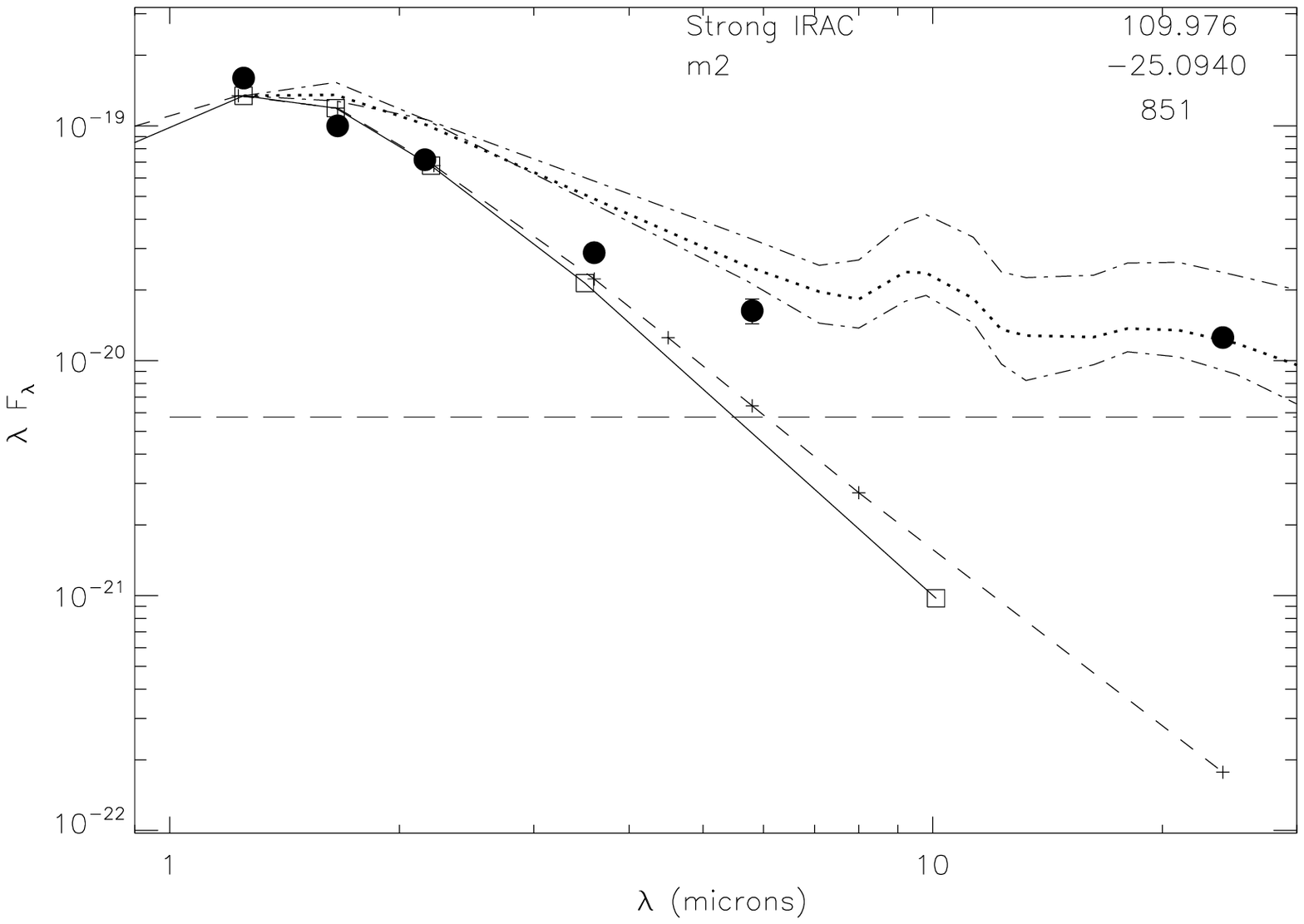}
\plottwo{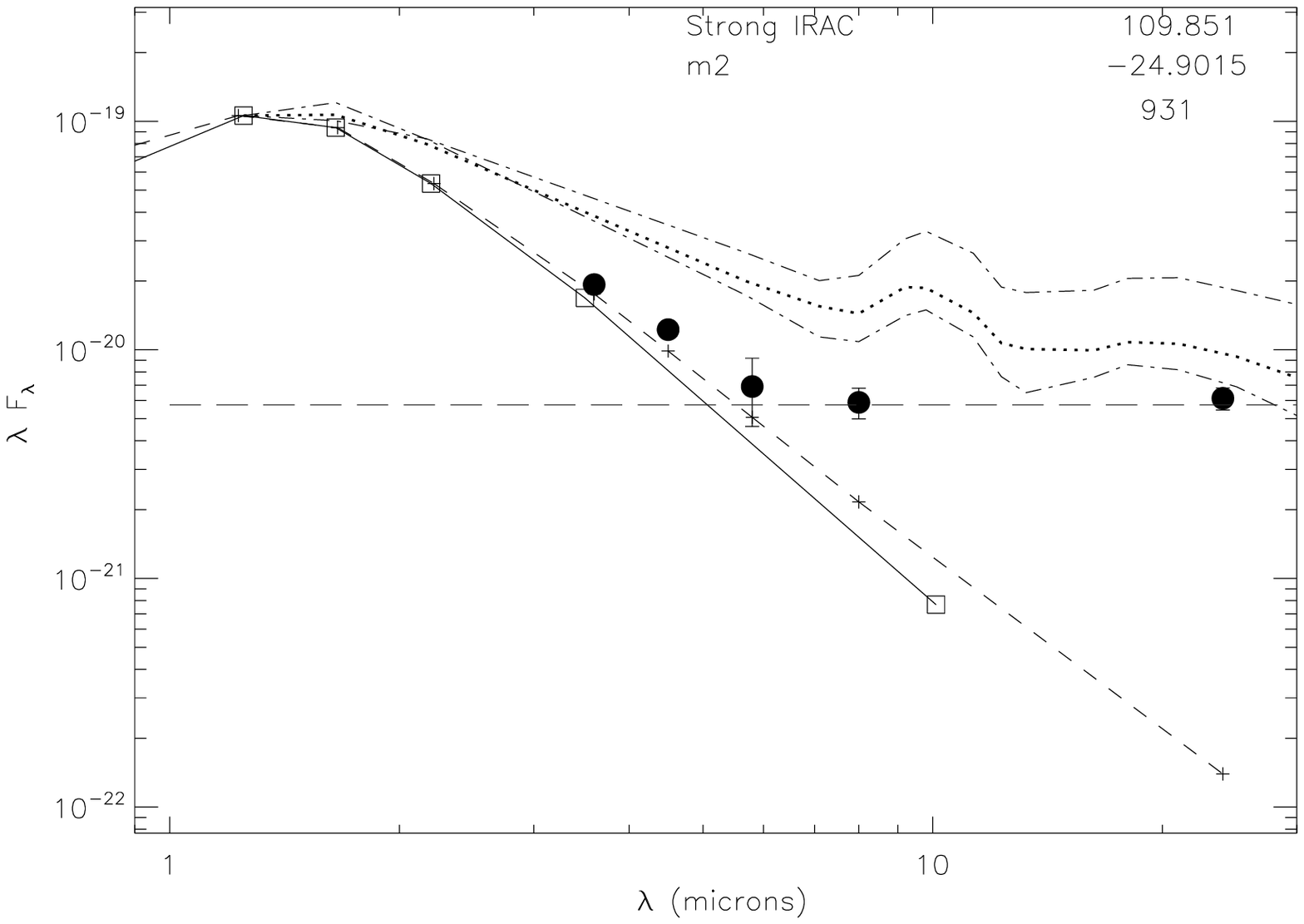}{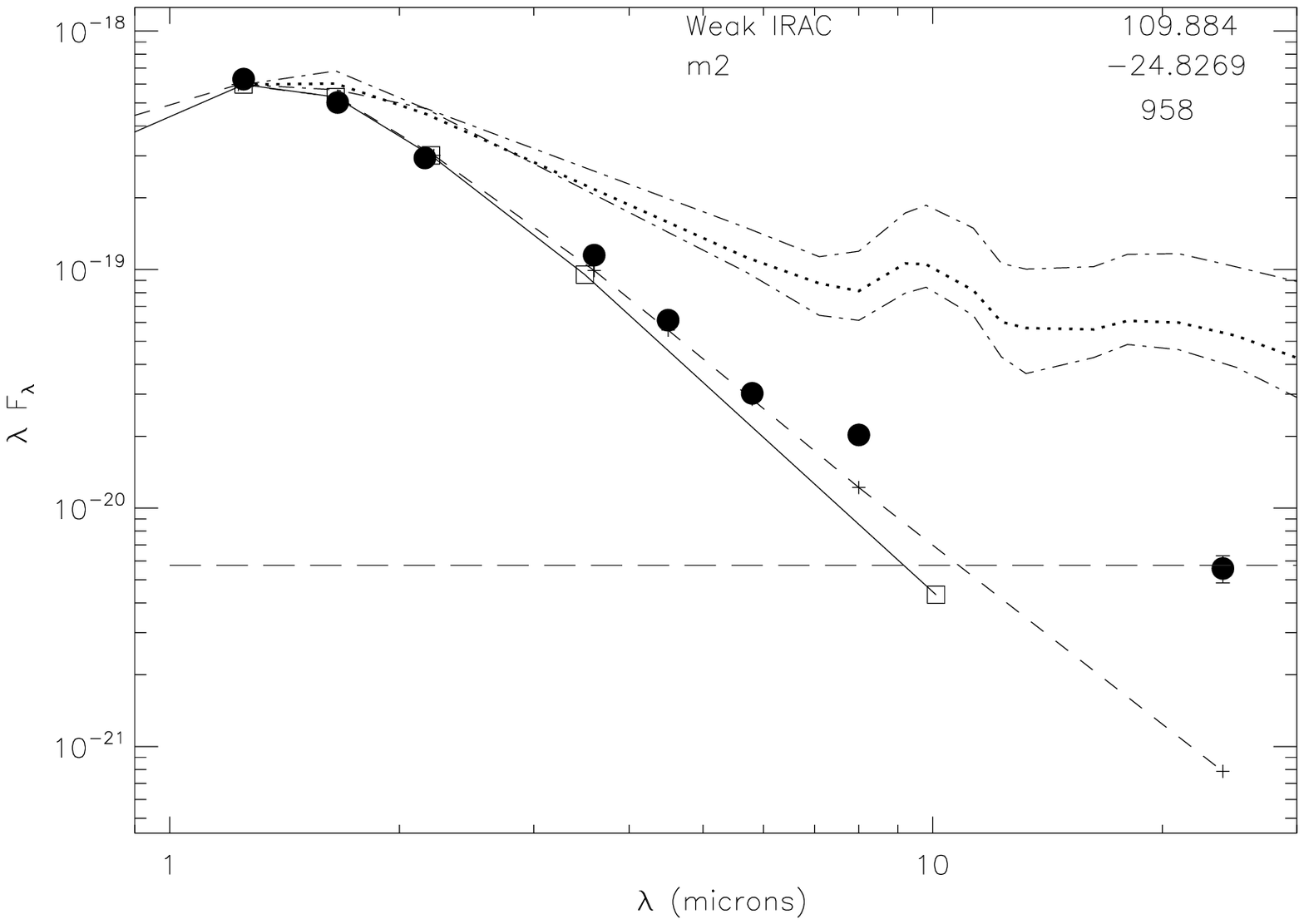}
\plottwo{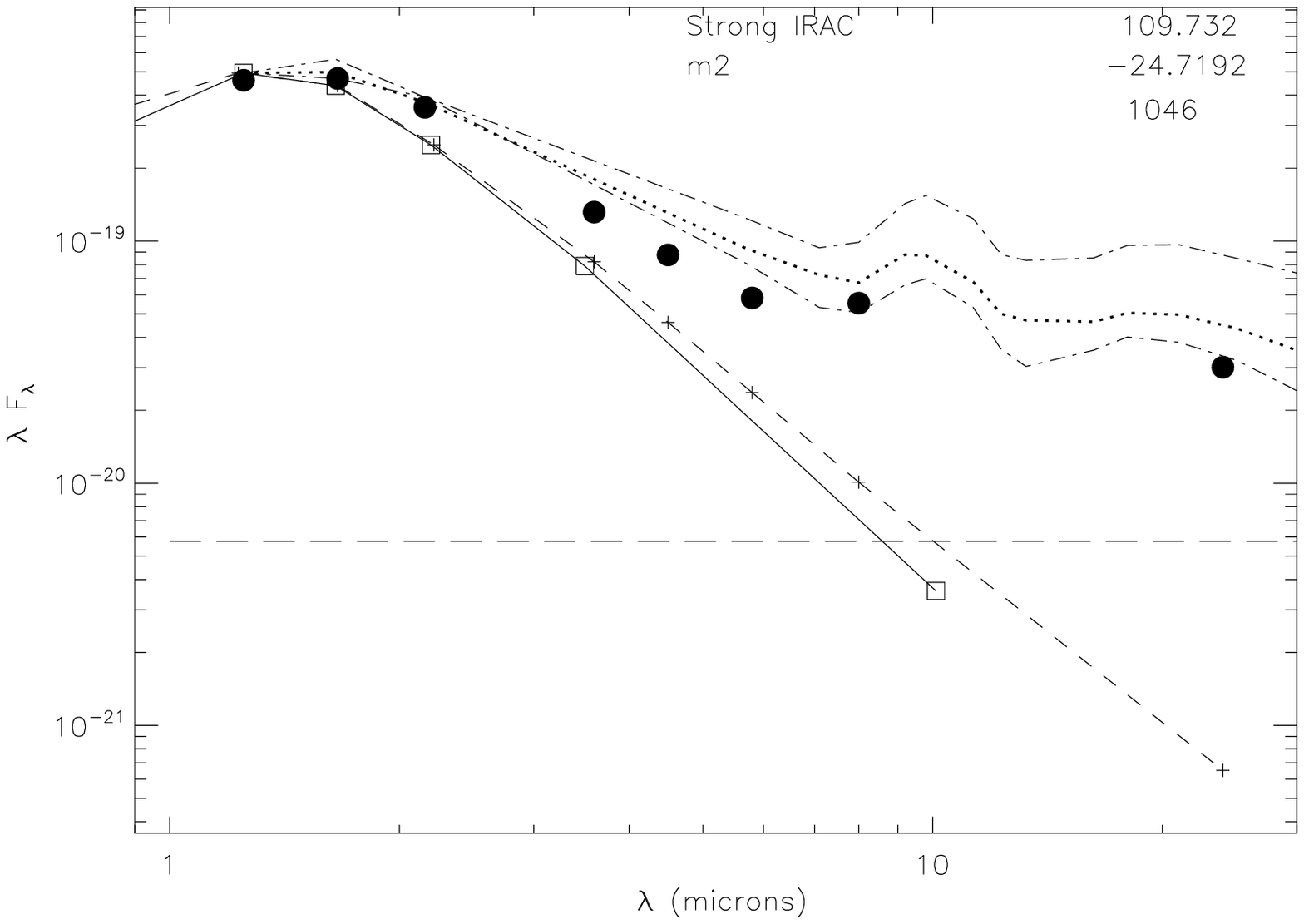}{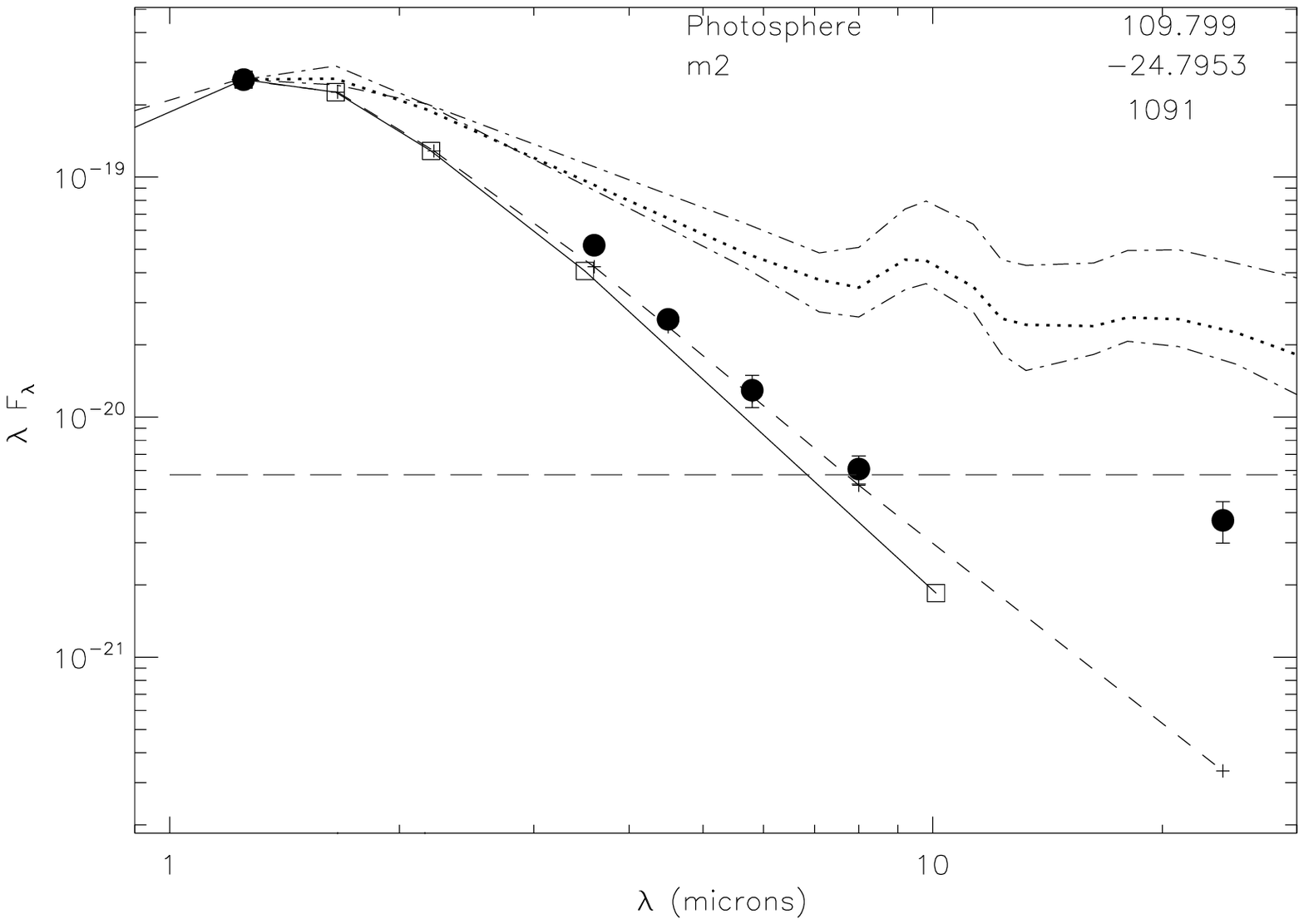}
\caption{Same as Figure \ref{sedsir1}.}
\label{sedsir2}
\end{figure}
\end{document}